\documentclass[submission, Phys,amsmath]{SciPost}

\usepackage{mathrsfs} \usepackage{amssymb} \usepackage{graphicx,color} \usepackage{bbm} \usepackage{multirow}
\usepackage{bbm} \usepackage{mathrsfs} \usepackage{commath}\usepackage{tikz}
\usetikzlibrary{shapes,arrows,positioning,calc}\usepackage{stmaryrd}\usepackage{bm}
\usepackage{mhchem,chemformula}

\hypersetup{
  colorlinks=true,
  linkcolor=blue,
  filecolor=magenta,
  citecolor=blue,
  urlcolor=blue,
}

\def\vS{\Vec{S}}

\newcommand{\ovS}{\overline{\Vec{S}}}

\newcommand{\eq}[1]{Eq.~(\ref{#1})}
\newcommand{\beq}{\begin{equation}} \newcommand{\eeq}{\end{equation}}
\newcommand{\beqn}{\begin{eqnarray}} \newcommand{\eeqn}{\end{eqnarray}}
\newcommand{\bmat}{\begin{mathdisplay}} \newcommand{\emat}{\end{mathdisplay}}
 
\newcommand{\Tr}{\mbox{Tr}}
\newcommand{\cD}{{\cal D}}

\renewcommand{\Vec}[1]{{\bf #1}}

\newcommand{\inti}{\int_{-\infty}^{\infty}}
\newcommand{\intii}{\int_{-i\infty}^{i\infty}}
\newcommand{\intS}{\int_{S}}

\newcommand{\bs}{\blacksquare}

\newcommand{\bc}{\begin{center}}
\newcommand{\ec}{\end{center}}

\begin{document}

\begin{center}{\Large \textbf{
      Disorder-free spin glass transitions and jamming
in exactly solvable mean-field models
}}\end{center}

\begin{center}
Hajime Yoshino\textsuperscript{1,2*}
\end{center}

\begin{center}
{\bf 1} Cybermedia Center, Osaka University, Toyonaka, Osaka 560-0043, Japan
\\
{\bf 2} Graduate School of Science, Osaka University, Toyonaka, Osaka 560-0043, Japan
\\
* yoshino@cmc.osaka-u.ac.jp
\end{center}

\begin{center}
\today
\end{center}


\section*{Abstract}
{\bf
    We construct and analyze a family of $M$-component vectorial spin systems which exhibit glass transitions
  and jamming within supercooled paramagnetic states without quenched disorder.
  Our system is defined on lattices with connectivity $c=\alpha M$ 
  and becomes exactly solvable in the limit of large number of components $M \to \infty$.
  We consider generic $p$-body interactions between the vectorial Ising/continuous
  spins with linear/non-linear potentials. The existence of self-generated
  randomness is demonstrated by showing that the random energy model is recovered from a $M$-component
  ferromagnetic $p$-spin Ising model in $M \to \infty$ and $p \to \infty$ limit.
  In our systems the quenched disorder, if present, and the self-generated disorder act additively.
  Our theory provides a unified mean-field theoretical framework for glass transitions
  of rotational degree of freedoms such as orientation of molecules in glass forming liquids,
  color angles in continuous coloring of graphs and
  vector spins of geometrically frustrated magnets. The rotational glass transitions
  accompany various types of replica symmetry breaking. In the case of repulsive hardcore
  interactions in the spin space, continuous the criticality of the jamming or SAT/UNSTAT transition
  becomes the same as that of hardspheres.
}

\vspace{10pt}
\noindent\rule{\textwidth}{1pt}
\tableofcontents\thispagestyle{fancy}
\noindent\rule{\textwidth}{1pt}
\vspace{10pt}

\section{Introduction}
\label{sec-Introduction}

Simple spin models often provide useful grounds to develop statistical mechanical approaches for various kinds of phase transitions.
For the glass transition \cite{angell2000relaxation,cavagna2009supercooled,berthier2011theoretical}, which is one of the most important open problem in physics,
a family of mean-field spinglass models called as the random energy model \cite{derrida1980random}
and $p$-spin spinglass models
\cite{derrida1980random,gross1984simplest,KT87,KW87b,KTW89,crisanti1992sphericalp,cugliandolo1993analytical,castellani2005spin}
 have played important roles. The concepts and techniques used in the spinglass theory have promoted substantial progress of the first principle theory for the glass transitions of supercooled liquids \cite{mezard1999first,parisi2010mean}. Most notably exact mean-field theory in the large dimensional limit was constructed recently for the hardspheres \cite{kurchan2012exact,kurchan2013exact,charbonneau2014exact,charbonneau2014fractal,YZ14,RUYZ14,rainone2016following} using the replica approach on the supercooled liquids \cite{mezard1999first,parisi2010mean}. 

There remains, however, a conceptual problem regarding the origin of the randomness.
The spinglass models\cite{edwards1975theory},
which have been developed originally by Edwards and Anderson to model a class of disordered and frustrated magnetic materials 
\cite{mydosh1993spin}, have quenched disorder which is apparently absent in glass forming liquids.
It is often emphasized in the studies of spinglass materials that both the quenched disorder and frustration
are important. However it is believed that somehow the disorder is self-generated in structural glasses which are born
out of supercooled liquid and thus the quenched disorder is not necessarily.
Early seminal works \cite{bouchaud1994self,marinari1994replica1,marinari1994replica2,lipowski2000slow,franz2001ferromagnet} have suggested
that self-generated randomness are actually realized in some spin models without quenched disorder. 
However a comprehensive understanding of the mechanism of the putative self-generated randomness and its possible relation to the quenched randomness in spinglass models is still lacking.

In order to shed a light on this issue, we explicitly develop and analyze a family of  mean-field
vectorial spin models. We show that they exhibit glass transitions within their supercooled paramagnetic phases without quenched disorder.
Our model consists of  $M$-component vectorial spins, which can take either the Ising $\pm 1$ or continuous values,  put on tree-like lattices
with connectivity $c=\alpha M$, which  becomes exactly solvable in the limit of large number of components $M \to \infty$.
We perform a unified study of the crystalline phase (e.g. ferromagnetic phase), supercooled paramagnetic phases and glassy phases
of the same model. We clarify the condition needed to ensure local stability of supercooled liquids and glasses
against crystallization.
We demonstrate in particular that the theoretical results of the random energy model \cite{derrida1980random}
and the $p$-spin spinglass models \cite{derrida1980random,gross1984simplest,crisanti1992sphericalp}
can be fully recovered from a  $M$-component $p$-spin models with
purely ferromagnetic interactions within their supercooled paramagnetic phases.
This proves the existence of the self-generated randomness in our models.
In a sense this observation strengthen the view that the $p$-spin spinglass models are good
caricature spin models for glass transitions \cite{cavagna2009supercooled,castellani2005spin} because the quenched disorder is actually not needed.
We show that the quenched disorder, if present, add on top of the self-generated randomness.

\begin{figure}[t]
  \bc
  \includegraphics[width=\textwidth]{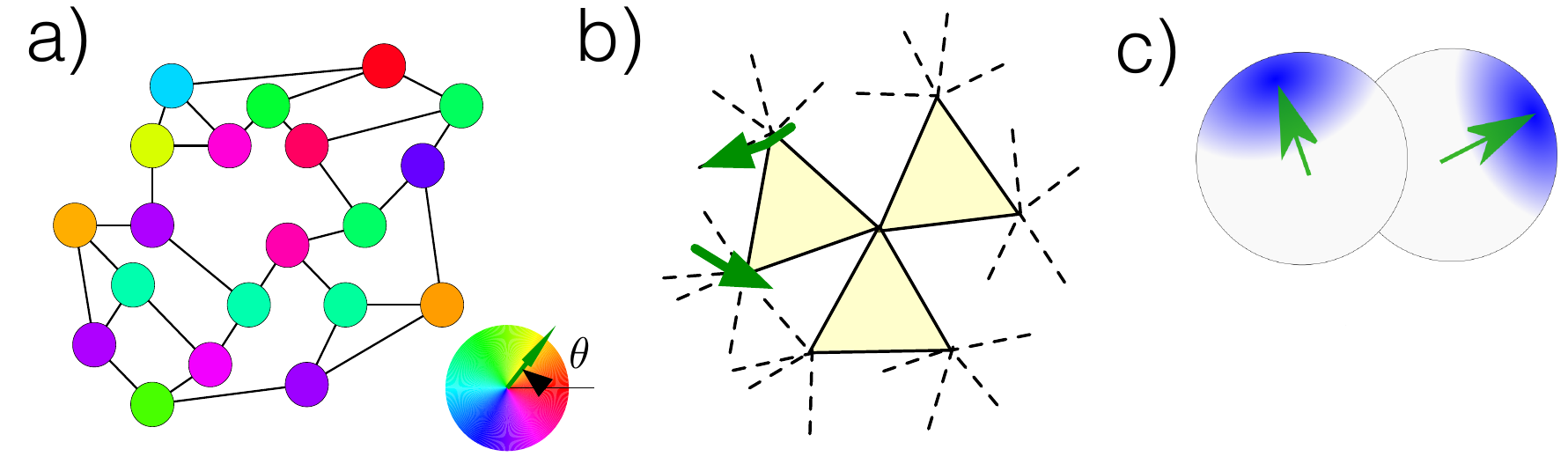}
  \ec
 \caption{Glassy systems carrying 'spins' representing rotational degree of freedoms.
   a) {\bf Continuous coloring of a graph}:
   The color angle $0 < \theta < 2\pi$, as in the standard HSV color map,
   can be represented by a XY spin, i.e.
    a vector with $M=2$ component (green arrow).
       The example shown here is a solution to the requirement
    that color angle on adjacent vertexes must be greater than
    or equal to $2\pi/3$.
 b) {\bf Geometrically frustrated magnets}:
    vectorial spins (green arrows) with anti-ferromagnetic couplings
    on adjacent vertexes on
    corner sharing triangles (e.g. kagome lattice), tetrahedra (e.g. pyrochlore lattice). The ground states are highly degenerate due to the loose connectivity of the lattices.
    c) {\bf Glass forming liquids of molecules or colloidal particles with 'spins'}:
    a simple molecule or a colloidal particle, like the Janus particle,
is symmetric under rotation around an axis whose direction can be specified by a spin.
 }
  \label{fig_rotational_glass}
\end{figure}

Glass transition of the rotational or spin degrees of freedom is an important problem by itself
and can be found not only in the spinglasses but also in many other real systems.
It should be noted first that most of the molecules and colloidal particles in glass-forming liquids
are not simply spherical but have rotational degrees of freedom because of their shapesq or patches on their surfaces (see Fig.~\ref{fig_rotational_glass}c)) and the rotational degree of freedom can exhibit glass transitions simultaneously or separately from that of the
translational degrees of freedom. Sometimes the rotational degrees of freedoms alone exhibit glassiness
on top of crystalline long-ranged order of the translational degrees of freedom.
This happens for instance in the so called plastic crystals where the rotations of molecules slow down and eventually
exhibit glass transitions \cite{suga1974thermodynamic}. Another important problem is the spinglass transition found in 
frustrated magnets but without quenched disorder (Fig.~\ref{fig_rotational_glass}b))\cite{schiffer1995frustration,gingras1997static}.
Possibilities of disorder-free spinglass transitions have been a matter of long debate in the field of frustrated magnets.
We expect our results provide a useful basis to tackle these problems theoretically.

Within our formalism we consider $p$-body interactions through generic non-linear potentials.
In particular we apply the scheme to the case of a $M$-component continuous spins interacting with each other 
through a hardcore potential which enables jamming transition of the vectorial spins. Here jamming means to loose thermal fluctuations by tightening the constraints.
This is relevant in the continuous constrained satisfaction problems such as the circular coloring of graphs or periodic scheduling \cite{zhu2001circular} (Fig.~\ref{fig_rotational_glass} a)): the problem is to put continuous colors parametrized by ``color angle'' $0 < \theta < 2\pi$  on the vertexes of a given graph such that angles on adjacent vertexes are sufficiently separated from each other. This is exactly a continuous version of the usual coloring problem where one is allowed to use only discrete colors like red, green and blue \cite{zdeborova2007phase,mezard2009information}. Remarkably
a recent study has shown that a discretized version of the circular coloring problem exhibits a complex free-energy landscape reminiscent of continuous replica symmetry breaking  \cite{schmidt2016circular}.

Increasing the coordination number $c$ of the graph,
the solution space exhibit clustering transition (glass transition) and eventually SAT/UNSAT transition (jamming) above which one cannot find a solution which satisfies the constraints.
Given the continuous variables, an interesting question is the universality class of the SAT/UNSAT transition.
Closely following the analysis done on hardspheres in the $d \to \infty$ limit
\cite{charbonneau2014exact},
we will show that the jamming criticality of our model belong indeed to the same universality of the hardspheres. Our result extends the result on the perceptron problem \cite{franz2016simplest,franz2015universal,franz2017universality} which can be regarded as a special case $p=1$ of our models.

The organization of this paper is as follows. In sec. \ref{sec-model} we introduce a family of large $M$-component vectorial Ising/continuous spin models with a generalized $p$-body interaction described by linear/non-linear  potentials.
We introduce a disorder-free model that has no quenched disorder and also a model which interpolates between the disorder-free model
and a fully disordered spinglass model.
In sec. \ref{sec-Supercooled-spin-liquid states} we discuss possible crystalline orderings in our disorder-free models and possibility to realize supercooled
paramagnetic states, which are crucial as the basis for glass transitions to take place without the quenched disorder.
In sec. \ref{sec-REM} we show that the random energy model can be recovered from a $M$-component $p$-spin Ising ferromagnetic model
with a linear potential in the limit $M \to \infty$ and $p \to \infty$. This demonstrates the  presence of self-generated randomness in our models.
In sec.  \ref{sec-Replicated-system} we derive the replicated free-energy functional in terms of the glass and crystalline order parameters.
We also discuss stability of the supercooled paramagnetic state and the glassy states against crystallization.
In sec.  \ref{sec-linear} we establish the connection between our model with linear potential and the standard $p$-spin spinglass models.
Then in the subsequent sections, we turn to study glassy phases of our model with non-linear potentials limiting our selves to the case of continuous spins.
In sec.  \ref{sec-RS} we discuss some general results within the replica symmetric (RS) ansatz.
In sec.  \ref{sec-RSB} we discuss some general results within 1 step and continuous replica symmetry breaking (RSB) ansatz.
In sec.  \ref{sec-quadratic-potential} we analyze the model with a quadratic potential as the simplest case of non-linear potential.
In sec.  \ref{sec-hardcore} we analyze in detail the model with a hardcore potential which exhibit jamming.
Finally in sec.  \ref{sec-Conclusions} we conclude this paper with some summary and remarks.
Some technical details are reported in the appendices.

\section{Vectorial spin model}
\label{sec-model}

\subsection{Generic model}
\label{sec-generic-model}

Let us now introduce the models that we study in this paper.
We consider vectorial spins with $M$ components $\vS_{i}=(S_{i}^{1},S_{i}^{2},\dots,S_{i}^{M})$ ($i=1,2,\ldots,N$) normalized such that
\beq
|\vS|^{2}=\sum_{\mu=1}^{M}(S^{\mu})^{2}=M.
\label{eq-spin-normalization}
\eeq
More specifically we consider two types of spins,
\begin{itemize}
\item {\bf Ising spin}

  $M$-component Ising spin with $S_{i}^{\mu} \in (-1,1)$ for $\mu=1,2,\ldots,M$.

\item {\bf Continuous spin}
  $M$-component continuous spin with length $|\vS|=\sqrt{M}$ which can continuously rotate in the $M$-dimensional space.
  It is known in some models that this case is closely related to
  the 'spherical model' which has just $M=1$ component spins $S_{i}$
  normalized by a global constraint $\sum_{i=1}^{N} S^{2}_{i}=N$
  \cite{stanley1968spherical,de1978infinite}.

\end{itemize}

\begin{figure}[t]
  \bc
  \includegraphics[width=\textwidth]{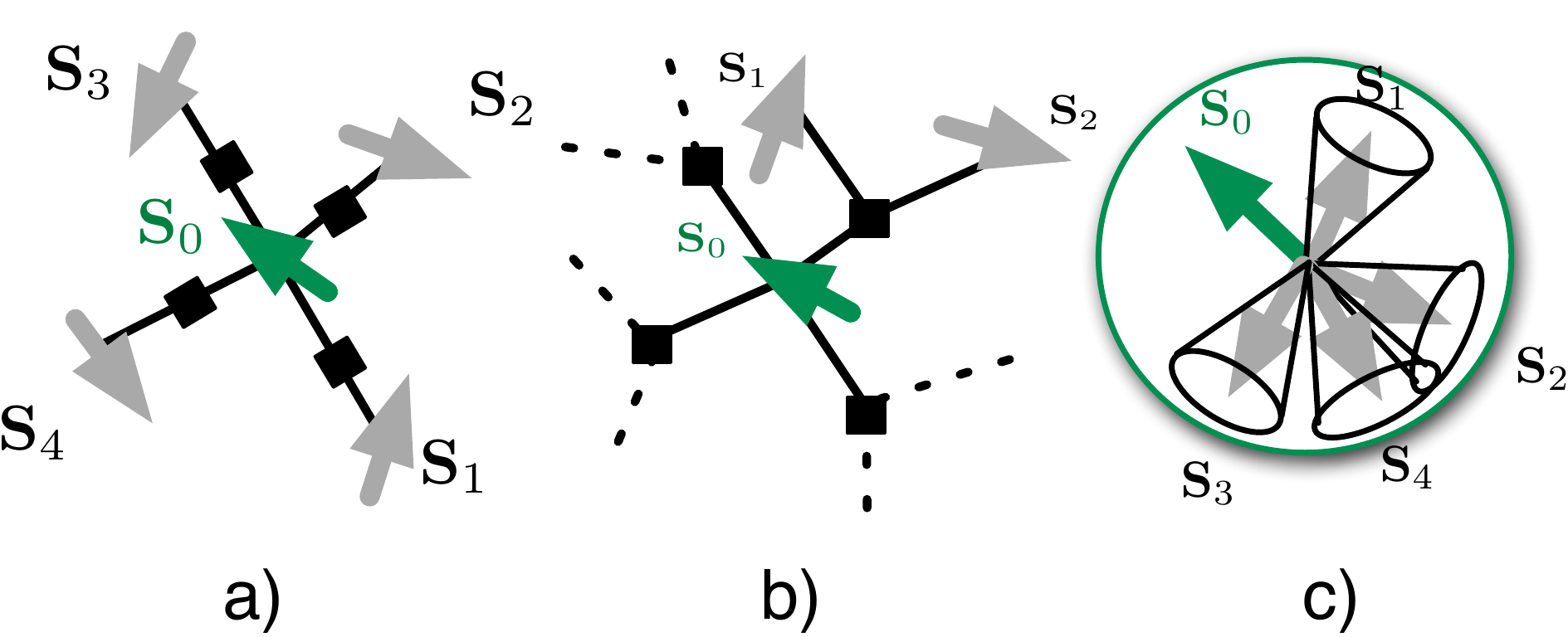}
  \ec
  \caption{A schematic figure of the model.
    Panel a) is for the cases of
    $p=2$ and b) is for the case of $p=3$-body
    interaction on a graph with connectivity $c=4$.
    Vectorial spins with $M$ components, in this example $M=3$ (Heisenberg spins), are put on the vertexes of a lattice or a graph as shown in the left panel a).
    The filled square represents the interaction nodes each of which
    connects a set of $p$ spins on the vertexes (variable nodes) interacting with each other.
    For the hardcore potential given by \eq{eq-hardcore-potential}
     the spin $\vS_{0}$ in panel c)
    is excluded from the cones around each of the neighboring
    spins $\vS_{1}$,$\vS_{2}$,$\vS_{3}$.
    (Note that, for instance, $\vS_{2}$ and
    $\vS_{4}$ can overlap if they are not directly connected by a link).
    The size of the cones grows with decreasing the parameter $\delta$.
    Thus the excluding volume effect becomes larger by decreasing $\delta$
    or increasing the connectivity $c$.
}
\label{fig-model}
\end{figure}

The spins are put on the vertexes of
lattices (graphs) which are locally tree-like with no closed loops as shown in Fig.~\ref{fig-model}.
Spins are involved in $p$-body interactions represented by factor nodes (interaction node)
$\bs$ in the figure. Each spin is involved in $c=\alpha M$
$p$-tuples. Thus the number of the $p$-tuples is given by,
\beq
N_{\bs}=NM(\alpha/p)
\label{eq-number-of-interactions}
\eeq

In the present paper we take not only the thermodynamic limit $N \to \infty$ but also
the limit of large number of spin components $M \to \infty$, which scales independently of $N$. 
As we will find below this brings about important consequences.
Later we will consider a special limit $p,\alpha \to \infty$  with the ratio $\gamma=\alpha/p$ fixed to a constant
of $O(1)$ in sec. \ref{sec-REM}. Otherwise the parameters $\alpha$ and $p$ are both constants of $O(1)$.

The interaction between the spins is given by a generalized $p$-body interaction,
\beq
H=\sum_{\bs=1}^{N_{\bs}}V \left(r_{\bs}\right).
\label{eq-hamiltonian}
  \eeq
  where
  \beq
  r_{\bs}=\delta- \frac{1}{\sqrt{M}} \sum_{\mu=1}^{M} X^{\mu}_{\bs}
  S^{\mu}_{1(\bs)}S^{\mu}_{2(\bs)}\cdots S^{\mu}_{p(\bs)}
  \label{eq-p-body-gap}
  \eeq
  Here $1(\bs),2(\bs),\ldots,p(\bs)$ represent the spins
involved in a given $p$-tuple $\bs$.
The function $V(r)$ represents a generic interaction potential.
We will call the argument variable $r_{\bs}$ as 'gap',
whose meaning will become clear later, 
with $\delta \in \mathcal{R}$ being  a control parameter.

In the present paper we mainly study models without quenched disorder (disorder-free model) but we also
discuss models with quenched disorder (disordered model).
\begin{itemize}
\item  {\bf disorder-free model}
  \beq
   X^{\mu}_{\bs}=1
   \label{eq-X-disorder-free}
   \eeq
 \item {\bf disordered model}
   \beq
   X^{\mu}_{\bs}=
  \frac{\lambda}{\sqrt{M}}
  +\sqrt{1-\left(\frac{\lambda}{\sqrt{M}}\right)^{2}}
  \xi^{\mu}_{\bs} \qquad  \left ( 0 \leq \frac{\lambda}{\sqrt{M}} \leq 1 \right)
\label{eq-X-disordered} 
\eeq
Here  $\xi^{\mu}_{\bs}$s are mutually independent, quenched random variables 
which obey the Gaussian distribution with zero mean and unit variance.
The parameter $\lambda$ represents the strength of the 'disorder-free' part in the disordered model. Note that the disorder-free model is recovered by choosing $\lambda/\sqrt{M}=1$. In the other limit $\lambda/\sqrt{M}=0$ we have
completely disordered, spinglass model. Thus we have a smooth interpolation between the two limits with this parametrization.
\end{itemize}

The free-energy $F$ of the system can be written as,
\beq
-\beta F =  \log Z
\eeq
where $\beta$ is the inverse temperature. The partition function $Z$
is defined as,
\beqn
&&  Z =\left(\prod_{i} \Tr_{\vS_i}\right)  \prod_{\bs} e^{-\beta V(r_{\bs})} \nonumber \\
&&  =\prod_{\bs} \left \{\inti \frac{d\kappa_{\bs}}{2\pi}
 Z_{\kappa_{\bs}} e^{i\kappa_{\bs} \delta}\right \}
\left( \prod_{i} \Tr_{\vS_i}  \right)
\exp \left [  \frac{1}{\sqrt{M}}\sum_{\mu=1}^{M} \sum_{\bs}
(-i\kappa_{\bs}) X_{\bs} S^{\mu}_{1(\bs)}S^{\mu}_{2(\bs)}\cdots S^{\mu}_{p(\bs)}
  \right ] \;\;\;
\label{eq-partition-function}
\eeqn
here $\Tr_{\vS}$ represents a trace over the spin space of the spin $\vS$,
\beqn
\mbox{(Ising)}  &\qquad & \Tr_{\vS}=\prod_{\mu=1}^{M} \sum_{S^{\mu}=\pm 1}
\label{eq-spin-trace-Ising}
 \\
\mbox{(Continuous)} &\qquad &\Tr_{\vS}=\int d\vS
= \left (\prod_{\mu=1}^{M} \inti dS^{\mu} \right) 
M \int_{-i\infty}^{i\infty}  \frac{d\lambda}{2\pi}e^{\lambda (M -\sum_{\mu=1}^{M} (S^{\mu})^{2})} \nonumber \\
&&=M \int_{-i\infty}^{i\infty} \frac{d\lambda}{2\pi}  e^{M \lambda} \prod_{\mu=1}^{M}
\inti dS^{\mu}e^{-\lambda(S^{\mu})^{2}}
\label{eq-spin-trace-continuous}
\eeqn
where $\int d{\bf S}_{i}$ is an integration over the surface of the $M$-dimensional sphere with diameter $\sqrt{M}$.
We have also introduced a Fourier transform of the Boltzmann's factor,
\beq
Z_{\kappa} \equiv \inti dh e^{-i\kappa h} e^{-\beta V(h)}.
\label{eq-def-z-kappa}
\eeq

Lastly let us note the similarity of our model
to the so called $M-p$ spinglass model
\cite{parisi1999continuous,larson2010numerical,caltagirone2011ising}.
In the $M-p$ spinglass model, one considers
$M$-component Ising spins on each vertex much as in our model.
Then $p \geq 2$-body interactions are introduced
between a pair of sites, say $i$ and $j$, taking
possible $p$-tuples using the $2M$ components of the spins. 
The model becomes exactly solvable in the $M \to \infty$ limit
\cite{caltagirone2011ising} much as in our model.
Moreover the model is very useful to study finite dimensional effects
\cite{parisi1999continuous,larson2010numerical}.
Although it is slightly different from our model, we anticipate
that much of the analysis we perform in the following could be done
also in the geometry of the $M-p$ spinglass model.

\subsection{Linear and non-linear potentials}

The most simple potential is the linear potential,
\beq
\mbox{(linear potential)} \qquad  V(x)=Jx  \qquad J > 0.
\label{eq-linear-potential}
\eeq
This is a $p$-spin ferromagnetic model. We will use this potential in order to establish connections to the
random energy model (sec. \ref{sec-REM}) and the $p$-spin spinglass models (sec. \ref{sec-linear}).

As a simplest non-linear potential, we will consider briefly
in sec. \ref{sec-quadratic-potential} the quadratic potential,
\begin{equation}
\mbox{(quadratic potential)} \qquad V(x)=\frac{\epsilon}{2} x^{2} \qquad \epsilon>0.
\end{equation}
We will study in detail in sec. \ref{sec-hardcore} the case of more strongly non-linear potential,
\beq
\mbox{(soft/hardcore potential)} \qquad V(x)=\epsilon x^{2}\theta(-x).
\label{eq-soft-hardcore-potential}
\eeq
The hardcore potential is obtained in the $\epsilon \to \infty$ limit.
This amount to bring in an excluded volume effect in the spin space
similarly to the interaction between the hardspheres (See Fig.~\ref{fig-model} c)).
With $p=2$ body interaction it can be used for the continuous coloring problem shown in Fig. \ref{fig_rotational_glass} a)): spins representing the color angles on adjacent vertexes are forced to be
separated in angle {\it larger than} $\cos^{-1}(\delta /\sqrt{M})$ for the hardcore potential (See Fig.~\ref{fig-model} c)).
In the case of $p=1$, and in the presence of quenched disorder $\xi^{\mu}$'s in \eq{eq-p-body-model-disorder},
the problem becomes the perceptron problem \cite{gardner1988space} \cite{franz2016simplest}. The case $p=2$ was also studied in part by a seminal work \cite{cugliandolo1996mean}. In the present paper we study the cases of $p \geq 2$.

\subsection{Pressure, distribution of gaps and isostaticity}
\label{sec-pressure-gr-isostaticity}

With the soft/hardcore  potential given by \eq{eq-soft-hardcore-potential},
the system becomes
more constrained as we decrease the parameter $\delta$ much as
an assembly of hardspheres becomes more constrained as the diameter of the spheres increase
so that the volume fraction increases.
This motivates us to introduce 'pressure' as an analogue of that in particulate systems,
\beq
\Pi = -\frac{1}{N_{\bs}}\frac{\partial \beta F}{\partial \delta}.
\label{eq-def-pressure}
\eeq
The normalization factor $N_{\bs}$ is simply the number of interaction links
in the system which is given by \eq{eq-number-of-interactions}.
Then it is also useful to introduce the  distribution function of the gap,
\beqn
g(r) & \equiv &  \left \langle \delta \left( r -  r(\Vec{S}_{i_{1}}, \ldots, \Vec{S}_{i_{p}})\right) 
\right \rangle \\
&=& \frac{1}{N_{\bs}}\frac{\delta (-\beta F)}{\delta \ln e^{-\beta V(r)}}
\label{eq-g-of-r}
\eeqn
In the 1st equation $\langle \ldots \rangle$ is the thermal average.
In the 2nd equation $\delta/\delta \ln e^{-\beta V(r)}$ is a functional derivative.
Apparently the distribution function of the gap $g(r)$
is analogous to the radial distribution function in the particulate systems.
The pressure given by \eq{eq-def-pressure} can be rewritten
using $\partial(-\beta F)/\partial \delta= \inti dr \frac{\delta (-\beta F)}{\delta \ln e^{-\beta V(r)}} (\ln e^{-\beta V(r)})'$ and $g(r)$ defined above as,
\beq
\Pi=\inti  dr g(r) (\ln e^{-\beta V(r)})'=\inti dr g(r) (-\beta V'(r)).
\label{eq-virial-pressure}
\eeq
This is the analogue of the virial equation for the pressure
in the liquid theory \cite{hansen1990theory}.

       Given $N$ spins $\vS_{i}$ ($i=1,2,\ldots, N$) with $M$ components, which are normalized such that $|\vS_{i}|^{2}=M$, the total number of
       the degrees of freedom is $N(M-1)$.  Each spin is 
       involved in $c=\alpha M$ sets of $p$-body interactions
       (See Fig. \ref{fig-model}). We say the gap associated with
       such an interaction is {\it closed} if
       $r(\Vec{S}_{i_{1}}, \ldots, \Vec{S}_{i_{p}})<0$.
The fraction of the interactions or contacts whose gaps are closed can be written as
\beq
f_{\rm closed}=\lim_{\epsilon \to 0}\int_{-\infty}^{\epsilon}dr g(r)
\label{eq-f-closed}
\eeq
where $g(r)$ is the distribution function of the gap
defined in \eq{eq-g-of-r}.
This means there are $N_{\bs}f_{\rm closed}$ constrains. Then isostaticity implies,
\beq
N(M-1)=N_{\bs}\lim_{\epsilon \to 0}\int_{-\infty}^{\epsilon}dr g(r).
\eeq
or
\beq
1=\frac{\alpha}{p}\lim_{\epsilon \to 0}\int_{-\infty}^{\epsilon}dr g(r).
\label{eq-isostaticity}
\eeq
in the $M \to \infty$ limit.

\section{Supercooled spin liquid states, crystalline states and their stability}
\label{sec-Supercooled-spin-liquid states}

In this section we focus on the crystallization and possibility of super-cooling,
~i.~e. realization of supercooled paramagnetic state which is at least locally
stable against crystallization.
This is an important step toward realization of glasses without quenched disorder.
In the present section we consider the disorder-free model given by \eq{eq-X-disorder-free}.
The effect of quenched disorder will be discussed in sec.   \ref{sec:quenched_disorder}.

\subsection{Crystalline order parameter and the free-energy functional}
\label{subsec:crystal}

Our disorder-free models given by the Hamiltonian \eq{eq-hamiltonian},
\eq{eq-p-body-gap} and \eq{eq-X-disorder-free} have the following global symmetries.
In sec. \ref{sec-generic-model} we introduced two types of spins: Ising and continuous spins.
In the cases of Ising spins $S_{i}^{\mu}=\pm 1$,
and for even $p$, the system has a global symmetry
with respect to $S^{\mu} \to -S^{\mu}$ for each component $\mu$.
Such symmetry is absent for the cases of odd $p$.
In the cases of continuous spins $S_{i}^{\mu} \in \mathcal{R}$,
and for $p=2$, the system has a global continuous symmetry with respect to
rotations of spins in the $M$-dimensional spin space. The continuous rotational symmetry
is lost for $p>2$ \footnote{Suppose that a rotation is defined by a $M \times M$ matrix $\hat{R}$, which is
   orthogonal $\hat{R}^{t}=\hat{R}^{-1}$. Vectors are transformed by the rotation as
  $S^{\mu} \to \sum_{\nu=1}^{M}R^{\mu\nu}S^{\nu}$. For instance, it can be easily checked that $\sum_{\mu=1}^{M}S_{i}^{\mu}S_{j}^{\mu}$
  remains invariant under the rotation but $\sum_{\mu=1}^{M}S_{i}^{\mu}S_{j}^{\mu}S_{k}^{\mu}S_{l}^{\mu}$ does not.}
and the residual global symmetries become just the
same as those in the Ising cases.

To be more specific, suppose that the system has
a ferromagnetic ground state $\Vec{S}_{i}=(1,1,\ldots,1)$ for $\forall i$.
This is achieved for example by choosing the
linear potential $V(x)=J x$ with $J > 0$ in \eq{eq-linear-potential}.
Because of the global symmetries mentioned above, there can be other
equivalent ground states, e.~g. $\Vec{S}_{i}=(-1,-1,\ldots,-1)$ for $\forall i$
(for even $p$). In order to study the possibility of
spontaneous symmetry breaking which select one ground state out of the equivalent ones (if they exist),
we may apply an external field of strength $h >0$
parallel to the ground state $(1,1,\ldots,1)$,
\beq
\beta H=-\sum_{\bs=1}^{N_{\bs}} \beta V \left(r_{\bs}\right)-h \sum_{i=1}^{N} \sum_{\mu=1}^{M}
S_{i}^{\mu}
\label{eq-hamiltonian-with-h}
\eeq
and examine the behavior of an order parameter,
\beq
m=\lim_{h \to 0}\lim_{N \to \infty} \frac{1}{NM}\sum_{i=1}^{N}\sum_{\mu=1}^{M}
\langle S_{i}^{\mu} \rangle_{h}.
\label{eq-ferromagnetic-order-parameter}
\eeq
where $\langle \cdots \rangle_{h}$ represents a thermal average
in the presence of the symmetry breaking field. The standard procedure to analyze the problem is as follows.
1) One first construct a free-energy $-\beta G(h)$ in the presence of
the field $h$ and then perform a Legendre transform to obtain $-\beta F(m)=-\beta G(m)+Nmh$
and then 2) seek for a solution $m$ which solves $\partial_{m} (-\beta F(m))=h=0$.

In addition, since we are considering to take the limit of large number of components $M \to \infty$,
we may also define a {\it local order parameter},
\beq
m_{i}=\lim_{h \to 0}\lim_{M \to \infty} \frac{1}{M}\sum_{\mu=1}^{M}
\langle S_{i}^{\mu} \rangle_{h}
\qquad (i=1,2,\ldots,N)
\label{eq-ferromagnetic-order-parameter-2}
\eeq
Let us emphasize again that $M$ scales independently of $N$, which will bring about important consequences below.

\subsubsection{Spin trace}

The above discussion motivates us to introduce an identity,
\beq
1= \inti dm_{i} \delta (Mm_{i}-\sum_{\mu=1}^{M} S_{i}^{\mu})
=M \int_{-i\infty}^{i\infty} \frac{dh_{i}}{2\pi} \int_{-\infty}^{\infty} dm_{i} e^{h_{i} (Mm_{i}-\sum_{\mu=1}^{M}S^{\mu}_{i})}
\qquad (i=1,2,\ldots,N).
\eeq
The integration over $h$ and $m$ corresponds to the steps 1) and 2) mentioned above.
Using the identity spin traces can be expressed formally in the $M \to \infty$ limit as,
\beq
\Tr_{\vS} \cdots = 
M \int_{-i\infty}^{i\infty} \frac{dh}{2\pi} \int_{-\infty}^{\infty}dm
\exp
\left
[
M h m+\ln
\Tr_{\vS} e^{-h\sum_{\mu=1}^{M}S^{\mu}}
\right ]
 \cdots
=
\int_{-\infty}^{\infty} dm e^{M s_{\rm ent}(m)}
 \prod_{\mu} \langle \cdots \rangle_{\mu}
 \label{eq-spin-trace-ferromagnet}
 \eeq
 Here the integration over $h$ can be done (formally) by the saddle point method
 in the limit $M \to \infty$.  The saddle point $h^{*}(m)$ is given by
 the saddle point equation,
 \beq
   m=\left. \frac{\Tr_{\vS} e^{-h^{*}\sum_{\mu=1}^{M}S^{\mu}}S^{\mu}}{
     \Tr_{\vS} e^{-h^{*}\sum_{\mu=1}^{M}S^{\mu}}} \right |_{h^{*}=h^{*}(m)}
   \label{eq-sp-m}
   \eeq
   and we find,
 \beq
 s_{\rm ent}(m)=h^{*}m+\ln \Tr_{\vS} e^{-h^{*}\sum_{\mu=1}^{M}S^{\mu}}
 \qquad
 \langle \cdots \rangle_{\mu}
 =\frac{  \Tr_{\vS} e^{-h^{*}S^{\mu}}\ldots}{  \Tr_{\vS} e^{-h^{*}S^{\mu}}}
 \eeq
 where $h^{*}=h^{*}(m)$ is given by \eq{eq-sp-m}.
  Using \eq{eq-spin-trace-ferromagnet} we find, for example,
\beq
\Tr_{\vS} S^{\mu}= \int_{-\infty}^{\infty} dm e^{M s_{\rm ent}(m)} m
\label{eq-simple-average}
\eeq

More specifically, by taking the spin traces explicitly we obtain the following expressions
for the Ising and continuous spin systems,
 \begin{itemize}
 \item {\bf Ising spin}

   We find  using \eq{eq-spin-trace-Ising},
 \beqn
s_{\rm ent}(m)=-m\tanh^{-1}(m)+\ln[2\cosh(\tanh^{-1}m)] 
\qquad h^{*}=\tanh^{-1}(m))
\label{eq-spin-trace-Ising-2}
\eeqn

\item {\bf Continuous spin}

  We find using \eq{eq-spin-trace-continuous},
\beqn
&& s_{\rm ent}(m)=\frac{1}{2}+\frac{1}{2}\ln (2\pi)+\frac{1}{2}\ln (1-m^{2})
\nonumber \\
&& \langle \cdots \rangle_{\mu} \equiv
\frac{\inti dS^{\mu}e^{-\lambda^{*}(S^{\mu})^{2}-h^{*}S^{\mu}}\cdots }{
  \inti dS^{\mu}e^{-\lambda^{*}(S^{\mu})^{2}-h^{*}S^{\mu}}}
\qquad h^{*}=-2m\lambda^{*} \qquad \lambda^{*}=\frac{1}{2(1-m^{2})}\;\;\;\;\;\;
\label{eq-spin-trace-continuous-2}
\eeqn
Here we performed the integrations $\inti dS^{\mu}$
assuming $\lambda>0$. Then we performed integrations over 
$\lambda$ and $h$ by the saddle point method.

 \end{itemize}

In  \eq{eq-spin-trace-ferromagnet}
we notice that different spin components $\mu$ are decoupled
in the average $\prod_{\mu}\langle \ldots \rangle_{\mu}$.
Then we obtain the following cumulant expansion which will become very useful
in the following,
\beqn
&&\ln \langle e^{\frac{1}{\sqrt{M}}\sum_{\mu=1}^{M}A_{\mu}} \rangle
=\frac{1}{\sqrt{M}}\sum_{\mu=1}^{M} \langle A_{\mu}\rangle
+\frac{1}{2!} \left(\frac{1}{\sqrt{M}}\right)^{2}\sum_{\mu=1}^{M} (\langle A^{2}_{\mu}\rangle-\langle A_{\mu}\rangle^{2})\nonumber \\
&&+\frac{1}{3!} \left(\frac{1}{\sqrt{M}}\right)^{3}\sum_{\mu=1}^{M} (\langle A^{3}_{\mu}\rangle-3\langle A^{2}_{\mu}\rangle^{2}\langle A_{\mu}\rangle+2\langle A_{\mu}\rangle^{3} ) \nonumber \\
&+&\frac{1}{4!} \left(\frac{1}{\sqrt{M}}\right)^{4}\sum_{\mu=1}^{M}
(\langle A^{4}_{\mu}\rangle
-4\langle A^{3}_{\mu}\rangle\langle A_{\mu}\rangle
-3\langle A^{2}_{\mu}\rangle^{2}
+12\langle A^{2}_{\mu}\rangle\langle A_{\mu}\rangle^{2}
-6\langle A_{\mu}\rangle^{3} )
+\cdots \;\;\;
\label{eq-cumulant-expansion}
\eeqn
Here we just used the fact that $\langle A^{\mu}A^{\nu} \rangle = \langle A^{\mu} \rangle \langle A^{\nu} \rangle$ holds
for $\mu \neq \nu$. 

\subsubsection{Evaluation of the free-energy}
\label{subsubsec:crystal-free-energy}

Using  \eq{eq-spin-trace-ferromagnet}, \eq{eq-simple-average}
and the cumulant expansion \eq{eq-cumulant-expansion} we find,
\beqn
&& \prod_{i} \Tr_{\vS_{i}}
\exp \left [  \frac{1}{\sqrt{M}}\sum_{\mu=1}^{M} \sum_{\bs}
(-i\kappa_{\bs}) S^{\mu}_{1(\bs)}S^{\mu}_{2(\bs)}\cdots S^{\mu}_{p(\bs)}
  \right ] \nonumber \\
&& \xrightarrow[M \to \infty]{}
\left( \prod_{i} \inti dm_{i} \right) e^{M \sum_{i} s_{\rm ent}(m_{i})}
\exp \left [  \sqrt{M} \sum_{\bs}
  (-i\kappa_{\bs})
  m_{1(\bs)}
  m_{2(\bs)}
  \cdots
  m_{p(\bs)} 
  \right ] \qquad
\eeqn
Now the partition function given by \eq{eq-partition-function} can be rewritten formally in the $M \to \infty$ limit as,
\beqn
 Z &=&
\left( \prod_{i}  \inti dm_{i} \right) e^{M \sum_{i} s_{\rm ent}(m_{i})}
\prod_{\bs} \left \{\inti \frac{d\kappa_{\bs}}{2\pi}
Z_{\kappa_{\bs}} e^{i\kappa_{\bs} \delta}\right \}
\exp \left [  \sqrt{M} \sum_{\bs}
  (-i\kappa_{\bs})
  m_{1(\bs)}
  m_{2(\bs)}
  \cdots
  m_{p(\bs)} 
  \right ]\nonumber \\
&=&
\left( \prod_{i=1}^{N} \inti dm_{i} \right)  e^{NM s(\{m_{i}\})}
 \label{eq-partition-function-ferromagnet}
\eeqn
where we defined
\beqn
s(\{m_{i}\})=
 \frac{1}{N}\sum_{i=1}^{N} \left \{ s_{\rm ent} (m_{i}) -
 \frac{1}{pM}\sum_{\bs \in \partial_{i}}
 \beta  V(\delta - \sqrt{M}m_{1(\bs)}m_{2(\bs)}\cdots m_{p(\bs)})
 \right \}
   \label{eq-action-ferromagnet}
   \eeqn
   where $\partial_{i}$ represents the set of interactions which involve $\vS_{i}$.
Now we are left with the integrations over $m_{i}$s in \eq{eq-partition-function-ferromagnet}
which can be done by the saddle point method in the $M  \to \infty$ limit.
The saddle point equation reads as,
 \beq
 0=\left. \frac{\partial s(\{m_{i}\})}{\partial m_{j}} \right |_{\{m_{i}=m_{i}^{*}\}} \qquad \mbox{for} \qquad (j=1,2,\ldots,N).
 \eeq

Since the system is regular and every vertex is exactly equivalent to each other
in our system, it is natural to expect a uniform solution $m_{i}^{*}=m$ for $\forall i$.
Moreover, since each spin is connected to $c=\alpha M$ neighbors which is a large number,
one can show that the effect of possible site-to-site fluctuation 
of $m_{i}$ can be neglected in the $M \to \infty$ limit.
In addition, possible small fluctuations of the coordination number $c$ can
also be neglected for the same reason.

We obtain the free-energy associated with such a uniform saddle point as,
 \beq
 -\beta\frac{F}{NM} =s(m)
 \label{eq-free-ene-ferro}
 \eeq
 with
\beq
s(m) \equiv s_{\rm ent} (m) -
\frac{\alpha}{p} \beta  V(\delta - \sqrt{M}   m^{p})
 \label{eq-sn-ferro}
 \eeq
 where $m$ must satisfy the saddle point equation
 \beq
 0=\frac{ds(m)}{dm}
 \label{eq-saddle-point-ferro}
 \eeq
It is also required to satisfy the stability condition,
 \beq
 d^{2}s(m)/dm^{2}  \leq 0
  \label{eq-saddle-point-ferro-stability}
  \eeq

\subsection{Possibilities of the crystalline states}

So far we have just considered a ferromagnetic phase
with the ground state $\Vec{S}_{i}=(1,1,\ldots,1)$  for $\forall i$
but we can also consider other crystalline states.
For example, suppose that there is a crystalline ground state in which
the spin configuration can be represented by
some configuration ${S}^{\mu}_{i}=(\sigma_{i})_{0} \in (-1,1)$ which is independent of $\mu$ but depends
on the vertex $i$. Just for simplicity we are limiting ourselves
to the cases that the ground
state configuration have the collinear spin structure, i.~e. spin configuration on different vertexes are either parallel or anti-parallel to each other.
The ferromagnetic case discussed in sec. \ref{subsec:crystal}
corresponds to $(\sigma_{i})_{0}=1$ for $\forall i$.  
Then it is useful to perform a gauge transformation
\beq
S_{i}^{\mu} \to \tilde{S}_{u}^{\mu}\equiv \sigma_{i}S_{i}^{\mu}.
\eeq
The crystalline order parameter $m$ can be defined again as 
\eq{eq-ferromagnetic-order-parameter} but replacing the spins $S$ by
the gauge transformed ones $\tilde{S}$.
Here the spins ${S}^{\mu}_{i}$ can be either the Ising type or continuous
type. The gauge transformation defined above does not change the character of the spins including the spin normalization \eq{eq-spin-normalization}
which reads $\sum_{\mu=1}^{M} (S_{i}^{\mu})^{2}=M$.

By the same gauge transformation the gap given by \eq{eq-p-body-gap} (with $X_{\bs}^{\mu}=1$)
is transformed to,
\beq
r_{\bs} \to
\tilde{r}_{\bs}=\delta-\frac{\eta_{\bs}}{\sqrt{M}}\sum_{\mu} 
\tilde{S}^{\mu}_{1(\bs)}\tilde{S}^{\mu}_{2(\bs)}\cdots \tilde{S}^{\mu}_{p(\bs)}
\label{eq-gap-gauge-transform}
\eeq
where we defined
\beq
\eta_{\bs}=\sigma_{1(\bs)}\sigma_{2(\bs)}\cdots \sigma_{p(\bs)}
\eeq
The variable $\eta_{\bs}$
takes $\pm 1$ values. For simplicity we limit ourselves to the
ground states such that it is a constant $\eta_{\bs}=\eta$ for all the interactions $\bs$.
Then the results in the previous section given by
\eq{eq-free-ene-ferro}-\eq{eq-saddle-point-ferro-stability} holds
just by changing the argument of the potential as:
\beq
V(\delta-\sqrt{M}m^{p}) \to V(\delta-\eta \sqrt{M}m^{p})
\eeq

  The simplest  example  is  $p=2$ model with the linear potential $V(x)=J x$ but with $J < 0$.
  Obviously the ground state is the anti-ferromagnetic one : $\sigma_{\bs}$ alternates
  the sign
across each of the interactions (note that we are considering tree-like lattices with no loops).
 In this case $\eta_{\bs}=\sigma_{1(\bs)}\sigma_{2(\bs)}=-1$ so that it becomes essentially the same as
 a ferromagnetic model with $J >0$ after the gauge transformation.

  \subsection{Crystalline transitions and possibility of super-cooling}
      \label{subsec-crystalline-transitions-and-supercooling}

 The saddle point equation given by \eq{eq-saddle-point-ferro}
 and the stability condition   given by \eq{eq-saddle-point-ferro-stability}
becomes, including the factor $\eta=\pm 1$ discussed above as the following:
 \begin{itemize}
 \item {\bf Ising spin}

   The saddle point equation given by \eq{eq-spin-trace-Ising-2}  becomes,
 \beq
 m=\tanh\left[ \eta \alpha  \sqrt{M} m^{p-1} \beta  V'(\delta-\eta\sqrt{M}m^{p})
   \right].
    \label{eq-saddle-point-p-spin-ferro-ising}
 \eeq
The stability condition  becomes,
  \beqn
  \frac{d^{2}s(m)}{dm^{2}}
  &=&-\frac{1}{1-m^{2}}+\eta\alpha \sqrt{M} (p-1)m^{p-2}\beta  V'(\delta-\eta\sqrt{M}m^{p})\nonumber \\
  &&  -\alpha p M m^{2(p-1)} \beta  V''(\delta - \eta\sqrt{M}m^{p}) < 0
  \label{eq-stability-para-ising}
  \eeqn

\item {\bf Continuous spin}

  The saddle point equation given by \eq{eq-spin-trace-continuous-2} becomes,
 \beq
 0=m \left( -\frac{1}{1-m^{2}}
 +\eta\alpha  \sqrt{M} \beta V'(\delta-\eta\sqrt{M} m^{p})m^{p-2}
 \right)
   \eeq
The stability condition becomes,
  \beqn
  \frac{d^{2}s(m)}{dm^{2}}
  &=&-\frac{1}{1-m^{2}}+\frac{2m^{2}}{(1-m^{2})^{2}}
  +\eta\alpha \sqrt{M} (p-1)m^{p-2} \beta V'(\delta-\eta\sqrt{M}m^{p}) \nonumber \\
  & & -\alpha p   M m^{2(p-1)}\beta V''(\delta - \eta\sqrt{M}m^{p}) < 0
    \label{eq-stability-para-continuous}
  \eeqn

  \end{itemize}
  
 It can be seen that the paramagnetic solution $m=0$ always verify the saddle point equations. We are especially interested with the possibility that the paramagnetic state with
 $m=0$ remains as a metastable state
after the crystalline transitions take place so that glass transitions within the
paramagnetic phase become possible.

 \begin{itemize}
 \item   $p=2$ case:
   \begin{itemize}
   \item if $|V'(\delta)| >0$, a 2nd order ferromagnetic transition takes place
     at a critical temperature
     \beq
     k_{\rm B} T_{c}=\alpha \sqrt{M} |V'(\delta)|
     \label{eq-Tc}
     \eeq
     below which the paramagnetic
  solution $m=0$ becomes unstable and the ferromagnetic or anti-ferromagnetic order with $ |m | > 0$
  emerges continuously. If $V'(\delta)$ is positive (negative) the ordering is
  ferromagnetic (anti-ferromagnetic) and we should choose $\eta=1$ ($\eta=-1$).
  Since the paramagnetic state $m=0$ is unstable
  below $T_{\rm c}$, supper-cooled paramagnetic state is absent
  and thus glass transitions is not possible without suppressing
  crystalline states by quenched disorder.

  \item If $V'(\delta)=0$, there will be no ferromagnetic nor anti-ferromagnetic
    phase transitions at finite temperatures. The $m=0$ solution remains stable at all
    finite temperatures,
  \beq
  \left.  \frac{d^{2}s(m)}{dm^{2}} \right |_{m=0}=-1 < 0.
  \label{eq-ultrastable-paramagnet}
  \eeq
  This is a very interesting situation where the crystallization is totally suppressed
  opening possibilities of glass transitions without quenched disorder.
   \end{itemize}
   
 \item  $p>2$ case:

   The paramagnetic solution $m=0$
   remains locally stable at all temperatures in the sense of \eq{eq-ultrastable-paramagnet}.
  Thus in this case supercooled paramagnetic state exist opening
  possibilities of glass transitions  without quenched disorder.
 \end{itemize}

\subsubsection{Linear potential: $p$-spin ferromagnetic model}
   \label{subsubsec-ferromagnetic-p-spin}

As a simplest example let us consider the case of the
linear potential $V(x)=Jx$ where $J > 0$, which means $V'(\delta) > 0$.
It is a ferromagnetic model so we choose $\eta=1$. The saddle point equation becomes
for the Ising spins,
  \beq
\mbox{(Ising)} \qquad  m=
  \tanh\left[ \alpha  \sqrt{M} \beta J m^{p-1}  \right]
   \label{eq-saddle-point-p-spin-ferro-ising-linear}
   \eeq
   and for the continuous spins,
   \beq
\mbox{(Continuous)} \qquad     0=m \left(-\frac{1}{1-m^{2}}+\alpha \sqrt{M} \beta J m^{p-2}
   \right)
   \eeq
  We see $m=0$ always verifies the saddle point equations as it should,
  \begin{itemize}
\item   For $p=2$ case, a 2nd order ferromagnetic transition takes place
  at a critical temperature $k_{\rm B} T_{c}/J=\alpha \sqrt{M}$.
  Supper-cooled paramagnet and thus glass transitions without quenched disorder are not possible as discussed above.
\item
  For $p>2$, a 1st order ferromagnetic transition take place
  at $k_{\rm B}T_{c}/J=  O(\alpha \sqrt{M})$.
  On the other hand the paramagnetic state $m=0$ remains locally stable at all temperatures
  as discussed above.

  Quite interestingly in \cite{franz2001ferromagnet} a $p=3$ Ising ferromagnet 
  with  $M=1$ component was studied  via cavity method and Monte Carlo simulations and 
  the supercooled paramagnetic state and the glass transition were discovered. Our result is consistent with this observation.

\item In the $p \to \infty$ limit with the Ising spins, the exact solution can be easily obtained.
  The saddle point equation given by \eq{eq-saddle-point-p-spin-ferro-ising-linear}
  admits only $m=\pm 1$ except for $m=0$. The paramagnetic free-energy (free-entropy)
  is obtained as $s(m=0)=\ln 2$ while that for the ferromagnetic
  phase is obtained as $s(m=1)=\gamma \beta J \sqrt{M}$. Here we introduced a parameter
  \beq
  \gamma=\frac{\alpha}{p}.
  \label{eq-def-gamma}
  \eeq
  Let us consider the $p \to \infty$ limit with $\gamma$ fixed.
  Then we easily see that a 1st order ferromagnetic phase transition takes place  at
  \beq
  k_{\rm B}T_{\rm c}/J=\gamma \sqrt{M}/\ln 2
  \label{eq-large-p-spin-ferromagnetic-Tc}
  \eeq
  In the next section \ref{sec-REM} we will find that the system becomes equivalent to the
  random energy model (REM) \cite{derrida1980random} by excluding the ferromagnetic state.
  \end{itemize}

  \subsubsection{Non-linear potentials with flatness}
    \label{sec-Non-linear-potentials-with-flatness}

  If the potential $V(x)$ has a flat part where $V'(x)=0$, it tends to
  suppress crystallization and thus enhances the possibility to realize glass
  transitions inside the paramagnetic phase.

 The simplest example may be the quadratic potential,
    \beq
  V(x)=\frac{\epsilon}{2} \frac{x^{2}}{2} \to V'(x)=\epsilon x
  \eeq
  Thus for $p=2$ and $\delta=0$, the system should remain paramagnetic
  at all finite temperatures.

More interesting case is the soft/hard core potential given by \eq{eq-soft-hardcore-potential}
which is completely flat for $x>0$,
  \beq
  V(x)=\epsilon x^{2}\theta(-x) \to  V'(x)=2\epsilon x\theta(-x)
  \label{eq-dev-V-softcore}
  \eeq
Let us consider again $p=2$ case.  
  For $\delta >0$, $V'(\delta)=0$ so that
  the system is paramagnetic at all finite temperatures.
  On the other hand for $\delta <0$, $V'(x)<0$
  anti-ferromagnetic phase emerges via 2nd order transition. Then we choose $\eta=-1$.
  The transition temperature given by \eq{eq-Tc} is found as,
  \beq
  k_{\rm B}T_{\rm c}(\delta)=2\alpha \sqrt{M}\epsilon (-\delta)\theta(-\delta)
  \label{eq-antiferro-softcore}
  \eeq
  In the hardcore limit $\epsilon \to \infty$, the anti-ferromagnetic transition
  takes place as $\delta \to 0^{+}$.

  \section{Self-generated randomness: connection to the random energy model in $p \to \infty$ ferromagnetic Ising model with $M \to \infty$}
  \label{sec-REM}

  Let us start looking for possible
  glass transitions within the supercooled paramagnetic phase.
  In this section we study the ferromagnetic $p$-spin $M$-component Ising model, with the linear potential $V(x)=J x$ with $J>0$
  discussed in sec \ref{subsubsec-ferromagnetic-p-spin}.
There we have seen that supercooled paramagnetic states with $m=0$ exist for $p>2$ below the ferromagnetic transition temperatures $k_{\rm T}T_{c}/J \sim O(\alpha\sqrt{M})$.
In the following we will find that the system becomes essentially
identical to the random energy model (REM) \cite{derrida1980random}
in the $p \to \infty$ limit as far as the supercooled states are concerned.
This proves the existence of the self-generated randomness.

The hamiltonian is given by
\beq
H_{\{S\}} = - \frac{J}{\sqrt{M}}\sum_{\bs} \sum_{\mu=1}^{M}
S_{1(\bs)}^{\mu}S_{2(\bs)}^{\mu}\cdots S_{p(\bs)}^{\mu}
\qquad S_{i}^{\mu} \in (-1,1)
\eeq
with $J >0$. Here we are especially interested with the $p \to \infty$ limit with
$\gamma=\alpha/p$ introduced in \eq{eq-def-gamma} fixed.
As we discussed in sec \ref{subsec-crystalline-transitions-and-supercooling}
it exhibits a ferromagnetic phase transition at
$k_{\rm B}T_{\rm c}=\gamma J \sqrt{M}/\ln 2$.

We examine the distribution of the energies of the disorder-free model
performing a similar analysis done for the $p$-spin Ising spinglass model {\it with} quenched disorder
in the original work by Derrida \cite{derrida1980random}.
To this end let us first introduce a flat average over the $2^{NM}$ spin configurations,
 \begin{displaymath}
 \langle \ldots \rangle_{S} \equiv
 \frac{\prod_{i=1}^{N} \prod_{\mu=1}^{M}\sum_{S_{i}^{\mu}=\pm 1} \ldots}{2^{NM}}.
\end{displaymath}
Then the distribution of energy among all configurations is obtained as,
 \beqn
 P(E)&=&\langle \delta(E-H_{\{S\}})) \rangle_{S}
= \int_{-\infty}^{\infty} \frac{d\kappa}{2\pi}e^{i\kappa E}
\left \langle
\exp \left[ i\kappa  \frac{J}{\sqrt{M}}\sum_{\bs} \sum_{\mu=1}^{M}
  S_{1(\bs)}^{\mu}S_{2(\bs)}^{\mu}\cdots S_{p(\bs)}^{\mu}
  \right ]
\right  \rangle_{S} \nonumber \\
&&  \xrightarrow[M \to \infty]{}
\frac{e^{-\frac{E^{2}}{2N_{\bs}J^{2}}}}{\sqrt{2\pi N_{\bs}J^{2}}}
\eeqn
  where $N_{\bs}=NM(\alpha/p)$ is the number of interactions given by \eq{eq-number-of-interactions}.
Here we evaluated the expectation value $\langle \ldots \rangle_{S}$
by performing expansion in power series of $1/\sqrt{M}$
(see \eq{eq-cumulant-expansion} for the cumulant expansion),
 \beqn
&& \ln \left \langle
\exp \left[ i\kappa  \frac{J}{\sqrt{M}}\sum_{\bs} \sum_{\mu=1}^{M}
  S_{1(\bs)}^{\mu}S_{2(\bs)}^{\mu}\cdots S_{p(\bs)}^{\mu}
  \right ]\right \rangle_{S} \nonumber \\
&& =  \ln \left[ 1+
\frac{(i\kappa)^{2}}{2!} \left ( \frac{J}{\sqrt{M}} \right)^{2}
\left \langle \left (
\sum_{\bs}\sum_{\mu}
S_{1(\bs)}^{\mu}  S_{2(\bs)}^{\mu} \cdots S_{p(\bs)}^{\mu} \right)^{2}
\right \rangle_{S} \right. \nonumber \\
&& \left. + 
\frac{(i\kappa)^{4}}{4!} \left ( \frac{J}{\sqrt{M}} \right)^{4}
\left \langle \left (
\sum_{\bs}\sum_{\mu}
S_{1(\bs)}^{\mu}  S_{2(\bs)}^{\mu} \cdots S_{p(\bs)}^{\mu} \right)^{4}
\right \rangle_{S} +
\ldots  \right] \nonumber \\
&=&  \frac{(i \kappa)^{2}}{2!}\left(\frac{J}{\sqrt{M}} \right)^{2}N_{\bs}M
+\frac{\kappa^{4}}{4!}\left(\frac{J}{\sqrt{M}} \right)^{4}
\left ( N_{\bs} M\right)  + \ldots
  \xrightarrow[M \to \infty]{}   -\frac{\kappa^{2}}{2}N_{\bs}J^{2}
  \eeqn
  Here $N_{\bs}=NM (\alpha/p)=NM \gamma$ (see \eq{eq-number-of-interactions}),
since  we take the $p \to \infty$ limit with fixed  $\gamma$ as defined in \eq{eq-def-gamma}.

Next let us examine simultaneous distribution of energy $E_{1}$
associated with an arbitrary chosen spin configuration
$(\vS_{1},\vS_{2},\ldots,\vS_{N})$
and the energy $E_{2}$ of another configuration $(\vS'_{1},\vS'_{2},\ldots,\vS'_{N})$.
Here the latter is created from the former by flipping, say 
according to a deterministic rule, a fraction $(1-q)/2$ with  $0 < q < 1$
of the elements of the former.
In other words the overlap between the two configurations is 
$q=(1/(NM))\sum_{i=1}^{N}\sum_{\mu=1}^{M} S_{i}^{\mu}(S')_{i}^{\mu}$. We find,
\beqn
&&P(E,E')=\langle \delta(E-H_{\{S\}}))\delta(E'-H_{\{S'\}})) \rangle_{S} \nonumber \\
&=&
\int_{-\infty}^{\infty}\int_{-\infty}^{\infty} \frac{d\kappa}{2\pi}
\frac{d\kappa'}{2\pi}e^{i\kappa E+i\kappa' E'}
\left \langle
\exp \left[  \frac{J}{\sqrt{M}}\sum_{\bs} \sum_{\mu=1}^{M}
  \left (
i\kappa   S_{1(\bs)}^{\mu}S_{2(\bs)}^{\mu}\cdots S_{p(\bs)}^{\mu} \right. \right.\right. \nonumber \\
&& \left.\left. \left.  +i\kappa'
(S')_{1(\bs)}^{\mu}(S')_{2(\bs)}^{\mu}\cdots (S')_{p(\bs)}^{\mu}
\right)
  \right ]
\right  \rangle_{S} \nonumber \\
&& \xrightarrow[M \to \infty]{} \frac{1}{2}\frac{1}{\sqrt{N_{\bs}\pi J^{2}A_{+}}}\frac{1}{\sqrt{N_{\bs}\pi J^{2}A_{-}}}
\exp\left[-\frac{E_{+}^{2}}{NJ^{2}A_{+}}-\frac{E_{-}^{2}}{NJ^{2}A_{-}}\right]
\eeqn
with $A_{\pm}=\frac{1}{2}[1\pm q^{p}]$ and $E_{\pm}=(E \pm E')/2$.
Here we evaluated the expectation value $\langle \ldots \rangle_{S}$
by performing expansion in power series of $1/\sqrt{M}$,
\beqn
&& \lim_{M \to \infty} \ln \left \langle
\exp \left[  \frac{J}{\sqrt{M}}\sum_{\bs} \sum_{\mu=1}^{M}
  \left (
i\kappa   S_{1(\bs)}^{\mu}S_{2(\bs)}^{\mu}\cdots S_{p(\bs)}^{\mu}
+i\kappa' (S')_{1(\bs)}^{\mu}(S')_{2(\bs)}^{\mu}\cdots (S')_{p(\bs)}^{\mu}
\right)
  \right ]
\right  \rangle_{S} \nonumber \\
&=& \lim_{M \to \infty}
\frac{1}{2} \left ( \frac{J}{\sqrt{M}} \right)^{2}
\left \langle \left \{
\sum_{\bs}\sum_{\mu}
\left(
i \kappa S_{1(\bs)}^{\mu}  S_{2(\bs)}^{\mu} \cdots S_{p(\bs)}^{\mu}
+i \kappa' (S')_{1(\bs)}^{\mu}  (S')_{2(\bs)}^{\mu} \cdots (S')_{p(\bs)}^{\mu}
\right)
\right\}^{2}
\right \rangle_{S} 
\nonumber \\
&=& 
-\frac{1}{2}[\kappa^{2}+(\kappa')^{2}+2\kappa\kappa' q^{p}]N_{\bs}J^{2}
 \eeqn
Here we realize that in the $p \to \infty$ limit, the distribution function decouples,
 \beq
 P(E,E') \xrightarrow[p \to \infty]{}  P(E)P(E').
 \eeq
 because  $0 < q < 1$.

 The above observations imply that the present ferromagnetic system
 without any quenched disorder behaves
essentially as a REM in the $p \to \infty$ limit:
over the majority of the $2^{NM}$ spin configurations ,
excluding the negligible fraction of the spin configurations
close to the ferromagnetic ground state, 
it is as if each microscopic states is
assigned a random energy drawn from a Gaussian distribution with 
$0$ mean and variance $\sqrt{\gamma}J$. 
Here we notice that all the exact results of the standard version
of the REM \cite{derrida1980random}, which corresponds to $\gamma=1/2$,
can be used in the present system
just by replacing $J$ of the standard model by $\sqrt{2\gamma}J$.
Then we readily find that the system exhibit the Kauzmann transition, i. e.
ideal static glass transition at, 
\beq
k_{\rm B}T_{\rm K}/J=\sqrt{\gamma}/\sqrt{2\ln 2}
\label{eq-large-p-spin-ferromagnetic-Tk}
\eeq
within the supercooled paramagnetic states.
At temperatures below $T_{\rm K}$, the internal energy becomes stuck at
$E/NM = -\sqrt{2\gamma\ln 2} J$ among the disordered states,
while the ferromagnetic ground state
energy is given by $E_{\rm g}/NM=- \sqrt{M} \gamma J$.

Readers would have noticed that the derivation of the REM discussed above is quite similar
to the standard procedure to prove the central limit theorem (CLT).
The underlying reason can be traced back to the tree-like structure of our system.

\section{Replicated system}
\label{sec-Replicated-system}

In this section we setup a formalism to study the glass transitions using the replica method.
We first develop a free-energy functional of the disorder-free model given by \eq{eq-X-disorder-free}. Then we also consider the model with quenched disorder given by \eq{eq-X-disordered}.

\subsection{Disorder free model}
\label{subsec:glass-free-energy}

We consider a system of replicas $a=1,2,\ldots,n$ of the disorder-free model given by \eq{eq-X-disorder-free},
\beq
H=-\sum_{a=1}^{n}\sum_{\bs=1}^{N_{\bs}}V \left(r^{a}_{\bs}\right)
\eeq
where
\beq
r^{a}_{\bs}=\delta- \frac{1}{\sqrt{M}} \sum_{\mu=1}^{M}
(S^{a})^{\mu}_{1(\bs)}(S^{a})^{\mu}_{2(\bs)}\cdots (S^{a})^{\mu}_{p(\bs)}.
\label{eq-p-body-model-replica}
\eeq
The free-energy of the replicated system can be expressed as,
\beq
-\beta F =  \log Z \nonumber = \left. \partial_{n}Z^{n} \right |_{n=0}
\eeq
with the replicated partition function
\beqn
&&  Z^{n} =\left( \prod_{i=1}^{N} \prod_{a=1}^{n} \Tr_{\vS_{i,a}} \right) \prod_{\bs} e^{-\sum_{a=1}^{n}\beta V(r^{a}_{\bs})}
  =\prod_{a=1}^{n}\prod_{\bs} \left \{\inti \frac{d\kappa_{\bs}}{2\pi}
 Z_{\kappa_{\bs,a}} e^{i\kappa_{\bs,a} \delta}\right \}
\nonumber \\
&& \left( \prod_{a} \prod_{i}\Tr_{\vS_{i,a}} \right)
\exp \left [  \frac{1}{\sqrt{M}}\sum_{a=1}^{n}\sum_{\mu=1}^{M} \sum_{\bs}
(-i\kappa_{\bs,a}) (S^{a})^{\mu}_{1(\bs)}(S^{a})^{\mu}_{2(\bs)}\cdots (S^{a})^{\mu}_{p(\bs)}
  \right ] 
\label{eq-partition-function-replica}
\eeqn
where $\Tr^{a}_{i}$ represents a trace over the spin space in  replica $a$.

In order to detect the spontaneous glass transition, we can follow steps
analogous to the one we took in sec. \ref{subsec:crystal}
for the crystalline (ferromagnetic) transition.
Namely we can explicitly break the replica symmetry as \cite{parisi1989mechanism},
\beq
\beta H=-\sum_{a=1}^{n}\sum_{\bs=1}^{N_{\bs}}\beta V \left(r^{a}_{\bs}\right)
-\sum_{a < b} \epsilon_{ab} \sum_{i} \sum_{\mu} (S^{a})^{\mu}_{i} (S^{b})^{\mu}_{i}
\eeq
and study the behavior of the glass order parameter matrix $\hat{Q}$
\begin{equation}
Q_{ab}=\lim_{\epsilon_{ab} \to 0}\lim_{N \to \infty}\frac{1}{NM} \sum_{i=1}^{N}
\sum_{\mu=1}^{M}\langle (S^{a})^{\mu}_{i} (S^{b})^{\mu}_{i} \rangle_{\epsilon}
\label{eq-def-Qab}
\end{equation}
Here $\langle \ldots \rangle_{\epsilon}$ represents the thermal average
in the presence of the symmetry breaking field $\epsilon_{ab}$.
Although \eq{eq-def-Qab} is meat for $a\neq b$, it is convenient to
extend it to include the diagonal elements
\beq
Q_{aa}=1
\label{eq-Q-diag}
\eeq
to reflect the spin normalization \eq{eq-spin-normalization}.

Just as the case of ferromagnetic transition discussed in  sec. \ref{subsec:crystal},
we can consider the following steps to analyze the problem:
1) One first construct a free-energy $-\beta G(\hat{\epsilon})$
in the presence of
the field $\hat{\epsilon}$ and then perform a Legendre transform to obtain $-\beta F(\hat{Q})=-\beta G(\hat{Q})+N \sum_{a<b}\epsilon_{ab} Q_{ab}$
and then 2) seek for a solution which solves
$\partial_{Q_{ab}} (-\beta F(\hat{Q}))=\epsilon_{ab}=0$.

Since we are considering the $M \to \infty$ limit we may also define
a {\it local glass order parameter},
\beq
(Q_{i})_{ab}=\lim_{\epsilon_{ab} \to 0} \lim_{M \to \infty} \frac{1}{M} 
\sum_{\mu=1}^{M}\langle (S^{a})^{\mu}_{i} (S^{b})^{\mu}_{i} \rangle_{\epsilon}
\qquad (i=1,2,\ldots,N)
\label{eq-def-Qab-2}
\eeq

\subsubsection{Spin trace}

The above discussion motivate us to introduce an identity,
\beqn
&& 1= \inti d(Q_{ab})_{i} \delta \left(M (Q_{ab})_{i}-\sum_{\mu=1}^{M} (S^{a})_{i}^{\mu}(S^{b})_{i}^{\mu}\right) \\
&& =M \intii \frac{d(\epsilon_{ab})_{i}}{2\pi}
\inti d(Q_{ab})_{i}
e^{(\epsilon_{ab})_{i} (M (Q_{ab})_{i}-\sum_{\mu=1}^{M}(S^{a})^{\mu}_{i}(S^{b})^{\mu}_{i})}
\qquad (i=1,2,\ldots,N) \nonumber
\eeqn
for $a < b$.
The integration over $\epsilon_{ab}$ and $Q_{ab}$ corresponds to the steps 1) and 2) mentioned above.
Using the latters, spin traces can be expressed formally in the $M \to \infty$ limit as,
\beqn
\prod_{c=1}^{n}\Tr_{\vS_{c}}  &\cdots &=
 M  \left( \prod_{a < b} \inti \frac{d\epsilon_{ab}}{2\pi} \inti d(Q_{ab})\right)
 \exp
 \left[
M \sum_{a< b}\epsilon_{ab}  Q_{ab}
+\ln \prod_{c=1}^{n}\Tr_{\vS_{c}}
e^{-\epsilon_{ab}\sum_{\mu=1}^{M}(S^{a})^{\mu}(S^{b})^{\mu}}
\right ] \cdots \nonumber \\
&=&
\left(   \prod_{a < b}  \inti d(Q_{ab}) \right) e^{M s_{\rm ent}[\hat{Q}]}
\prod_{\mu} \langle \cdots \rangle_{\mu}
\label{eq-spin-trace-glass-bias}
\eeqn
Here the integration over $\epsilon_{ab}$ can be done by the saddle point
method in $M \to \infty$. The saddle point equation which determines the saddle point  $\epsilon_{ab}^{*}(\hat{Q})$ is given by,
\beq
   Q_{ab}=\left. 
   \frac{\prod_{c}\Tr_{S^{c}} e^{-\sum_{a < b}\epsilon_{ab} S^{a}S^{b}} S^{a}S^{b}}{\prod_{c}\Tr_{S^{c}} e^{-\sum_{a<b}\epsilon_{ab} S^{a}S^{b}}}
   \right |_{\epsilon_{ab}=\epsilon^{*}_{ab}(\hat{Q})}
   \label{eq-sp-q}
  \eeq
and we find,  
  \beq
  s_{\rm ent}[\hat{Q}]=\sum_{a<b}
  \epsilon^{*}_{ab}Q_{ab}+\ln \prod_{c=1}^{n}\Tr_{S^{c}}
  e^{-\sum_{a<b}\epsilon^{*}_{ab}S^{a}S^{b}}
  \qquad
    \langle \cdots \rangle_{\mu}= 
\frac{\prod_{c=1}^{n}\Tr_{S^{c}}
  e^{-\sum_{a<b}\epsilon_{ab}^{*} (S^{a})^{\mu} (S^{b})^{\mu}} \cdots
  }{
  \prod_{c=1}^{n}\Tr_{S^{c}}
      e^{-\sum_{a<b}\epsilon_{ab}^{*} (S^{a})^{\mu} (S^{b})^{\mu}}
}
\label{eq-spin-trace-Ising-2-replica}
\eeq
where $\epsilon^{*}_{ab}=\epsilon^{*}_{ab}(\hat{Q})$ determined by
\eq{eq-sp-q}.
Using \eq{eq-spin-trace-glass-bias} we find, for example,
\beq
\prod_{c}\Tr_{\vS_{c}} (S^{a})^{\mu}=0
\qquad
\prod_{c}\Tr_{\vS_{c}} (S^{a})^{\mu}(S^{b})^{\mu}
=\inti d(Q_{ab}) e^{M s_{\rm ent}[\hat{Q}]} Q_{ab}
\label{eq-simple-average-replica}
\eeq

More precisely, by taking the spin traces we obtain the following expressions
for the Ising and continuous spin systems,
\begin{itemize}
  \item {\bf Continuous spin}: We find using \eq{eq-spin-trace-continuous}
and introducing $\epsilon_{aa}=\lambda_{a}$
and $Q_{aa}=1$ (spin normalization, see \eq{eq-Q-diag}),
\beqn
&& s_{\rm ent}[\hat{Q}]=\sum_{a, b} 
\frac{1}{2}  \epsilon^{*}_{ab}Q_{ab}+
  \ln  \sqrt{\frac{(2\pi)^{n}}{{\rm det} (\hat{\epsilon})^{*}}}
  =\frac{n}{2}+\frac{n}{2} \ln (2\pi) + \frac{1}{2}\ln {\rm det} \hat{Q}
\nonumber \\
&& 
\langle \cdots \rangle_{\mu} \equiv
\frac{
  \inti dS^{\mu} e^{-\frac{1}{2}\sum_{a,b}\epsilon^{*}_{ab}(S^{a})^{\mu} (S^{b})^{\mu}}
    \ldots
  }{
      \inti dS^{\mu} e^{-\frac{1}{2}\sum_{a,b}\epsilon^{*}_{ab}(S^{a})^{\mu} (S^{b})^{\mu}}
}
\label{eq-spin-trace-continuous-2-replica}
\eeqn
Here we have performed integration over $\epsilon_{ab}$ by the saddle point method
which yields a saddle point
\beq
\hat{\epsilon}^{*}=\hat{Q}^{-1}
\eeq

\item {\bf Ising spin}: We find  using \eq{eq-spin-trace-Ising},
\beq
   s_{\rm ent}[\hat{Q}]=\sum_{a \neq b} 
\frac{1}{2}  \epsilon^{*}_{ab}Q_{ab}+
  \ln  
e^{-\sum_{a<b}\epsilon^{*}_{ab}\frac{\partial^{2}}{\partial h_{a}\partial h_{b}}}
\left. \prod_{a} 2\cosh (h_{a})  \right |_{\{h_{a}=0\}}
\eeq
Here we performed the spin trace formally as
\beqn
&& \Tr_{\vS_{c}}e^{-\sum_{a<b}\epsilon_{ab}S^{a}S^{b}}
=\Tr_{\vS_{c}}e^{-\sum_{a<b}\epsilon_{ab}\frac{\partial^{2}}{\partial h_{a}\partial h_{b}}}
\left. e^{\sum_{a}h_{a}S^{a}}  \right |_{\{h_{a}=0\}} \nonumber \\
&& = e^{-\sum_{a<b}\epsilon_{ab}\frac{\partial^{2}}{\partial h_{a}\partial h_{b}}}
\left. \prod_{a} 2\cosh (h_{a})  \right |_{\{h_{a}=0\}}
\nonumber
\eeqn
For the integration over $\epsilon_{ab}$, the saddle point $\epsilon_{ab}^{*}=\epsilon^{*}_{ab}[\hat{Q}]$ is obtained formally as,
\beqn
&& Q_{ab}=-\left. \frac{\delta}{\delta \epsilon_{ab}}\ln e^{-\sum_{a<b}\epsilon_{ab}\frac{\partial^{2}}{\partial h_{a}\partial h_{b}}} 
\left. \prod_{a} 2\cosh (h_{a})  \right |_{\{h_{a}=0\}}
\right |_{\epsilon_{ab}=\epsilon^{*}_{ab}[\hat{Q}]}
\eeqn

\end{itemize}

\subsubsection{Evaluation of the free-energy}

In  \eq{eq-spin-trace-glass-bias}
we notice again that different spin components $\mu$ are decoupled
in the average $\prod_{\mu}\langle \ldots \rangle_{\mu}$.
Then we can evaluate the spin trace in the replicated
partition function given by \eq{eq-partition-function-replica}
in the $M \to \infty$ limit using
the cumulant expansion
given by \eq{eq-cumulant-expansion} and \eq{eq-simple-average-replica} as,
\beqn
&& \prod_{a}\prod_{i} \Tr_{\vS_{i,a}}
\exp \left [  \frac{1}{\sqrt{M}} \sum_{a}\sum_{\mu=1}^{M}\sum_{\bs}
(-i\kappa^{a}_{\bs}) (S^{a})^{\mu}_{1(\bs)}(S^{a})^{\mu}_{2(\bs)}\cdots (S^{a})^{\mu}_{p(\bs)}
  \right ]  \\
&& \xrightarrow[M \to \infty]{}
\left( \prod_{i} \prod_{a < b} \inti d(\hat{Q}_{i})_{ab} \right) e^{M \sum_{i} s_{\rm ent}[\hat{Q}_{i}]}
\exp \left [ \sum_{\bs} \sum_{a,b}\frac{(i\kappa_{a})(i\kappa_{b})}{2}
  (Q_{1(\bs)})_{ab}(Q_{2(\bs)})_{ab} \ldots (Q_{p(\bs)})_{ab}
  \right ]
\;\;\; \nonumber
\eeqn
In the exponent we assume $Q_{aa}=1$ (see \eq{eq-Q-diag}) for the diagonal terms. The above expression is a crucial result because it reveals the self-generated randomness in our 'disorder-free' model.

Collecting the above results, the partition function
given by \eq{eq-partition-function-replica}
can be rewritten formally in the $M \to \infty$ limit as,
\beqn
 Z &=&
\left(  \prod_{i} \prod_{a< b}\inti d(\hat{Q}_{i})_{ab} \right) e^{M \sum_{i} s_{\rm ent}[\hat{Q}_{i}]}
\prod_{\bs,a} \left \{\inti \frac{d\kappa^{a}_{\bs}}{2\pi}
Z_{\kappa^{a}_{\bs}} e^{i\kappa^{a}_{\bs} \delta}\right \} \nonumber \\
&& \hspace*{5cm}  \exp \left[ \sum_{\bs} \sum_{a,b}\frac{(i\kappa_{a})(i\kappa_{b})}{2}
  (Q_{1(\bs)})_{ab}(Q_{2(\bs)})_{ab} \ldots (Q_{p(\bs)})_{ab} \right]
\nonumber \\
  &=&
   \left( \prod_{i} \prod_{a< b}\inti d(\hat{Q}_{i})_{ab} \right) e^{NM s[(\hat{Q}_{i})]}
 \label{eq-partition-function-glass}
\eeqn
where we defined
\beqn
&& s[\hat{Q}_{i}]
= \frac{1}{N}\sum_{i=1}^{N} \left \{ s_{\rm ent} [\hat{Q}_{i}]
+\frac{1}{pM}\sum_{\bs \in \partial_{i}}
e^{ \frac{1}{2}\sum_{a,b}\frac{\partial^{2}}{\partial h_{a}\partial h_{b}}
  (Q_{1(\bs)})_{ab}(Q_{2(\bs)})_{ab} \ldots (Q_{p(\bs)})_{ab}
}
\left. \prod_{a=1}^{n}e^{-\beta V(\delta + h_{a})}  \right |_{\{h_{a}=0\}}
\right \} \qquad \nonumber \\
   \label{eq-action-glass}
   \eeqn
   To derive the last expression we used \eq{eq-def-z-kappa}
   and performed integrations by parts (see \eq{eq-replicated-Mayer-result} for the same calculation.).

The integrations over
each $(Q_{i})_{ab}$ can be performed in the $M \to \infty$ limit by
the saddle point method. The saddle point equation read,
 \beq
 0=\left. \frac{\partial s[\hat{Q}_{i}]}{\partial (Q_{j})_{ab}} \right |_{
   [\hat{Q}_{i}=\hat{Q}^{*}_{i}]} \qquad \mbox{for} \qquad (j=1,2,\ldots,N)
 \eeq
 Now repeating the same argument as in sec.\ref{subsubsec:crystal-free-energy},
we can assume that the equations
admit a uniform solution $\hat{Q}_{i}^{*}=\hat{Q}$ for $\forall i$
since in our system every vertex is equivalent to each other.
 As the result we obtain the free-energy associated with such a saddle point as,
 \beq
 -\beta\frac{F}{NM} =\partial_{n} \left . s_{n}[\hat{Q}] \right |_{n=0}
 \label{eq-F-functional-glass}
 \eeq
 with
\beqn
s_{n}[\hat{Q}] &\equiv &s_{\rm ent} [\hat{Q}]
-\frac{\alpha}{p}  {\cal F}_{\rm int}[\hat{Q}] \nonumber \\
-{\cal F}_{\rm int}[\hat{Q}] &\equiv &  \exp\left(\frac{1}{2}\sum_{a,b=1}^{n}Q_{ab}^{p}\frac{\partial^{2}}{\partial h_{a}\partial h_{b}} \right) \left. \prod_{a=1}^{n}e^{-\beta V(\delta +h_{a})} \right |_{h_{a}=0}
\label{eq-F-int-glass}
\eeqn
where $-{\cal F}_{\rm int}$ represents the interaction part of the free-energy (free-entropy). Importantly $Q_{ab}$ must satisfy the saddle point equations,
 \beq
 0=\frac{\partial s[\hat{Q}]}{ \partial Q_{ab}}
 \label{eq-saddle-point-glass}
 \eeq
 It is also required to satisfy the stability condition, i.e.
the eigen values of the the Hessian matrix,
 \beq
H_{(ab),(cd)} \equiv -\frac{\partial^{2}s_{n}[\hat{Q}]}{\partial Q_{ab} \partial Q_{cd} }
  \label{eq-replica-hessian}
  \eeq
  in the $n \to 0$ limit,  must be non negative.

  The exact free-energy functional
given by  \eq{eq-F-functional-glass} with \eq{eq-F-int-glass}
  can be derived also using
a density functional approach as we show 
in appendix  \ref{appendix_density_functional}
for the case of continuous spins.
It is done closely following the strategy used in the
recent replicated liquid theory for
hardsphere glass in the large dimensional limit \cite{kurchan2012exact,kurchan2013exact,charbonneau2014exact}.

  \subsection{Interpolation between disorder-free and completely disordered model}
  \label{sec:quenched_disorder}

In the present paper we are most concerned with systems 
where the disorder is self-generated.
However it is very instructive to consider also
the model with quenched disorder given by \eq{eq-X-disordered},
\begin{equation}
r_{\bs}=\delta- \frac{1}{\sqrt{M}} \sum_{\mu=1}^{M}
\left[
  \frac{\lambda}{\sqrt{M}}
  +\sqrt{1-\left(\frac{\lambda}{\sqrt{M}}\right)^{2}}
  \xi^{\mu}_{\bs}\right]
S^{\mu}_{1(\bs)}S^{\mu}_{2(\bs)}\cdots S^{\mu}_{p(\bs)}
\qquad  \left ( 0 \leq \frac{\lambda}{\sqrt{M}} \leq 1 \right)
\label{eq-p-body-model-disorder}
\end{equation}
Here $\xi^{\mu}_{\bs}$ is a random variable 
with Gaussian distribution with zero mean and unit variance.
With this parametrization, we have
a continuous
interpolation between the disorder-free model $\lambda/\sqrt{M}=1$
and completely disordered, spinglass model $\lambda/\sqrt{M}=0$.

Analysis of this model is useful for the following reasons.
\begin{itemize}
\item
  We can show that the self-generated disorder and the quenched disorder act additively.

\item In the disorder-free model given by \eq{eq-X-disorder-free},
  which corresponds to $\lambda/\sqrt{M}=1$,
the energy scale of glass transition and crystalline transitions
are widely separated. For instance the $p$-spin ferromagnetic
Ising model in the $p \to \infty$ limit exhibit a ferromagnetic transition at
$k_{\rm B}T_{\rm c}/J =O(\sqrt{M})$  given by \eq{eq-large-p-spin-ferromagnetic-Tc}
while the glass transition takes place in the supercooled paramagnetic sector
at $k_{\rm B}T_{\rm K}/J =O(1)$ given by  \eq{eq-large-p-spin-ferromagnetic-Tk}.
Actually the fact that $T_{\rm K}$ is lower than $T_{\rm c}$ is natural by itself but they become too much separated
in the disorder-free model in the $M \to \infty$ limit.
With the choice of the disordered model given by \eq{eq-X-disordered} we can bring the energy scales
of two transitions much closer. This is achieved in two steps:
(i) reduce the interaction energy scale down to order $O(1/\sqrt{M})$
of the original 'disorder-free' model 
(ii) then add additional quenched disorder such that the effective
energy scale for the glass transition is brought back to the original level
of $O(1)$. Bringing the crystallization and glass transition temperatures closer, it becomes easier to investigate competitions between liquid, glass and crystalline phases. Note that similar treatments for the energy scales
are considered in standard spinglass models with ferromagnetic biases \cite{kirkpatrick1978infinite}. 
\end{itemize}

Averaging over the quenched disorder of the replicated partition function
given by \eq{eq-partition-function-replica}, we obtain,
\beqn
&&  \overline{Z^{n}}^{\xi}
=
\left( \prod_{i}\prod_{a=1}^{n}\Tr^{a}_{i} \right)
\prod_{\bs} \left \{  \prod_{a} \inti  \frac{d\kappa_{\bs,a}}{2\pi}
Z_{\kappa_{\bs,a}} e^{i\kappa_{\bs,a} \delta}
e^{{\cal L}_{\bs}} \right \}
\label{eq-partitionfunction-disorder}
\eeqn
where the overline denotes the average over the disorder and we introduced,
\beqn
&& {\cal L}_{\bs} \equiv \lambda \sum_{a=1}^{n}(-i\kappa_{\bs,a})
\frac{1}{M}\sum_{\mu=1}^{M} 
(S^{a})^{\mu}_{1(\bs)}(S^{a})^{\mu}_{2(\bs)}\cdots (S^{a})^{\mu}_{p(\bs)} \nonumber \\
&&+ \sum_{a,b=1}^{n}\frac{ (-i\kappa_{\bs,a})  (-i\kappa_{\bs,b})}{2}
\left(
1-\left(\frac{\lambda}{\sqrt{M}} \right)^{2}
  \right)
\frac{1}{M}\sum_{\mu=1}^{M} 
(S^{a})^{\mu}_{1(\bs)}(S^{b})^{\mu}_{1(\bs)}
\cdots (S^{a})^{\mu}_{p(\bs)}(S^{b})^{\mu}_{p(\bs)} \;\;\; \;\;\; \;\;\;
\label{eq-partition-function-replica-quenched-disorder}
\eeqn

For the order parameters we may consider both the glass
order parameters  given by \eq{eq-def-Qab} and
the crystalline one given by \eq{eq-ferromagnetic-order-parameter}.
Since we are considering the $M \to \infty$ limit we naturally define the following
set of local order parameters,
\beqn
(Q_{i})_{ab} & = & \lim_{\epsilon_{ab} \to 0} \lim_{M \to \infty} \frac{1}{M} 
\sum_{\mu=1}^{M}\langle (S^{a})^{\mu}_{i} (S^{b})^{\mu}_{i} \rangle_{\epsilon} \\
m_{a} & = & \lim_{h_{a} \to 0} \lim_{M \to \infty} \frac{1}{M} 
\sum_{\mu=1}^{M}\langle (S^{a})^{\mu}_{i}  \rangle_{h}
\label{eq-def-Qab-ma}
\eeqn
and the corresponding identities,
\beqn
1 &= &\inti d(Q_{ab})_{i} \delta \left(M (Q_{ab})_{i}-\sum_{\mu=1}^{M} (S^{a})_{i}^{\mu}(S^{b})_{i}^{\mu}\right) \nonumber \\
&=& M\intii \frac{d(\epsilon_{ab})_{i}}{2\pi} \inti d(Q_{ab})_{i}
e^{(\epsilon_{ab})_{i} (M (Q_{ab})_{i}-\sum_{\mu=1}^{M}(S^{a})^{\mu}_{i}(S^{b})^{\mu}_{i})} \;\;\;\\
1 &=& \inti d(m_{a})_{i} \delta (M(m_{a})_{i}-\sum_{\mu=1}^{M} S_{i}^{\mu})
=M \intii \frac{dh^{a}_{i}}{2\pi} \inti d(m_{a})_{i} e^{h^{a}_{i} (M(m_{a})_{i}-\sum_{\mu=1}^{M}S^{\mu}_{i})} \qquad 
\eeqn

\subsubsection{Spin trace}

Using the identities shown above spin traces can be expressed in the $M \to \infty$ limit as,
\beqn
&& \prod_{c=1}^{n} \Tr_{\vS_{c}}  \cdots =
\left( \prod_{a < b} \inti dQ_{ab} \right)\left( \prod_{a} \inti dm_{a} \right) e^{M s_{\rm ent}[\hat{Q},\hat{m}]}
\prod_{\mu} \langle \cdots \rangle_{\mu}
\label{eq-spin-trace-glass-ferro-bias}
\eeqn
where
\beqn
&& s_{\rm ent}[\hat{Q},\hat{m}]=\sum_{a<b}
  \epsilon^{*}_{ab}Q_{ab}+\sum_{a} h_{a}^{*}m_{a}+\ln \prod_{c=1}^{n}\Tr_{S^{c}}
  e^{-\sum_{a<b}\epsilon^{*}_{ab}S^{a}S^{b}-\sum_{a}h^{*}_{a}S^{a}}
  \nonumber \\
  && \langle \cdots \rangle_{\mu} =
\frac{\prod_{c=1}^{n}\Tr_{\vS_{c}}
  e^{{\cal L}^{\rm ent}_{\mu}}\cdots}{
  \prod_{c=1}^{n}\Tr_{\vS_{c}}
  e^{{\cal L}^{\rm ent}_{\mu}} }
\qquad
{\cal L}^{\rm ent}_{\mu} =
-\sum_{a<b}\epsilon_{ab}^{*} (S^{a})^{\mu} (S^{b})^{\mu}
-\sum_{a}h^{*}_{a}(S^{a})^{\mu}
    \nonumber \\
&& Q_{ab}=
    \frac{\prod_{c} \Tr_{\vS_{c}}e^{{\cal L}^{\rm ent}} S^{a}S^{b}}{\prod_{c}\Tr_{\vS_{c}} e^{{\cal L}^{\rm ent}}}
\qquad m_{a}=\frac{\prod_{c}\Tr_{\vS_{c}} e^{{\cal L}^{\rm ent}}S^{a}}{
  \prod_{c}\Tr_{\vS_{c}}e^{{\cal L}^{\rm ent}}} 
\label{eq-spin-trace-2-replica}
\eeqn
Here we introduced a short hand notation $\hat{m}=(m_{1},m_{2},\ldots,m_{n})$.
The last equations are the saddle point equations
for the integrations over $\epsilon_{ab}$  and  $h_{a}$ 
which fix the saddle points $\epsilon_{ab}^{*}=\epsilon^{*}_{ab}[\hat{Q},\hat{m}]$
and  $h_{a}^{*}=h^{*}_{a}[\hat{Q},\hat{m}]$.
Using \eq{eq-spin-trace-glass-ferro-bias} we find, for example,
\beqn
&& \prod_{c}\Tr_{\vS_{c}} (S^{a})^{\mu}=
\left( \prod_{a<b}\inti d(Q_{ab})\inti \prod_{a}d(m_{a}) \right) e^{M s_{\rm ent}[\hat{Q},\hat{m}]} m_{a}
\nonumber \\
&& \prod_{c}\Tr_{\vS_{c}} (S^{a})_{i}^{\mu}(S^{b})^{\mu}
=\left(\prod_{a<b}\inti d(Q_{ab})\right) \left(\prod_{a}\inti d(m_{a}) \right)e^{M s_{\rm ent}[\hat{Q},\hat{m}]} Q_{ab}
\qquad 
\label{eq-simple-average-replica-with-ferro}
\eeqn

More precisely, by taking the spin traces we obtain the following
expressions for the Ising and continuous spin systems,
\begin{itemize}

\item {\bf Continuous spin}:  We find similarly to
\eq{eq-spin-trace-continuous-2-replica},
\beqn
 s_{\rm ent}[\hat{Q},\hat{m}]&=&\frac{1}{2}\sum_{a,  b} 
 \epsilon^{*}_{ab}Q_{ab}+\sum_{a}h_{a}^{*}m_{a}
+ \ln  \sqrt{\frac{(2\pi)^{n}}{{\rm det} \epsilon^{*}}}
+\frac{1}{2}\sum_{ab}h^{*}_{a}(\epsilon^{*})^{-1}_{ab}h^{*}_{b}
\nonumber \\
&  = & \frac{n}{2}+\frac{n}{2} \ln (2\pi) + \frac{1}{2}\ln {\rm det}
(\hat{Q}-\hat{m}^{T}\hat{m})
\label{eq-sent-q-and-m-spherical}
\eeqn
and
\beqn
\langle \cdots \rangle_{\mu}  &\equiv&
\frac{
  \inti dS^{\mu} e^{{\cal L}_{\mu}}
    \ldots
  }{
  \inti dS^{\mu} e^{{\cal L}_{\mu}}      
} \qquad
{\cal L}_{\mu} =
-\frac{1}{2}\sum_{a,b}\epsilon_{ab}^{*} (S^{a})^{\mu} (S^{b})^{\mu}
-\sum_{a}h^{*}_{a}(S^{a})^{\mu}
\eeqn
Here we have performed integration over $\epsilon_{ab}$ and $h_{a}$ by the saddle point method which yield a saddle point
\beq
(\hat{\epsilon}^{*})_{ab}^{-1}=Q_{ab}-m_{a}m_{b}
\qquad 
h^{*}_{a}=-\sum_{b}\epsilon_{ab}^{*}m_{b}.
\eeq

\item {\bf Ising spin}: We find similarly to \eq{eq-spin-trace-Ising-2-replica},
  \beq
   s_{\rm ent}[\hat{Q},\hat{m}]=\sum_{a \neq b} 
   \frac{1}{2}  \epsilon^{*}_{ab}Q_{ab}
   +\sum_{a}h_{a}^{*}m_{a}
   +
\ln
e^{-\sum_{a<b}\epsilon^{*}_{ab}\frac{\partial^{2}}{\partial h_{a}\partial h_{b}}}
\left. \prod_{a} 2\cosh (h_{a})\right |_{\{h_{a}=h^{*}_{a}\}}
\label{eq-sent-q-and-m-ising}
\eeq
where we performed the spin trace formally as
\beq
{\rm Tr}_{\vS_{c}}e^{-\sum_{a<b}\epsilon_{ab}S^{a}S^{b}-\sum_{a}h_{a}S^{a}}
= e^{-\sum_{a<b}\epsilon_{ab}\frac{\partial^{2}}{\partial h_{a}\partial h_{b}}}
\prod_{a} 2\cosh (h_{a}) 
\eeq
For the integration over $\epsilon_{ab}$ and $h_{a}$,
the saddle points $\epsilon_{ab}^{*}=\epsilon_{ab}^{*}[\hat{Q},\hat{m}]$
and $h_{a}^{*}=h_{a}^{*}[\hat{Q},\hat{m}]$ are obtained formally as,
\beqn
&&
Q_{ab}=- \frac{\delta}{\delta \epsilon_{ab}}\ln e^{-\sum_{a<b}\epsilon_{ab}\frac{\partial^{2}}{\partial h_{a}\partial h_{b}}} 
\left. \prod_{a} 2\cosh (h_{a})
\right |_{\{\epsilon_{ab}=\epsilon^{*}_{ab}[\hat{Q},\hat{m}],h_{a}=h^{*}_{a}[\hat{Q},\hat{m}]\}} \nonumber \\
\qquad
&& m_{a}=- \frac{\delta}{\delta h_{a}}\ln e^{-\sum_{a<b}\epsilon_{ab}\frac{\partial^{2}}{\partial h_{a}\partial h_{b}}} 
\left. \prod_{a} 2\cosh (h_{a})
\right |_{\{\epsilon_{ab}=\epsilon^{*}_{ab}[\hat{Q},\hat{m}],h_{a}=h^{*}_{a}[\hat{Q},\hat{m}]\}} \nonumber \\
\label{eq-spin-trace-Ising-2-replica-with-ferro}
\eeqn

  \end{itemize}

\subsubsection{Evaluation of the free-energy}

In  \eq{eq-spin-trace-glass-ferro-bias}
we notice again that different spin components $\mu$ are decoupled
in the average $\prod_{\mu}\langle \ldots \rangle_{\mu}$.
Then we can evaluate the spin trace in the replicated
partition function given by \eq{eq-partition-function-replica}
in the $M \to \infty$ limit using
the cumulant expansion
given by \eq{eq-cumulant-expansion} and \eq{eq-simple-average-replica-with-ferro} as,
\beqn
&& \prod_{a}\prod_{i} \Tr_{\vS_{i,a}} \prod_{\bs} e^{{\cal L}_{\bs}}
\nonumber \\
&& \xrightarrow[M \to \infty]{} 
\left( \prod_{i} \prod_{a < b} \inti d(Q_{i})_{ab} \prod_{a} \inti d(m_{i})_{a}\right)
\exp
\left [
  \lambda \sum_{a=1}^{n}(-i\kappa_{\bs,a})
  (m_{1(\bs)})_{a}(m_{2(\bs)})_{a}\cdots (m_{p(\bs)})_{a}\right.
  \nonumber \\
  &&+ \left. \sum_{a,b=1}^{n}\frac{ (-i\kappa_{\bs,a})  (-i\kappa_{\bs,b})}{2}
  (Q_{1(\bs)})_{ab}
  (Q_{2(\bs)})_{ab}
  \cdots
  (Q_{p(\bs)})_{ab}
  \right]
\eeqn
Here we point out that the last term in the exponent of the last equation
is the result of a summation of the contributions of two different different kinds of disorder:
(1) quenched disorder of amplitude $1-(\lambda/\sqrt{M})^{2}$
(see the 2nd term in \eq{eq-partition-function-replica-quenched-disorder})
(2) self-generated disorder of amplitude $(\lambda/\sqrt{M})^{2}$.
Now it is clear that parametrization given by \eq{eq-X-disordered} is chosen such that
the energy scale of the glass transition does not change between
the disorder-free limit ($\lambda/\sqrt{M}=1$) and completely disordered limit $\lambda/\sqrt{M}=0$.

Collecting the above results, the disorder averaged replicated
partition function given by \eq{eq-partitionfunction-disorder}
can be rewritten formally in the $M \to \infty$ limit as,
\beqn
\overline{Z^{n}}^{\xi}
  &=&
\prod_{i} \left ( \prod_{a< b} \inti d(Q_{i})_{ab} \prod_{a}\inti d(m_{i})_{a} \right)
e^{NM s[\hat{Q}_{i},\hat{m}_{i}]}
 \label{eq-partition-function-glass-ferro}
\eeqn
with
\beqn
&& s[\hat{Q}_{i},\hat{m}_{i}]
= \frac{1}{N}\sum_{i=1}^{N} 
\Bigg \{ s_{\rm ent} [\hat{Q}_{i},\hat{m}_{i}]    
  \label{eq-action-glass-ferro}
\\
&& \left.
+\frac{1}{pM}\sum_{\bs \in \partial_{i}} 
  \frac{1}{2}\sum_{a,b}\frac{\partial^{2}}{\partial h_{a}\partial h_{b}}
  (Q_{1(\bs)})_{ab}
  \cdots (Q_{p(\bs)})_{ab}
  \prod_{a=1}^{n}e^{-\beta V(\delta -
    \lambda (m_{1(\bs)})_{a}
    \ldots   (m_{p(\bs)})_{a} +
    h_{a})}  \right |_{\{h_{a}=0\}} \Bigg\}
  \;\;
\nonumber
   \eeqn

The integrations over
each $(Q_{i})_{ab}$ and $(m_{i})_{a}$ can be performed in the $M \to \infty$ limit by
the saddle point method. The saddle point equations read,
 \beqn
&&  0=\left. \frac{\partial s[\hat{Q}_{i},\hat{m}_{i}]}{\partial (Q_{j})_{ab}} \right |_{
   [\hat{Q}_{i}=\hat{Q}^{*}_{i},\hat{m}_{i}=\hat{m}^{*}_{i}]} \qquad \mbox{for} \qquad (j=1,2,\ldots,N) \nonumber \\
 &&  0=\left. \frac{\partial s[\hat{Q}_{i},\hat{m}_{i}]}{\partial (m_{j})_{a}} \right |_{
   [\hat{Q}_{i}=\hat{Q}^{*}_{i},\hat{m}_{i}=\hat{m}^{*}_{i}]} \qquad \mbox{for} \qquad (j=1,2,\ldots,N)
 \eeqn

After the average over the quenched disorder every vertex has become
again identical to each other. Then we can repeat the same argument
as in sec. \ref{subsubsec:crystal-free-energy} and 
assume uniform solutions : $\hat{Q}_{i}^{*}=\hat{Q}$ and
 $\hat{m}_{i}^{*}=\hat{m}$ for $\forall i$.
 As the result we obtain the free-energy associated with such a saddle point as,
 \beq
 -\beta\frac{F}{NM} =\partial_{n} \left . s_{n}[\hat{Q},\hat{m}] \right |_{n=0}
 \label{eq-free-entropy-Q-and-m}
 \eeq
 with
\beqn
s_{n}[\hat{Q}] &\equiv &s_{\rm ent} [\hat{Q},\hat{m}]
-\frac{\alpha}{p}  {\cal F}_{\rm int}[\hat{Q},\hat{m}] \nonumber \\
-{\cal F}_{\rm int}[\hat{Q},\hat{m}] &\equiv &  \exp\left(\frac{1}{2}\sum_{a,b=1}^{n}(\hat{Q})_{ab}^{p}\frac{\partial^{2}}{\partial h_{a}\partial h_{b}} \right) \left. \prod_{a=1}^{n}e^{-\beta V(\delta-\lambda m^{p}_{a} +h_{a})} \right |_{h_{a}=0}
\label{eq-F-int-glass-with-m}
 \eeqn
 where $Q_{ab}$ and $m_{a}$ must satisfy the saddle point equations
 \beq
 0=\frac{\partial s[\hat{Q},\hat{m}]}{\partial Q_{ab}} \qquad
  0=\frac{\partial s[\hat{Q},\hat{m}]}{\partial m_{a}}
 \label{eq-saddle-point-glass-with-ferro}
 \eeq
 It is also required to satisfy the stability condition, i.e.
the eigen values of the the Hessian matrix,
 \beq
 H_{(ab),(cd)} \equiv -\frac{\partial^{2}s_{n}[\hat{Q},\hat{m}]}{\partial Q_{ab} \partial Q_{cd} }
 \qquad
   H_{(ab),c} \equiv -\frac{\partial^{2}s_{n}[\hat{Q},\hat{m}]}{\partial Q_{ab} \partial m_{c} }
 \qquad 
  H_{a,b} \equiv -\frac{\partial^{2}s_{n}[\hat{Q},\hat{m}]}{\partial m_{a} \partial m_{b} }
  \label{eq-replica-hessian-with-ferro}
  \eeq
  in the $n \to 0$ limit,  must be non negative.

 Finally let us note again that the disorder-free model can be recovered by choosing $\lambda/\sqrt{M}=1$
 in the above expressions.  For the disorder-free model discussed in previous sections,
 we gave free-energy functional in terms of the crystalline order parameter $m$ in \eq{eq-sn-ferro} and that in terms of the glass order parameter $Q_{ab}$ in \eq{eq-F-int-glass} separately
just to simplify the presentations.
In any case here we now have complete free-energy functional where both the crystalline and glass order parameters are present.

\subsection{Stability against crystallization}
\label{sec-stability-against-crystalization}

Given the complete free-energy functional in term of both the crystalline and glass order parameters,
we can now investigate the stability of glassy phases against crystallization
extending the analysis
in sec. \ref{subsec-crystalline-transitions-and-supercooling} which
was limited to the liquid phase.
Here we limit ourselves with a glassy phase without crystalline order parameter $m=0$ and do not consider
possible 'glassy crystals' with $m > 0$. First we note that,
\beq
\left.   H_{(ab),c} \right |_{\{m_{a}=0\}}
= \left.-\frac{\partial^{2}s_{n}[\hat{Q},\hat{m}]}{\partial Q_{ab} \partial m_{c} }
\right |_{\{m_{a}=0\}}=0
\label{eq-decouling-between-q-and-m-for-m=0}
\eeq
holds. This can be checker by taking the derivatives explicitly.
     For the entropic part  we find,
     \beq
\left. \frac{\partial^{2}s_{\rm ent}[\hat{Q},\hat{m}]}{\partial Q_{ab} \partial m_{c} }
\right |_{\{m_{a}=0\}}
=\left. \frac{\partial}{\partial Q_{ab}} h^{*}[\hat{Q},\hat{m}] \right |_{\{m_{a}=0\}}=0
\eeq
The last equation follows from
the last equation of \eq{eq-spin-trace-2-replica}
which implies that $m_{a}=0$ requires $h^{*}_{a}=0$. For the interaction part
of the free-energy we find,
\beqn
&& - \left. \frac{\partial^{2}{\cal F}_{\rm int}[\hat{Q},\hat{m}]}{\partial Q_{ab} \partial m_{c} }
      \right |_{\{m_{a}=0\}} \nonumber \\
     &=& 
     pQ_{ab}^{p-1}\frac{\partial^{2}}{\partial h_{a}\partial h_{b}}
     e^{\frac{1}{2}\sum_{a,b=1}^{n}(\hat{Q})_{ab}^{p}\frac{\partial^{2}}{\partial h_{a}\partial h_{b}}}
(-\lambda p m_{c}^{p-1}) (-\beta V'(\delta-\lambda m_{c}^{p}+h_{c}))
      \prod_{a=1}^{n}e^{-\beta V(\delta-\lambda m^{p}_{a} +h_{a})}
     \Bigg|_{\{h_{a},m_{a}=0\}}
     \nonumber \\
&=&  0
     \eeqn
     The last equation holds for $p>1$.
     Thus \eq{eq-decouling-between-q-and-m-for-m=0} must hold.

     Then the local stability of the glassy phase with $m=0$ against
     crystallization is solely determined by the matrix,
     \beqn
&&    H_{a,b}= \left. -\frac{\partial^{2}s_{n}[\hat{Q},\hat{m}]}{\partial m_{a} \partial m_{b} }
    \right |_{\{m_{a}=0\}} \nonumber \\
&&    =Q_{ab}^{-1}
     -\delta_{ab} \lambda \alpha (p-1) \left . m_{a}^{p-2}\right|_{m_{a}=0}
          e^{\frac{1}{2}\sum_{a,b=1}^{n}(\hat{Q})_{ab}^{p}\frac{\partial^{2}}{\partial h_{a}\partial h_{b}}}
     (-\beta V'(\delta+h_{a})
          \left. \prod_{a=1}^{n}e^{-\beta V(\delta +h_{a})} \right |_{h_{a}=0}
          \qquad
     \label{eq-Hab}
     \eeqn
     where we assumed $p>1$ and we used
\beq
\left. -\frac{\partial^{2}s_{\rm ent}[\hat{Q},\hat{m}]}{\partial m_{a} \partial m_{b} }
\right |_{\{m_{a}=0\}}=Q_{ab}^{-1}.
\eeq
which follows from \eq{eq-spin-trace-2-replica}. \footnote{Taking $\partial_{m_{b}}$
on both sides of the last equation of \eq{eq-spin-trace-2-replica} we find
$\delta_{ab}=\sum_{c}\frac{\partial h_{c}^{*}}{\partial m_{b}}(Q_{ab}-m_{a}m_{b})$
thus $\left. -\frac{\partial^{2}s_{\rm ent}[\hat{Q},\hat{m}]}{\partial m_{a} \partial m_{b} } \right |_{\{m_{a}=0\}}=\left. \frac{\partial h_{a}^{*}}{\partial m_{b}}\right |_{\{m_{a}=0\}}=Q^{-1}_{ab}$.}

In the liquid phase we have $m_{a}=0$ and $Q_{ab}=\delta_{ab}$
and thus $Q^{-1}_{ab}=\delta_{ab}$.
due to spin normalization (see \eq{eq-Q-diag}).
There one can check that non-negativeness
of the eigenvalues of the matrix \eq{eq-Hab}
in $n \to 0$ limit 
becomes equivalent to the stability conditions
\eq{eq-stability-para-ising} and \eq{eq-stability-para-continuous}
of the paramagnetic solution $m=0$ as it should.

Here we see that the 2nd term on the r.h.s of \eq{eq-Hab},
which is due to the interaction part of the free-energy,
vanishes in two cases
(i) $ p> 2$ (ii) $p=2$ with non-linear potential with the flatness $V'(\delta)=0$.
Remarkably in these cases the matrix becomes independent of $\lambda$ 
and its the eigen values are simply the inverse of the eigen values of the matrix $Q_{ab}$.
We expect the latters are positive for physical solutions.
\footnote{
For instance note that in the case of the continuous spins
the entropic part of the free-energy has a term $\ln {\rm det}(\hat{Q})$
with $m_{a}=0$ (see \eq{eq-sent-q-and-m-spherical}).
Thus the eigenvalues of the matrix $Q_{ab}$ are needed to be positive.}
This is very interesting because including the regime of
large enough $\lambda$, especially the disorder-free case $\lambda=\sqrt{M}$,
where we naturally expect crystalline order as the true equilibrium phase,
paramagnetic phase $m=0$ (for which $Q_{ab}=\delta_{ab}$ due to the spin normalization) and also the glassy phase with $m=0$ (for which $Q_{ab} \neq \delta_{ab}$)
remains locally stable against
crystallization for the two cases: (i) and (ii).

Contrarily, in the case of $p=2$ without the flat potential
the $m=0$ solution cannot be stable against
crystallization if $\lambda/\sqrt{M}$ is finite
in $M \to \infty$ limit, including
the in particular the disorder-free case $\lambda=\sqrt{M}$.
Thus in these cases the quenched disorder is necessary
to realize the glass phases.
The range of the stability of the liquid $Q_{ab}=\delta_{ab}$
and glass phase $Q_{ab} \neq \delta_{ab}$ with
a given $\lambda$ must be examined analyzing
the eigenvalues of \eq{eq-Hab}.


Finally we note that the situation can change in systems with
finite connectivity. The supercooled paramagnetic phase can
disappear for sufficiently large $\lambda$.
In the context of statistical inference problems this is an important issue
because one has to find the hidden crystalline
state (ground truth) in the immense sea of wrong solutions
(glasses) \cite{zdeborova2016statistical}.

  \section{Linear potential: connection to the standard $p$-spin Ising/spherical spinglass models and the random energy model}
  \label{sec-linear}

Let us discuss here the simplest case, the linear potential given by \eq{eq-linear-potential} which reads,
\begin{equation}
V(x)=Jx.
\end{equation}
The interaction part of the free-energy given by \eq{eq-free-entropy-Q-and-m}
becomes,
\beq
-{\cal F}_{\rm int}[\hat{Q},\hat{m}]=
e^{\lambda (\beta J) \sum_{a} m^{p}_{a}+\frac{(\beta J)^{2}}{2}\sum_{a,b=1}^{n}Q_{ab}^{p}}
\eeq
then we find
\beq
-\frac{\beta F}{NM}=\partial_{n} s_{n} |_{n=0}
\qquad
s_{n}=s_{\rm ent}(\hat{Q},\hat{m})
+\lambda (\beta J) \sum_{a}m^{p}_{a}
+\frac{(\beta J)^{2}}{2}\frac{\alpha}{p} \sum_{a,b}Q_{ab}^{p}
\label{eq-p-spin-q-and-m}
\eeq
where $s_{\rm ent}(\hat{Q},\hat{m})$ is given in
\eq{eq-sent-q-and-m-spherical} for the continuous spin case
and \eq{eq-sent-q-and-m-ising} for the Ising case.
This is  exactly the same as those of the standard $p$-spin Ising/spherical spinglass models with $M=1$ but with global couplings \cite{derrida1980random,crisanti1992sphericalp,crisanti2013exactly} by choosing $\alpha/p=1/2$.
Note that such a correspondence has been known for the case of a $p=2$ continuous spin model with global coupling \cite{de1978infinite}.

Let us summarize below some important known results of the $p$-spin spinglass models. The case $p=2$ with the Ising spin corresponds to
the SK (Sherrington-Kirkpatrick) model\cite{kirkpatrick1978infinite}.
It exhibits a continuous phase transition from the paramagnetic
to the spinglass phase accompanying the continuous replica symmetry breaking (RSB) \cite{parisi1979infinite} while the spherical version of it exhibits a continuous phase transition but without RSB \cite{kosterlitz1976spherical}. The SK model is the standard mean-field model for spinglasses \cite{binder1986spin,mydosh1993spin}.
On the other hand $p>2$ system exhibit 1 step RSB \cite{derrida1980random,gross1984simplest,crisanti1992sphericalp} with a discontinuous transition from the paramagnetic
to the spinglass phase. These models show the essence of the glass phenomenology such as the dynamical and static glass transitions so that they are regarded as prototypical theoretical model to capture the physics of structural glasses \cite{kirkpatrick1987dynamics,KTW89,cavagna2009supercooled,castellani2005spin}. Among the latter models those with the Ising spin exhibit yet another glass transition to enter the continuous RSB phase at lower temperatures \cite{Ga85}. In the $p \to \infty$ limit of the Ising case, the random energy model is recovered \cite{derrida1980random}.

We emphasize that the result given by \eq{eq-p-spin-q-and-m}
is valid also in the disorder-free limit $\lambda/\sqrt{M}=1$.
Indeed in sec. \ref{sec-REM} we have shown that the random energy model is
recovered in the disorder-free limit. Thus the disorder-free model
have sufficient amount of
self-generated disorder to realize glass transitions.
The supercooled paramagnetic state and the glass phase which emerge
there are stable against crystallization for $p >2$ as discussed in
sec. \ref{sec-stability-against-crystalization}.
In the case $p=2$, however, we have
to invoke the quenched disorder
to suppress the crystalline (ferromagnetic) states.
(This amount to yield nothing but the SK model for the Ising spins and spherical SK model for the continuous spins mentioned above.)

\section{Replica symmetric (RS) ansatz}
\label{sec-RS}

For the rest of the present paper we study glass transitions of our model, 
which emerge within the supercooled paramagnetic phase with
no crystalline order $m=0$. And we limit our selves with the continuous spin models
for the rest of the present paper.
In the present section and
in the next section we derive some generic results within
the replica symmetric (RS) and replica symmetry breaking (RSB) ansatz.
We apply these schemes to systems with non-linear potentials in later sections.

Our starting point is the free-energy functional
given by \eq{eq-F-functional-glass} which reads,
 \beq
-\beta f[\hat{Q}]=  -\beta\frac{F[\hat{Q}]}{NM} =\partial_{n} \left . s_{n}[\hat{Q}] \right |_{n=0}
\label{eq-F-functional-glass-ref}
 \eeq
with \eq{eq-F-int-glass} which reads,
\beqn
s_{n}[\hat{Q}] &\equiv &
\frac{1}{2}\ln {\rm det}
(\hat{Q})
-\frac{\alpha}{p}  {\cal F}_{\rm int}[\hat{Q}] \nonumber \\
-{\cal F}_{\rm int}[\hat{Q}] &\equiv &  \exp\left(\frac{1}{2}\sum_{a,b=1}^{n}Q_{ab}^{p}\frac{\partial^{2}}{\partial h_{a}\partial h_{b}} \right) \left. \prod_{a=1}^{n}e^{-\beta V(\delta +h_{a})} \right |_{\{h_{a}=0\}}
\label{eq-s-glass-ref}
\eeqn
Here ${\cal F}_{\rm int}$ represents the interaction part of the free-energy.
For the entropic part in \eq{eq-F-int-glass}
we used the expression given by \eq{eq-spin-trace-continuous-2-replica}
and we omitted irrelevant constants $\frac{n}{2}+\frac{n}{2} \ln (2\pi)$ for simplicity.

The pressure given by \eq{eq-def-pressure} can be computed as,
\beq
\Pi = -\frac{p}{\alpha}\frac{\partial \beta f[\hat{Q}]}{\partial \delta}
=
-\frac{\partial }{\partial \delta} \left. \partial_{n} {\cal F}_{\rm int} \right|_{n=0}.
\label{eq-pressure-replica}
\eeq
and similarly the distribution function of the gap given by \eq{eq-g-of-r} as,
\beq
g(r)=-
\frac{p}{\alpha}\frac{\delta \beta f[\hat{Q}]}{\delta (-\beta V(r))}
=
-\frac{\delta }{\delta (-\beta V(r))} \left. \partial_{n} {\cal F}_{\rm int} \right|_{n=0}.
\label{eq-g-of-r-replica}
\eeq

Before passing let us recall the discussion in  sec \ref{sec-Non-linear-potentials-with-flatness} and 
sec. \ref{sec-stability-against-crystalization} that the supercooled paramagnetic $m=0$ states and glassy states of the
model is locally stable against crystallization if $p> 2$. But for the case $p=2$ we must have the flatness $V'(\delta) =0$
or quenched disorder. Although we may not mention these points often in the following, we must keep these in our minds.

\subsection{Formulation}

In the replica symmetric (RS) ansatz we assume the following form of the
overlap matrix parametrized by a single parameter $q$,
\beq
Q^{\rm RS}_{ab}=(1-q)\delta_{ab}+q.
\label{eq-RS-ansatz}
\eeq
Note that diagonal part $Q_{aa}=1$ reflects the spin normalization.

\subsubsection{Free-energy}

First let us compute the free-energy
given by \eq{eq-F-functional-glass-ref}
-\eq{eq-s-glass-ref}
within the RS ansatz.
Using \eq{eq-RS-ansatz} we find,
\beq
\ln {\rm det}{\hat Q}^{\rm RS}=\ln [1+(n-1)q]+(n-1)\ln (1-q)
\eeq
so that the entropic part of the free-energy is obtained as
\beq
\left. \frac{1}{2}\partial_{n} \ln {\rm det}{\hat Q}^{\rm RS}\right|_{n=0}
=\frac{1}{2}\left(\frac{q}{1-q}+\ln(1-q) \right).
\eeq
The interaction part of the free-energy is obtained as,
\begin{eqnarray}
   - {\cal F}_{\rm int}[\hat{Q}^{\rm RS}]&=&\left. \exp \left(
    \frac{1}{2}\sum_{a,b=1}^{n} [(1-q^{p})\delta_{ab}+q^{p}]
    \frac{\partial^{2}}{\partial h_{a}\partial h_{b}} \right)
    \prod_{a=1}^{n}e^{-\beta V(\delta+h_{a})}\right |_{\{h_{a}=0\}}\nonumber \\
    &=& \left. \exp \left(\frac{1}{2}q^{p}\sum_{a,b=1}^{n}
    \frac{\partial^{2}}{\partial h_{a}\partial h_{b}} \right)
    \prod_{a=1}^{n}
   \left  \{
    \exp\left(\frac{1}{2}(1-q^{p})\frac{\partial^{2}}{\partial h^{2}_{a}}
    \right)e^{-\beta V(\delta+h_{a})} 
    \right \}\right |_{h_{a}=0} \nonumber \\
        &=& \gamma_{q^{p}} \otimes 
(\gamma_{1-q^{p}} \otimes e^{- \beta V(\delta)})^{n}
\end{eqnarray}
where we used the formula
\beq
\exp\left(\frac{a}{2}\frac{\partial^{2}}{\partial h^{2}}\right)A(h)=\gamma_{a} \otimes A(h)
\label{eq-formula1}
\eeq
and the following short hand notations: $\gamma_{a}(x)$ is a Gaussian with
zero mean and variance $a$\cite{duplantier1981comment},
\beq
\gamma_{a}(x)=\frac{1}{\sqrt{2\pi a}}e^{-\frac{x^{2}}{2a}},
\eeq
by which we write a convolution of a function $A(x)$ with the Gaussian as,
\beq
\gamma_{a}\otimes A(x) \equiv \int dy \frac{e^{-\frac{y^{2}}{2a}}}{\sqrt{2\pi a}}A(x-y)=\int {\cal D}zA(x-\sqrt{a}z)
\label{eq-formula2}
\eeq
where
\beq
\int \cD z \ldots \equiv \int
dz \frac{e^{-\frac{z^{2}}{2}}}{\sqrt{2\pi}} \cdots
\eeq

Collecting the above results we obtain the variational free-energy
given by \eq{eq-F-functional-glass-ref}--\eq{eq-s-glass-ref} within the RS ansatz as
\beq
-\beta f_{\rm RS}(q)= \partial_{n} s_{\rm RS}(q)|_{n=0}=\frac{1}{2}\left( \frac{q}{1-q}+\ln (1-q)\right)
+\frac{\alpha}{p}
\int \cD z_{0}
\ln \int \cD z_{1} e^{-\beta V(\delta-\sqrt{1-q^{p}}z_{1}-\sqrt{q^{p}}z_{0})}
\label{eq-free-energy-RS}
\eeq

\subsubsection{The saddle point equation}

The saddle point equation for the order parameter $q$ is obtained as,
\beqn
0&=&\frac{\partial (-\beta  f_{\rm RS}(q))}{\partial q}=
\frac{1}{2}\frac{q}{(1-q)^{2}}
-\frac{\alpha}{p}
\frac{pq^{p-1}}{2}
\int \cD z_{0}
\left .
\left (
\frac{\int \cD z_{1} (e^{-\beta V(x)})'}{\int \cD z_{1}e^{-\beta V(x)}}
\right)^{2}
\right |_{x=\delta - \sqrt{1-q^{p}}z_{1}-\sqrt{q^{p}}z_{0}} \nonumber \\
&=& \frac{1}{2}\frac{q}{(1-q)^{2}}{\cal G}(q)
\label{eq-RS-saddlepoint}
\eeqn
where we introduced
\beq
    {\cal G}(q) \equiv  1-\alpha (1-q)^{2}q^{p-2}
    \int \cD z_{0}
    \left .
\left (
\frac{\int \cD z_{1} (e^{-\beta V(x)})'}{\int \cD z_{1}e^{-\beta V(x)}}
\right)^{2}
\right |_{x=\delta - \sqrt{1-q^{p}}z_{1}-\sqrt{q^{p}}z_{0}}
\label{eq-def-function-Gq}
\eeq

\subsubsection{Pressure and distribution of gap}

       Using \eq{eq-free-energy-RS} we obtain
       the pressure \eq{eq-def-pressure} as,
\beq
\Pi= \int \cD z_{0}
\left .
\frac{\int \cD z_{1} (e^{-\beta V(x)})'}{\int \cD z_{1} e^{-\beta V(x)}}
\right |_{x=\delta-\sqrt{1-q^{p}}z_{1}-\sqrt{q^{p}}z_{0}}
\label{eq-pressure-RS}
\eeq
and similarly the distribution of the gap given by \eq{eq-g-of-r-replica} as,
\beqn
g(r)
&=& \int \cD z_{0}
\frac{\left .\int \cD z_{1} \delta (x-r) e^{-\beta V(x)}
\right |_{x=\delta-\sqrt{1-q^{p}}z_{1}-\sqrt{q^{p}}z_{0}}
}{\left . \int \cD z_{1} e^{-\beta V(x)}\right |_{x=\delta-\sqrt{1-q^{p}}z_{1}-\sqrt{q^{p}}z_{0}}} \nonumber \\
& = & e^{-\beta V(r)} \int \cD z_{0}
\frac{\gamma_{1-q^{p}}(\delta-\sqrt{q^{p}}z_{0}-r)}
     {\left . \int \cD z_{1} e^{-\beta V(x)}\right |_{x=\delta-\sqrt{1-q^{p}}z_{1}-\sqrt{q^{p}}z_{0}}}
     \label{eq-g-of-r-RS}
     \eeqn
     which is properly normalized such that $\int dr g(r)=1$.
     One can also check easily that the 'virial equation'
     given by \eq{eq-virial-pressure} for the pressure
     is satisfied, as it should be.

     \subsection{The liquid phase : $q=0$ solution }
     \label{section-The-liquid phase-:-q=0-solution}

Apparently $q=0$ representing the liquid state
is always a solution of the RS saddle point equation
given by \eq{eq-RS-saddlepoint} for $p \geq 2$.
The stability of the solution must be examined by studying the eigenvalues of
the Hessian matrix reported in the appendix \ref{sec-Hessian-RS}.

\subsubsection{$p=2$ case}

From the results in appendix \ref{sec-Hessian-RS} we find for the $q=0$ solution
for $p=2$, 
\beqn
M_{1}&=&2-2\alpha \left(\frac{\gamma_{1}\otimes
  (e^{-\beta V(\delta)})'}{\gamma_{1}\otimes e^{-\beta V(\delta)}}\right)^{2}
  \\
M_{2}&=&M_{3}=0
\eeqn
from which we find the eigenvalues
of the Hessian matrix as (see \eq{eq-replicon-RS}),
\beqn
&& \lambda_{R}=\lambda_{L}=\lambda_{A}=
2-2\alpha \left(\frac{\gamma_{1}\otimes
  (e^{-\beta V(\delta)})'}{\gamma_{1}\otimes e^{-\beta V(\delta)}}\right)^{2}
\eeqn
which vanishes at,
\beq
\alpha_{\rm c}(\delta)=
\left(\frac{\gamma_{1}\otimes
  (e^{-\beta V(\delta)})'}{\gamma_{1}\otimes e^{-\beta V(\delta)}}\right)^{-2}
=\left (\frac{\int \frac{dz}{\sqrt{2\pi}}e^{-z^{2}/2} (e^{-\beta V(\delta-z)})'}{
    \int \frac{dz}{\sqrt{2\pi}}e^{-z^{2}/2}
  e^{-\beta V(\delta-z)}} \right)^{-2}
\label{eq-critical-point-rs}
\eeq
For $\alpha < \alpha_{\rm c}(\delta)$, the eigenvalues are positive so that
the $q=0$ solution is stable but it becomes unstable for $\alpha > \alpha_{\rm c}(\delta)$.

Interestingly we see that at the critical point
$\alpha=\alpha_{\rm c}(\delta)$, $q=0$ solves also ${\cal G}(q)=0$
(see \eq{eq-def-function-Gq}).
Since $q\neq 0$ solution must solve
${\cal G}(q)=0$, this suggests a possibility of continuous phase transition
at the critical point such that $q\neq 0$ solution emerges continuously.
The situation is similar to the Sherrington-Kirkpatrick model
which exhibits a continuous spinglass transition at
the d'Almeida-Thouless (AT) instability \cite{de1978stability} line.

\subsubsection{$p>2$ case}

For $p>2$, using the results reported in appendix \ref{sec-Hessian-RS}, 
we find $\lambda_{R}=\lambda_{L}=\lambda_{A}=2 >0$  (see \eq{eq-replicon-RS}) so that the $q=0$
solution is always stable.
Thus contrary to the $p=2$ model, we find the liquid
phase described by the $q=0$ RS solution is 
always (meta)stable. The situation is very similar to the usual $p$-spin spherical spinglass models \cite{crisanti1992sphericalp}. Then we are naturally lead to consider the possibility of a discontinuous glass phase represented by 1 step replica symmetry breaking (1RSB) much as the usual $p$-spin SG models
(see sec \ref{sec-linear})
including the random energy model (see sec. \ref{sec-REM})).
The latter is the standard  random first order (RFOT) scenario \cite{kirkpatrick1987dynamics,KTW89} which is established theoretically for the hardspheres in the $d \to \infty$ limit \cite{kurchan2012exact}.

\section{Replica symmetry breaking (RSB) ansatz}
\label{sec-RSB}

Here we continue the previous section and obtain
some generic results based on the replica symmetry
breaking (RSB) ansatz for the $M$-component continuous vector spin system
with generic potential $V(x)$.

\subsection{Parisi's ansatz}
\label{subsec-Parisi-ansatz}

\begin{figure}[h]
  \includegraphics[width=\textwidth]{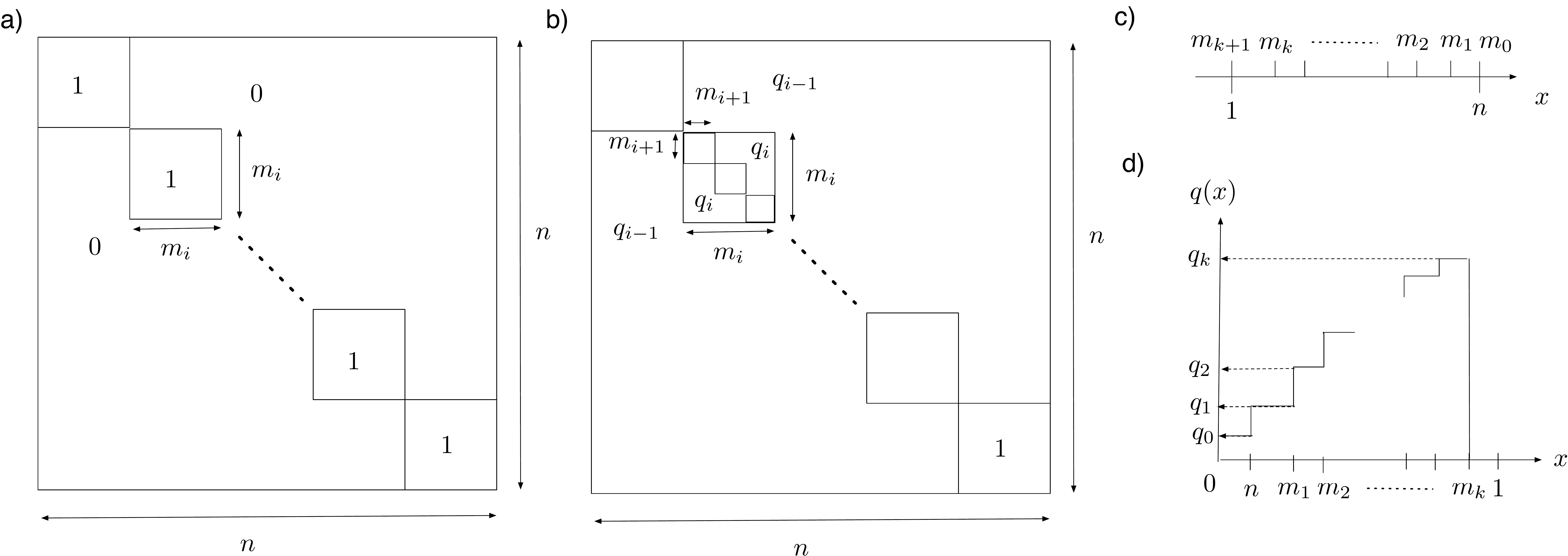}
  	       \caption{Parametrization of the Parisi's matrix  a) the 'fat' identity matrix
	       $I_{ab}^{m_{i}}$ b)  Parisi's order parameter matrix given by \eq{eq:parisi-matrix}
	         c) the hierarchy of the sizes $m_{i}$ of the sub-matrices d) the $q(x)$
                 function with $0 < n < 1$.}
	       \label{fig:parisi-matrix}
\end{figure}

We assume the following structure of the glass order parameter in the glass phase which is the Parisi' ansatz \cite{parisi1979infinite}.
It reads,
       \beq
       Q^{k-\rm{RSB}}_{ab}=q_{0}+\sum_{i=1}^{k+1} (q_{i}-q_{i-1}) I_{ab}^{m_{i}}
       =\sum_{i=0}^{k+1} q_{i}(I_{ab}^{m_{i}}-I_{ab}^{m_{i+1}}) 
       \label{eq:parisi-matrix}
       \eeq
       where
 $I^{m}_{ab}$ is a kind of generalized ('fat') identity matrix
       of size $n \times n$ composed of blocks of size $m \times m$.
       (see Fig.~\ref{fig:parisi-matrix})
       The matrix elements in the diagonal blocks,
       are all $1$ while those in the off-diagonal blocks are all $0$.
       The Parisi's matrix has a hierarchical structure such that
       \beq
       1=m_{k+1} < m_{k} < \ldots < m_{1} < m_{1}  < m_{0}=n
       \label{eq-m-convention}
       \eeq
       which becomes
              \beq
              0= m_{0} < m_{1} < \ldots <  m_{k} < m_{k+1}=1
	             \label{eq-m-convention-2}
              \eeq
              in the $n \to 0$ limit.
              The expression given by \eq{eq:parisi-matrix} becomes valid also for the
              diagonal part by introducing
              \beq
              q_{k+1}=1,
              \eeq
              which reflects
              the normalization of the spins.
              Let us note that we may sometimes extend the labels $m$'s
in \eq{eq-m-convention-2}
introducing an additional label $m_{k+2}$ just for conveniences.
In the last equation of \eq{eq:parisi-matrix}, $I_{ab}^{m_{k+2}}=0$.

Note that the replica symmetric (RS) ansatz corresponds to $k=0$. Thus
we should be able to recover the results discussed in the previous section \ref{sec-RS} by taking $k=0$ in the following.

The Parisi's ansatz describes the hierarchical organization of the free-energy landscape in glass phases. The value of the matrix elements which belong to the off-diagonal part but most close to the diagonal, namely $q_{k}$, is interpreted as the self-overlap of the glassy states or the Edwards-Anderson order parameter $q_{\rm EA}$ \cite{edwards1975theory}. Within the RS ansatz discussed in sec \ref{sec-RS}, $q_{\rm EA}=q$.

In the $k \to \infty$ limit, the matrix elements can be parametrized by
a function $q(x)$ defined in the range $0 \leq x \leq 1$  (See Fig. \ref{fig:parisi-matrix} d)). It encodes the information of the self and mutual overlaps
among glassy metastable states. For instance, it is known that
\cite{parisi1983order,mezard1984nature} 
the probability distribution function $P(q)$ of the overlap $q$
between two independently sampled thermalized spin configurations,
say $S^{a}$ and $S^{b}$,
\beq
P(q)=\langle \delta(q-q_{ab}) \rangle \qquad 
q_{ab}=\frac{1}{NM}\sum_{i=1}^{N}\sum_{\mu=1}^{M} (S^{a})^{\mu}_{i}(S^{b})^{\mu}_{i} 
\eeq
where $\langle \ldots \rangle$ is meant for the thermal average,
can be related to the $q(x)$ function as,
\beq
P(q)=\frac{dx(q)}{dq}
\eeq
where $x(q)$ is the inverse of $q(x)$.
In general the Edwards-Anderson parameter $q_{\rm EA}$
appears as a plateau height $q(x_{1})=q_{\rm EA}$ of the $q(x)$ function
in some range  $x_{1}< x < 1$ (where $x_{1}$ corresponds to $m_{k}$,
See Fig. \ref{fig:parisi-matrix} d)). This leads to a delta-peak
$\propto \delta(q-q_{\rm EA})$ in the overlap distribution function $P(q)$.
On the other hand, the behavior of the $q(x)$ function at $0 < x < x_{1}$
encodes non-trivial features of the distribution function $P(q)$ of
the mutual overlaps between different glassy metastable states.
See \cite{MPV87} for more discussions on the physical consequences
of the Parisi's ansatz.

Finally let us note that 'Jamming' simply means disappearance of thermal fluctuations within glassy states due to tightening of constraints. This means the self-overlap saturates to $1$ ,  $q_{\rm EA}=q_{k}=q(x_{1}) \to 1$.

\subsection{Free-energy}
\label{subsec-free-energy-kRSB}

Let us evaluate the free-energy
given by \eq{eq-F-functional-glass-ref}
-\eq{eq-s-glass-ref}
using the above
ansatz.  To compute the entropic part of the free-energy one needs 
to evaluate $\ln {\rm det} \hat{Q}^{k-{\rm RSB}}$.  Given the hierarchical structure of the
Parisi's matrix, one can obtain it in a recursive fashion
and one finds \cite{mezard1991replica},
\beqn
\ln {\rm det} \hat{Q}^{k-{\rm RSB}}&=&
\ln \left(1+\sum_{j=0}^{k}(m_{j}-m_{j+1})q_{j} \right) \\
&+& n \sum_{i=0}^{k}\left (\frac{1}{m_{i+1}}-\frac{1}{m_{i}} \right)
\ln \left(1+\sum_{j=i}^{k}(m_{j}-m_{j+1})q_{j}-m_{i}q_{i}\right)
\label{eq-det-Q-k-RSB}
\eeqn
Remembering that $m_{0}=n$ we find,
\beqn
\left. \partial_{n}\ln {\rm det} \hat{Q}^{k-{\rm RSB}} \right |_{n=0}&=&
\frac{q_{0}}{G_{0}}
+ \frac{1}{m_{1}}\ln G_{0} +
\sum_{i=1}^{k}\left (\frac{1}{m_{i+1}}-\frac{1}{m_{i}} \right)\ln G_{i}
\label{eq-ln-det-Q}
\eeqn
with
\beqn
G_{i}&=&1+\sum_{j=i}^{k}(m_{j}-m_{j+1})q_{j}-m_{i}q_{i} \qquad i=0,1,\ldots,k
\label{eq-G-q}
\eeqn
which implies
\beqn
q_{i}&=& 1-G_{k}+\sum_{j=i+1}^{k}\frac{1}{m_{j}}(G_{j}-G_{j-1}) \qquad i=0,1,\ldots,k
\label{eq-q-G}
\eeqn
The interaction part of free-energy can also be evaluated in a recursive
fashion \cite{duplantier1981comment}. One finds,
\beqn
-{\cal F}_{\rm int}[\hat{Q}^{k-{\rm RSB}}]
  &=&
  \left.   \prod_{l=0}^{k+1} \exp \left(
  \frac{  \Lambda_{l}}{2}   \sum_{a,b=1}^{n} 
I_{ab}^{m_{l}}
  \frac{\partial^{2}}{\partial h_{a}\partial h_{b}} \right)
  \prod_{a=1}^{n}  e^{-\beta V(\delta+h_{a})}  \right |_{\{h_{a}=0\}}   \nonumber \\
    &=&
  \left.   \prod_{l=0}^{k} \exp \left(
  \frac{  \Lambda_{l}}{2}   \sum_{a,b=1}^{n} 
I_{ab}^{m_{l}}
\frac{\partial^{2}}{\partial h_{a}\partial h_{b}} \right)
\prod_{a=1}^{n}
g(m_{k+1},h_{a})
  \right |_{\{h_{a}=0\}}  
 \label{eq-Fint-k-RSB}
\eeqn
where we introduced
        \beqn
        \Lambda_{0} & \equiv & \lambda_{0}  \nonumber \\
        \Lambda_{i} & \equiv  &\lambda_{i}-\lambda_{i-1} \qquad i=1,2,\ldots,k+1
        \label{eq-def-Lambda}
        \eeqn
        with
        \beq
        \lambda_{i}  \equiv  q_{i}^{p}.
        \label{eq-def-lambda}
        \eeq
        In the 2nd equation of  \eq{eq-Fint-k-RSB} we used
        $I_{ab}^{m_{k+1}=1}=\delta_{ab}$ and introduced
        \beq
        g(m_{k+1},h) \equiv \gamma_{\Lambda_{k+1}} \otimes e^{-\beta V(h)}
        =\int \cD z_{k+1} e^{-\beta V(h-\sqrt{\Lambda_{k+1}}z_{k+1})}
        \label{eq-initial-condition-recursion-g}
        \eeq
where we used the formula given by \eq{eq-formula1}.

The expression given by \eq{eq-Fint-k-RSB} naturally motivates
a family of functions $g$'s which obey a recursion relation,
        \beqn
        g(m_{l},h)&=&e^{\frac{\Lambda_{l}}{2}\frac{\partial^{2}}{\partial h^{2}}}
        g^{\frac{m_{l}}{m_{l+1}}}(m_{l+1},h)
        =\gamma_{\Lambda_{l}} \otimes        g^{\frac{m_{l}}{m_{l+1}}}(m_{l+1},h)
                \label{eq-recursion-g-0}
        \\
        &=&
        \int \cD z_{l}  g^{\frac{m_{l}}{m_{l+1}}}(m_{l+1},h-\sqrt{\Lambda_{l}}z_{l})
        \label{eq-recursion-g}
        \eeqn
        for $l=0,1,\ldots,k$.
        Then it is easy to see that
        \beqn
        -{\cal F}_{\rm int}[\hat{Q}^{k-{\rm RSB}}]
                &=&\gamma_{\Lambda_{0}}\otimes \left(
\gamma_{\Lambda_{1}} \otimes (\ldots \gamma_{\Lambda_{k}}\otimes g^{m_{k}}(m_{k+1},\delta)\ldots)^{m_{1}/m_{2}}\right)^{m_{0}/m_{1}} \nonumber \\
        &=& \ldots  \nonumber \\
        &=&\gamma_{\Lambda_{0}}\otimes \left(
\gamma_{\Lambda_{1}} \otimes g^{m_{1}/m_{2}}(m_{2},\delta)
\right)^{m_{0}/m_{1}} \nonumber \\
&=&\gamma_{\Lambda_{0}}\otimes g^{m_{0}/m_{1}}(m_{1},\delta) \nonumber \\
&=&g(m_{0},\delta)
\label{eq-Fint-recursive-construction}
        \eeqn

        Equivalently we can introduce a related family of
        functions $f$'s,
\beq
f(m,h) \equiv -\frac{1}{m}\ln g(m,h)
\label{eq-def-f-g}
\eeq
which follows a recursion relation,
\beqn
e^{-m_{i}f(m_{i},h)}
&=&\gamma_{\Lambda_{i}} \otimes e^{-m_{i}f(m_{i+1},h)}
=\int \cD z_{i} e^{-m_{i}f(m_{i+1},h-\sqrt{\Lambda_{i}}z_{i})}
\label{eq-recursion-f}
        \eeqn
        for $i=0,1,\ldots,k+1$ with the boundary condition,
                \beq
        e^{-m_{k+1}f(m_{k+1},h)}= \int \cD z_{k+1} e^{-m_{k+1}\beta V(h-\sqrt{\Lambda_{k+1}}z_{k+1})}.
        \label{eq-boundary-f}
        \eeq
        where $m_{k+1}=1$.
        We may also express the boundary condition as,
        \beq
        f(m_{k+2},h)=\beta V(h).
        \label{eq-boundary-f-type2}
        \eeq
        by introducing $m_{k+2}$ just as an additional label for convenience.
                Remembering that $m_{0}=n$ we find the interaction part of the free-energy becomes,
    \beqn
    \left.    -\partial_{n}     {\cal F}_{\rm int}[\hat{Q}^{k-{\rm RSB}}] \right |_{n=0}
    &=&    \left.    \partial_{m_{0}}  g(m_{0},\delta) \right |_{m_{0}=n=0} =-f(m_{0}=0,\delta)\nonumber \\
   & =& \gamma_{\Lambda_{0}}\otimes (-f(m_{1},\delta))
    =-\int \cD z_{0} f(m_{1},\delta-\sqrt{\Lambda_{0}}z_{0})
    \label{eq-del-n-f}
\eeqn

Finally collecting the above results we obtain the free-energy within the $k$-RSB ansatz as,
\beqn
-\beta f_{k-{\rm RSB}}[\hat{Q}] &=&
\frac{1}{2}
\frac{q_{0}}{G_{0}}
+ \frac{1}{2}\frac{1}{m_{1}}\ln G_{0} +
\frac{1}{2}
\sum_{i=1}^{k}\left (\frac{1}{m_{i+1}}-\frac{1}{m_{i}} \right)\ln G_{i} \nonumber \\
&+&
\frac{\alpha}{p}
\int \cD z_{0}
(-f(m_{1},\delta-\sqrt{\Lambda_{0}}z_{0}))
\label{eq-free-ene-k-RSB}
\eeqn

        \subsubsection{$k=0$ case: RS}

        Let us check if $k=0$ case recovers the result we obtained previously for the
        replica symmetric (RS) ansatz.
        \beqn
-\beta f_{0-{\rm RSB}}(q_{0}) &=&
\frac{1}{2}
\frac{q_{0}}{(1-q_{0})}
+ \frac{1}{2}\ln (1-q_{0}) 
+\frac{\alpha}{p}
\int \cD z_{0}
\ln \int \cD z_{1} e^{-\beta V(\Xi(\delta,q_{0})}
\eeqn
with
\beq
\Xi(\delta,q_{0})=\delta-\sqrt{1-q_{0}^{p}}z_{1}-\sqrt{q_{0}^{p}}z_{0}
\eeq
where we used $G_{0}=1-(m_{1}-m_{0})q_{0}$  and that $m_{0}=0$ and $m_{1}=1$.
In the 2nd equation we used \eq{eq-boundary-f}.
The result agrees with \eq{eq-free-energy-RS} as it should.

\subsubsection{$k=1$ case: 1 RSB}
\label{subsubsec-1RSB}

For the $k=1$ RSB case we find,
\beqn
&& -\beta f_{1-{\rm RSB}}(q_{0},q_{1},m_{1}) =
 \nonumber \\
 &&
\frac{1}{2}
\frac{q_{0}}{1-m_{1}q_{0}+(m_{1}-1)q_{1}} 
 + \frac{1}{2}\frac{1}{m_{1}}\ln (1-m_{1}q_{0}+(m_{1}-1)q_{1}) +
\frac{1}{2}
\left (1-\frac{1}{m_{1}} \right)\ln (1-q_{1}) 
 \nonumber \\
&& +\frac{\alpha}{p}\frac{1}{m_{1}}
\int \cD z_0 \ln \int \cD z_{1}
\left[\int \cD z_{2} e^{-\beta V(\Xi(\delta,q_{0},q_{1})} \right]^{m_{1}}
\label{eq-free-ene-1-RSB}
\eeqn
with
\beq
\Xi(\delta,q_{0},q_{1})=\delta-\sqrt{q_{0}^{p}}z_{0}-\sqrt{q_{1}^{p}-q_{0}^{p}}z_{1}-\sqrt{1-q_{1}^{p}}z_{2}.
\eeq
where $q_{0}$ and $q_{1}$ must be determined through the saddle point equations 
which we discuss in sec. \ref{sec-1RSB-equations}.

An important quantity is the complexity or the configurations entropy $\Sigma(f)$,
which describes the exponentially large number of states
$\propto e^{N \Sigma(f)}df$ with free-energy density between $f$ and $f+df$
in the glass phase. Using  Monasson's prescription \cite{Mo95}, which is equivalent
to the approach based on the Franz-Paisi's potential\cite{FP95,FP97}, 
one can construct the complexity
function $\Sigma(f)$ treating $m=m_{1}$ as a parameter;
\beqn
\Sigma^{*}(m)= m^{2}\partial_{m} \beta f_{1{\rm -RSB}}(q_{0},q_{1},m)  \\
\beta f^{*}(m)=\partial_{m}(m \beta f_{1{\rm -RSB}})(q_{0},q_{1},m)
\eeqn
Thus extremization of the free-energy with respect to $m$,
$0=\partial_{m} \beta f_{1{\rm -RSB}}(q_{0},q_{1},m)$ amounts to
force  the complexity to vanish $\Sigma^{*}(m)$ \cite{Mo95}.

We readily find the following explicit expressions,
\beqn
\beta f^{*}(m_{1})&=&-\frac{1}{2}\frac{q_{0}}{1-m_{1}q_0+(m_1-1)q_1}-\frac{1}{2}\ln (1-q_1)\nonumber \\
&-&\frac{1}{2}\left (-\frac{m_1 q_0}{[1-m_1q_0 +(m_1-1)q_1]^2}+\frac{1}{1-m_1q_0 +(m_1-1)q_1} \right)(q_1-q_0) \nonumber \\
&-&\frac{\alpha}{p}\int \cD z_0
\frac{\int \cD z_1 \left[ \int \cD z_2 e^{-\beta V(\Xi(\delta,q_{0},q_{1}))}\right]^{m_{1}} \ln \left[ \int \cD z_2 e^{-\beta V(\Xi(\delta,q_{0},q_{1}))}\right]}{\int \cD z_1 \left[ \int \cD z_2 e^{-\beta V(\Xi(\delta,q_{0},q_{1}))}\right]^{m_{1}}} \qquad 
\eeqn
and
\beqn
\Sigma^{*}(m_{1})&=& \frac{1}{2}\ln [1-m_1 q_0 +(m_1 -1)q_1]-\frac{1}{2}\ln (1-q_1) \nonumber \\
&&+\frac{\alpha}{p}\int \cD z_0
\ln \left[ \int \cD z_1 \left[ \int \cD z_2 e^{-\beta V(\Xi(\delta,q_{0},q_{1}))}\right]^{m_{1}}\right] \nonumber \\
&+&m_{1} \left \{
-\frac{1}{2}\left (-\frac{m_1 q_0}{[1-m_1q_0 +(m_1-1)q_1]^2}+\frac{1}{1-m_1q_0 +(m_1-1)q_1} \right)(q_1-q_0)  \right. \nonumber \\
&& \left. 
-\frac{\alpha}{p}\int \cD z_0
\frac{\int \cD z_1 \left[ \int \cD z_2 e^{-\beta V(\Xi(\delta,q_{0},q_{1}))}\right]^{m_{1}} \ln \left[ \int \cD z_2 e^{-\beta V(\Xi(\delta,q_{0},q_{1}))}\right]}{\int \cD z_1 \left[ \int \cD z_2 e^{-\beta V(\Xi(\delta,q_{0},q_{1}))}\right]^{m_{1}}} 
\right\}
\;\;\;
\eeqn

\subsubsection{$k=\infty$ case: continuous RSB}

In the limit $k \to \infty$, the overlap matrix
$\hat{Q}$ is parametrized by a function $q(x)$ with $n< x < 1$.
Then \eq{eq-det-Q-k-RSB} becomes,
\beqn
\ln {\rm det} \hat{Q}^{\infty -{\rm RSB}}&=&
\ln \left(1 -\int_{n}^{1} dx q(x) \right) 
- n \int_{n}^{1} \frac{dx}{x^{2}}
\ln \left(1-\int_{x}^{1}dyq(y)-xq(x) \right) \;\;\;\
\eeqn
From the above expression  we find 
\beqn
\left. \partial_{n} \ln {\rm det} \hat{Q}^{\infty -{\rm RSB}} \right |_{n=0}
=\frac{q(0)}{G(0)}+\ln G(1)
+\int_{0}^{1}\frac{dx}{G(x)}
\eeqn
with
\beq
G(x) \equiv 1-\int_{x}^{1}dyq(y)-xq(x)
\label{eq-def-g-x}
\eeq

Then the free-energy given by \eq{eq-free-ene-k-RSB} can be written as,
\beqn
 -\beta f_{\infty-{\rm RSB}}[\hat{Q}] &= &
\frac{1}{2}\left[
\frac{q(0)}{G(0)}+\frac{1}{2}\ln G(1)
+\int_{0}^{1}\frac{dx}{G(x)}\right] \nonumber \\
&& + \frac{\alpha}{p}\int \cD z_{0} (-f(m(0),\delta-\sqrt{q^{p}(0)}z_{0}))
\label{eq-free-ene-full-RSB}
\eeqn
where the function $f(x,h)$ obeys,
\beq
\dot{f}(x,h)=
-\frac{1}{2}\dot{\lambda}(x)
\left [ f''(x,h)
-x \left (f'(x,h)  \right)^{2}
\right],
\label{eq-pdeq-f}
\eeq
with
\beq
\lambda(x) \equiv q^{p}(x).
\label{eq-def-lambda-x}
\eeq
Here and in the following we denote a partial derivative
with respect to the 1st argument by a dot, e.~g.
$\partial_{x}f(x,h)=\dot{f}(x,h)$ and
that with respect to the 2nd argument by a dash
e.~g. $\partial_{h}f(x,h)=f'(x,h)$.
The  partial differential equation  given by \eq{eq-pdeq-f}
is the continuous limit of recursion formula \eq{eq-recursion-f}.
The boundary condition given by \eq{eq-boundary-f} becomes,
\beq
f(1,h)=-\ln \int  \cD z e^{-\beta V(h-\sqrt{1-q^{p}(1)}z)}.
\label{eq-pdeq-f-boundary}
\eeq

\subsection{Variation of the interaction part of the free-energy}
\label{subsec-variation-of-Fint}

Later we will often meet the needs to consider variation of the free-energy.
This happens when we solve
the saddle point equation for the order parameters $q_{l}$'s
(sec. \ref{subsec:variational-equations}),
analyze the stability of the saddle point solutions
(sec. \ref{sec-stability-kRSB}),
compute the pressure $\Pi$ given by \eq{eq-pressure-replica}
and the distribution of the gap $g(r)$
given by \eq{eq-g-of-r-replica}.
We note that a variation scheme for continuous RSB
has been formulated originally in \cite{sommers1984distribution}.

\subsubsection{Some useful functions}
Here we consider a strategy make variation of the interaction
part of the free-energy given by \eq{eq-Fint-k-RSB}.
Within the Parisi's ansatz it becomes \eq{eq-del-n-f} which reads as,
$ \left.    -\partial_{n}     {\cal F}_{\rm int}[\hat{Q}^{k-{\rm RSB}}] \right |_{n=0}
=-f(m_{0}=n=0,\delta)
=-\int \cD z_{0} f(m_{1},\delta-\sqrt{\Lambda_{0}}z_{0})$.

As we discussed in sec. \ref{subsec-free-energy-kRSB},
the interaction part of the free-energy is is constructed in a recursive
way and the functions $f(m_{i},h)$
follows a recursion formula \eq{eq-recursion-f}.
This fact naturally motivates us to introduce for $0 \le i \le j \le k+1$,\beq
P_{i,j}(y,h)\equiv \frac{\delta f(m_{i},y)}{\delta f(m_{j},h)}.
\label{eq-def-pij}
\eeq
Using  the chain rule we can write,
\beq
P_{i,j}(y,z)=\int dx P_{i,j-1}(y,x)P_{j-1,j}(x,z).
\label{eq-pij-chain-rule}
\eeq
where
\beq
P_{j-1,j}(x,z)=\frac{\delta f(m_{j-1},x)}{\delta f(m_{j},z)}
=e^{m_{j-1}(f(m_{j-1},x)-f(m_{j},z)}
\frac{e^{-\frac{(x-z)^{2}}{2\Lambda_{j-1}}}}{\sqrt{2\pi \Lambda_{j-1}}}
\label{eq-pij-one-step}
\eeq
as one can easily find from the recursion relation given by \eq{eq-recursion-f}.
Then we find a recursion relation,
\beq
P_{ij}(y,z)=e^{-m_{j-1}f(m_{j},z)}
\gamma_{\Lambda_{j-1}} \otimes_{z} \frac{P_{i,j-1}(y,z)}{e^{-m_{j-1}f(m_{j-1},z)}}
\label{eq-recursion-pij}
\eeq
with the 'boundary condition'
\beq
P_{ii}(y,h)=\delta(y-h).
\label{eq-boundary-pij}
\eeq
Here $\otimes_{h}$ stands for a convolution with respect to the variable $h$.
A useful property to note is that the recursion relation given by \eq{eq-recursion-pij} preserves the 'normalization',
\beq
\int dh P_{i,j}(y,h)=1
\label{eq-normalization-p-ij}
\eeq
which can be easily proved using \eq{eq-recursion-f}.

For a convenience, let us introduce another quantity which is related to
$P_{ij}(y,h)$,
\beq
P(m_{j},h) \equiv \frac{\delta f(m_{0},\delta)}{\delta f(m_{j+1},h)}=P_{0,j+1}(\delta,h)=\int dx P_{0,1}(\delta,x)P_{1,j+1}(x,h)
=\int \cD z_{0} P_{1,j+1}(\delta-\sqrt{\Lambda_{0}}z_{0},h)
\label{eq-def-P}
\eeq
where we used the chain rule, \eq{eq-pij-one-step} and set $m_{0}=n \to 0$.
Clearly it follows the same recursion formula as \eq{eq-recursion-pij},
\beq
P(m_{j},h)=e^{-m_{j}f(m_{j+1},h)} \gamma_{\Lambda_{j}}\otimes_{h}
\frac{P(m_{j-1},h)}{e^{-m_{j}f(m_{j},h)}} \qquad j=1,2,\ldots,k+1
\label{eq-recursion-p}
\eeq
 with the 'boundary condition'
\beq
P(m_{0},h)=\frac{1}{\sqrt{2\pi \Lambda_{0}}}e^{-\frac{(\delta-h)^{2}}{2\Lambda_{0}}}
\label{eq-boundary-P}
\eeq
which follows from \eq{eq-def-P} and \eq{eq-boundary-pij}.
The functions $P(m_{i},h)$ is also normalized such that
\beq
\int dh P(m_{i},h)=1
\label{eq-normalization-p}
\eeq
reflecting \eq{eq-normalization-p-ij}.

In the $k \to \infty$ limit, the function $P(x,h)$ can be obtained by solving a differential equation,
\beqn
\dot{P}(x,h)=\frac{1}{2}\dot{\lambda}(x) \left[
  P''(x,h)-2x (P(x,h)\pi(x,h))'
  \right]
\label{eq-pdeq-P}
\eeqn
which is the continuous limit of \eq{eq-recursion-p}.
The boundary condition given by \eq{eq-boundary-P} becomes,
\beq
P(0,h)=\frac{1}{\sqrt{2\pi q^{p}(0)}}e^{-\frac{(\delta-h)^{2}}{2q^{p}(0)}}.
\label{eq-pdeq-P-boundary}
\eeq

\subsubsection{Distribution of the gap}

Now the distribution function of the gap $g(r)$ given by \eq{eq-g-of-r-replica} for the generic k-RSB ansatz is obtained
using \eq{eq-free-ene-k-RSB}, \eq{eq-boundary-f-type2}
which reads $\beta V(r)=f(m_{k+2},r)$, \eq{eq-def-pij},
\eq{eq-def-P} and \eq{eq-recursion-p} as,
\beqn
g(r)
&=&-
\frac{p}{\alpha}\frac{\delta \beta f_{k_RSB}[\hat{Q}^{*}]}{\delta (-\beta V(r))}
=\int \cD z_{0} \frac{\delta f(m_{1},\delta-\sqrt{\Lambda_{0}}z_{0})}{\delta f(m_{k+2},r)} \label{eq-g-of-r-kRSB} \\
&=& \int \cD z_{0} P_{1,k+2}(\delta-\sqrt{\Lambda_{0}}z_{0},r)= P(m_{k+1},r)
=e^{-\beta V(r)}\; \gamma_{\Lambda_{k+1}}
\otimes_{r}\frac{P(m_{k},r)}{e^{-f(m_{k+1}=1,r)}}
\nonumber 
\eeqn
We used $m_{k+1}=1$ in the last equation.
It can be seen that $g(r)$ is properly normalized such that $\int dr g(r)=1$
reflecting \eq{eq-normalization-p}.
The previous result in the RS case ($k=0$) \eq{eq-g-of-r-RS}
can be recovered using \eq{eq-boundary-P},
\eq{eq-recursion-f}, \eq{eq-boundary-f} in \eq{eq-g-of-r-kRSB} as expected.

\subsubsection{Pressure}

The pressure given by \eq{eq-pressure-replica} for the
generic $k$-RSB ansatz is obtained
using and \eq{eq-free-ene-k-RSB} as,
\beq
\Pi= -\frac{p}{\alpha}\frac{\partial \beta f_{\rm k-RSB}[\hat{Q}^{*}]}{\partial \delta}
=-\int \cD z_{0} \frac{\partial f(m_{1},\delta-\sqrt{\Lambda_{0}}z_{0})}{\partial \delta}
=\int \cD z_{0} \pi(m_{1},\delta-\sqrt{\Lambda_{0}}z_{0}).
\label{eq-pressure-kRSB}
\eeq
where we introduced,
\beq
\pi(m,h)\equiv -f'(m,h)
\label{eq-def-pi}
\eeq
Here and in the following the prime stands for taking a partial derivative
with respect to the 2nd argument $h$. 

The function $\pi(m,h)$ introduced above also follows a recursion formula which can be obtained using
\eq{eq-recursion-f} and \eq{eq-boundary-f} as,
\beq
\pi(m_{i},h)
=e^{m_{i}f(m_{i},h)}\gamma_{\Lambda_{i}}\otimes \pi(m_{i+1},h)e^{-m_{i}f(m_{i+1},h)}
\label{eq-recursion-pi}
\eeq
for $i=1,2,\ldots,k$ with the 'boundary condition',
\beq
\pi(m_{k+1},h)
=\frac{\int \cD z_{k+1} (e^{-\beta V(h-\sqrt{\Lambda_{k+1}}z_{k+1})})'}{\int \cD z_{k+1} e^{-\beta V(h-\sqrt{\Lambda_{k+1}}z_{k+1})}}
\label{eq-boundary-f-dash}
\eeq
The previous result in the RS case ($k=0$) given by \eq{eq-pressure-RS}
can be recovered using \eq{eq-boundary-f-dash} in \eq{eq-pressure-kRSB}.

In the $k \to \infty$ limit, the function
$
\pi(x,h)\equiv -f'(x,h)
$
obeys a differential equation,
\beq
\dot{\pi}(x,h)=-\frac{\dot{\lambda}(x)}{2}
  \left( \pi''(x,h)+2x \pi(x,h)\pi'(x,h)\right)
  \label{eq-pdeq-pi}
  \eeq
which is  the continuous limit of \eq{eq-recursion-pi} and
can be obtained from \eq{eq-pdeq-f}. The boundary condition for the latter 
is given by \eq{eq-boundary-f-dash} which reads,
\beq
\pi(1,h)=\frac{\int  \cD z (e^{-\beta V(h-\sqrt{1-q^{p}(1)}z)})'}{\int  \cD z e^{-\beta V(h-\sqrt{1-q^{p}(1)}z)}}.
\eeq

It is instructive to verify that the pressure given by \eq{eq-pressure-kRSB}
can be recovered through the virial equation for the pressure given by \eq{eq-virial-pressure}. In fact the pressure can be expressed as,
\beq
\Pi=\int dh P(m_{i-1},h)\pi(m_{i},h)
\label{eq-pi-krsb}
\eeq
with {\it any} $i=1,2,\ldots,k+2$ (
so that $\pi=\int dh P(x,h)\pi(x,h)$ for any
$0 < x< 1$ in the $k \to \infty$ limit). 
Using the recursion formulas given by \eq{eq-recursion-p} and \eq{eq-recursion-pi}
one can check that  $\int dh P(m_{i-1},h)\pi(m_{i},h)=\int dh' P(m_{i},h')\pi(m_{i+1},h')$ so that the r. h. s of the above equation does not depend on the level $i$ of the hierarchy. The case of the virial equation
for the pressure $\Pi=\int dr g(r)(-\beta V'(r))$
given by \eq{eq-virial-pressure} corresponds to the case
$i=k+2$ which can be seen by noting
$\pi(m_{k+2},h)=-f'(m_{k+2},h)=-\beta V'(h)$
(see \eq{eq-boundary-f}) and \eq{eq-g-of-r-kRSB}.
On the other hand, the case $i=1$ corresponds to the
expression given by \eq{eq-pressure-kRSB} which can be seen
using \eq{eq-boundary-P}.

\subsection{Saddle point equations for the order parameters}
\label{subsec:variational-equations}

Here we derive variational equations to determine $q_{i}$ for $i=0,1,2,\ldots,k$.
Since $q_{i}$'s are related linearly
to $G_{i}$'s through \eq{eq-q-G}, the saddle point equations can be written as,
\beqn
0&=&\frac{\partial (-\beta f_{k-RSB}[\hat{Q}])}{\partial G_{i}} \nonumber \\
&=& \frac{1}{2}
\left[
  -\frac{q_{0}}{G_{0}^{2}}\delta_{i,0}+\left(\frac{1}{m_{i+1}}-\frac{1}{m_{i}} \right)
\left(\frac{1}{G_{i}}-\frac{1}{G_{0}} \right)(1-\delta_{i,0}) \right]  \\
&+&\left [-\left(\frac{1}{m_{i+1}}-\frac{1}{m_{i}} \right)\sum_{j=0}^{i-1}
pq_{j}^{p-1}\frac{\partial}{\partial \lambda_{j}}  
-\frac{1}{m_{i+1}}
pq_{i}^{p-1}\frac{\partial}{\partial \lambda_{i}}
\right] \frac{\alpha}{p} \int \cD z_{0} (-f(m_{1},\delta-\sqrt{\Lambda_{0}}z_{0}))\nonumber
\eeqn
In the last equation we used  \eq{eq-def-lambda} and \eq{eq-def-Lambda}.
As we show in appendix \ref{sec-dev-f-lambda} we find,
\beq
-\frac{\partial}{\partial \lambda_{j}}f(m_{i},y)=
\frac{1}{2}(m_{j}-m_{j+1})\int dh P_{i,j+1}(y,h)\pi^{2}(m_{j+1},h)
\label{eq-dev-f-lambda}
\eeq
where $P_{i,j}(y,h)$ is defined in \eq{eq-def-pij}
and $\pi(m,h)$ is defined in \eq{eq-def-pi}

Collecting the above results we obtain the variational equations as
\beqn
\frac{q_{0}}{G_{0}^{2}} &=&\kappa_{0}   \nonumber \\
\frac{1}{G_{i}}-\frac{1}{G_{0}}&=&\sum_{j=0}^{i-1}(m_{j}-m_{j+1})\kappa_{j}+m_{i}\kappa_{i}
\qquad i=1,2,\ldots,k
\label{eq-saddle-point-k-RSB}
\eeqn
where we introduced
\beq
\kappa_{j}\equiv \alpha q_{j}^{p-1}
\int dh P(m_{j},h) \pi^{2}(m_{j+1},h) \qquad j=0,1,\ldots,k
\label{eq-def-kappa}
\eeq
Note that $P(m_{j},h)$ and $\pi(m_{i},h)$ used here can be obtained
by solving the recursion formulas given by \eq{eq-recursion-p}
and \eq{eq-recursion-pi} respectively together with their boundary conditions.

Finally we note that for $p > 1$, $q_{0}=0$ always solves the 1st equation
of \eq{eq-saddle-point-k-RSB}.

\subsubsection{$k=0$ case: RS}
\label{subsubsec:variational-equations-RS}

Let us check if $k=0$ case recovers the result we obtained previously for the
replica symmetric (RS) ansatz.
In this case we just need the 1st equation of \eq{eq-saddle-point-k-RSB}
which becomes,
\beqn
\frac{q_{0}}{(1-q_{0})^{2}} &=& \alpha q_{0}^{p-1}
\int dh P(m_{0},h) (\pi(m_{1},h))^{2} \nonumber \\
&=& \alpha q_{0}^{p-1}\int \cD z_{0}
\left .
\left (
\frac{\int \cD z_{1} (e^{-\beta V(x)})'}{\int \cD z_{1}e^{-\beta V(x)}}
\right)^{2}
\right |_{x=\delta - \sqrt{1-q_{0}^{p}}z_{1}-\sqrt{q_{0}^{p}}z_{0}}
\eeqn
where we used $G_{0}=1-(m_{1}-m_{0})q_{0}$  and that $m_{0}=0$ and $m_{1}=1$.
In the 2nd equation we used \eq{eq-boundary-f-dash} and \eq{eq-boundary-P}.
The result agrees with \eq{eq-RS-saddlepoint} as it should.

\subsubsection{$k=1$ case: 1RSB}
\label{sec-1RSB-equations}

For the $k=1$ case (1RSB) \eq{eq-saddle-point-k-RSB} becomes,
,
\beqn
\frac{q_{0}}{G_{0}^{2}}&=&\kappa_{0} \nonumber \\
\frac{1}{G_{1}}-\frac{1}{G_{0}}&=&m_{1}(\kappa_{1}-\kappa_{0})
\label{eq-for-G-1RSB}
\eeqn
with
\beqn
\kappa_{0}&=&\alpha q_{0}^{p-1}\int dh P(m_{0},h)\pi^{2}(m_{1},h) \nonumber \\
\kappa_{1}&=&\alpha q_{1}^{p-1}\int dh P(m_{1},h)\pi^{2}(m_{2},h)
\label{eq-kappa-1RSB}
\eeqn
After solving the above equations for $G_{0}$ and $G_{1}$,
the order parameters
$q_{0}$ and $q_{1}$ can be obtained as (See \eq{eq-q-G}),
\beqn
q_{0}&=&1-G_{1}+\frac{1}{m_{1}}(G_{1}-G_{0}) \nonumber \\
q_{1}&=&1-G_{1}
\label{eq-for-q-1RSB}
\eeqn

To evaluate $\kappa_{0}$ and $\kappa_{1}$ in \eq{eq-kappa-1RSB} we need
more information. Suppose that we are given some initial guess for
the values of $q_{0}$ and $q_{1}$. Then we can recursively obtain
functions $f(m_{2},h)$ and $f(m_{1},h)$
(see \eq{eq-boundary-f}) and \eq{eq-recursion-f}) as,
\beqn
e^{-m_{2}f(m_{2},h)}&=&\int \cD z_{2}
e^{-\beta V(h-\sqrt{1-q_{1}^{p}}z_{2})} \nonumber \\
e^{-m_{1}f(m_{1},h)}&=&\int \cD z_{1}
e^{-m_{1} f(m_{2},h-\sqrt{q_{1}^{p}-q_{0}^{p}}z_{1})}
\label{eq-for-f-1RSB}
\eeqn
where $m_{2}=1$. Similarly we can recursively obtain
functions $\pi(m_{1},h)$ and $\pi(m_{2},h)$
(See \eq{eq-recursion-pi} and \eq{eq-boundary-f-dash}) as,
\beqn
\pi(m_{2},h) &=& \frac{\int \cD z_{2} (e^{-\beta V(h-\sqrt{1-q_{1}^{p}}z_{2})})'}{\int \cD z_{2} e^{-\beta V(h-\sqrt{1-q_{1}^{p}}z_{2})}} \nonumber \\
\pi(m_{1},h) &=& e^{m_{1}f(m_{1},h)}
  \int \cD z_{1}
\left . \pi(m_{2},h')e^{-m_{1}f(m_{2},h')}
\right |_{h'=h-\sqrt{q^{p}_{1}-q^{p}_{0}}z_{1}}
\label{eq-for-pi-1RSB}
\eeqn
Next we can recursively obtain functions $P(m_{0},h)$ and $P(m_{1},h)$ (see
\eq{eq-recursion-p} and \eq{eq-boundary-P}) as,
\beqn
P(m_{0},h)&=&\frac{1}{\sqrt{2\pi q_{0}^{p}}}e^{-\frac{(\delta-h)^{2}}{2q_{0}^{p}}}\nonumber \\
P(m_{1},h)&=& e^{-m_{1}f(m_{2},h)}
  \int \cD z_{1}
\left . P(m_{0},h')e^{m_{1}f(m_{1},h')}
\right |_{h'=h-\sqrt{q^{p}_{1}-q^{p}_{0}}z_{1}}
\label{eq-for-P-1RSB}
  \eeqn
  With these we are now readily to evaluate $\kappa_{0}$ and $\kappa_{1}$ using \eq{eq-kappa-1RSB}.

  To sum up, we can evaluate the 1RSB solution numerically as follows: (0) make some initial guess for the values of $q_{0}$ and $q_{1}$ (1) obtain functions
  $f(m_{2},h) \to f(m_{1},h)$ using \eq{eq-for-f-1RSB}
  (2) obtain functions
  $\pi(m_{2},h) \to \pi(m_{1},h)$
 using \eq{eq-for-pi-1RSB}
  (3) obtain functions
 $P(m_{0},h) \to P(m_{1},h)$
 using \eq{eq-for-P-1RSB}
 (4) Compute $\kappa_{0}$ and $\kappa_{1}$ using \eq{eq-kappa-1RSB}
 (4) solve for $G_{0}$ and $G_{1}$
  using \eq{eq-for-G-1RSB} and \eq{eq-kappa-1RSB}
  (5) compute $q_{0}$ and $q_{1}$
  using \eq{eq-for-q-1RSB}
  (6) return to (1). The procedure has to be repeated until the solution converges.

We note that the parameter $m_{1}$ remains. In order to study the equilibrium
state $m_{1}$ is fixed by the condition of vanishing complexity $\Sigma(m_{1})=0$. (See sec. \ref{subsubsec-1RSB})

\subsubsection{$k>1$ case}
\label{sec-numerical-procedure-kRSB}

The saddle point equations for a generic finite $k$-RSB ansatz
with some fixed values of $0 < m_{1} < m_{2} < \ldots < m_{k} <1$
can be solved numerically generalizing the procedure explained above.

\begin{itemize}
\item[0.] Make some guess for the initial values of $q_{i}$ ($i=0,1,2,\ldots,k$).
  Then compute $G_{i}$ for $i=0,1,\ldots,k$ using \eq{eq-G-q}.
  \item[1.] Compute function $f(m_{i})$ recursively
  for $i=k+1,k,\ldots,0$ using \eq{eq-recursion-f} with the boundary condition given by \eq{eq-boundary-f-type2}.
  Compute also functions $\pi(m_{i},h)$ recursively  for $i=k,k-1,\ldots,2,1$ 
using  \eq{eq-recursion-pi} with the boundary condition given by \eq{eq-boundary-f-dash}.
\item[2.] Compute functions $P(m_{i},h)$ recursively for $i=1,\ldots,k+1$
using  \eq{eq-recursion-p} with the boundary condition given by \eq{eq-boundary-P}.
\item[3.] Compute $\kappa_{i}$ for $i=0,1,\ldots,k$ using \eq{eq-def-kappa},
  $G_{i}$ for $i=0,1,\ldots,k+1$ using using \eq{eq-saddle-point-k-RSB},
  $q_{i}$ for $i=0,1,2,\ldots,k$ using \eq{eq-q-G} and finally
  $\Lambda_{i}$ for $i=0,1,\ldots,k+1$ using \eq{eq-def-Lambda} and \eq{eq-def-lambda}.
\item[4.] Return to 1.
\end{itemize}
The above procedure 1-4 must be repeated until the solution converges.

\subsubsection{$k=\infty$ case: continuous RSB}

In the limit $k \to \infty$, the variational equations
given by \eq{eq-saddle-point-k-RSB} become,
\beqn
\frac{q(0)}{G^{2}(0)} &=&\kappa(0)   \\
\frac{1}{G(x)}-\frac{1}{G(0)} &=&
-\int_{0}^{x} dy \kappa(y)+x \kappa(x)
\label{eq-saddle-point-continuous-RSB}
\eeqn
with
\beq
\kappa(x) \equiv \alpha q^{p-1}(x) \int dh P(x,h) \pi^{2}(x,h)
\label{eq-def-kappa-x}
\eeq

From the above equations we can derive some exact identities which
become useful later. Taking a derivative with respect to $x$
on both sides of \eq{eq-saddle-point-continuous-RSB}
and using \eq{eq-def-kappa-x}, \eq{eq-def-g-x},
\eq{eq-pdeq-P}, \eq{eq-pdeq-pi}, we find after some
integrations by parts,
\beq
1=\alpha(p-1)q^{p-2}(x)G^{2}(x)\int dh P(x,h)\pi^{2}(x,h)
+\alpha p q^{2(p-1)}(x)G^{2}(x)\int dh P(x,h) (\pi'(x,h))^{2}
\label{eq-exact-identity-fullRSB-1}
\eeq
Then taking another derivative on both sides of the above equation we
find after some integrations by parts,
\beqn
0&=&(p-1)q^{p-3}(x)[(p-2)G^{2}(x)-2q(x)xG(x)]\int dh P(x,h)\pi^{2}(x,h) \nonumber \\
&+&3p(p-1)q^{2p-3}(x)G^{2}(x)\int dh P(x,h)(\pi'(x,h))^{2} \nonumber \\
&+& pq^{2(p-1)}(x)(-2xG(x))\int dh P(x,h)(\pi'(x,h))^{2} \nonumber \\
&+& p^{2}q^{3(p-1)}(x)G^{2}(x)\int dh P(x,h)\left [ (\pi''(x,h))^{2}-2x (\pi'(x,h))^{3}\right].
\label{eq-exact-identity-fullRSB-2}
\eeqn

\subsection{Stability of the kRSB solution}
\label{sec-stability-kRSB}

Stability of the $k( \geq 1)$-RSB ansatz must be examined by studying
the eigenvalues of the Hessian matrix.
As we note in appendix \ref{subsec-Hessian-kRSB},  there is a residual
replica symmetry within each of the inner-core part of the replica groups.
Here we do not study the complete spectrum of the eigen-modes
of the Hessian matrix but focus on the so called replicon eigenvalue $\lambda_{R}$ which is responsible for the replica symmetry breaking of the residual replica symmetry.

\subsubsection{$k=1$ case: 1RSB}

For the $k=1$ case we find from \eq{eq-replicon-kRSB},
\beq
\lambda_{\rm R}=
\frac{2}{(1-q_{1})^{2}}
-2\frac{\alpha}{p}
\int dh P(m_{1},h)
  \left[
           p(p-1)q^{p-2}
          (\pi(m_{2},h))^{2}
           +(pq^{p-1})^{2}
           (\pi'(m_{2},h))^{2}  \right]
  \label{eq-replicon-1RSB}
  \eeq
  where $m_{2}=1$.
  Here we used $\pi(x,h)\equiv -f'(x,h)$ defined in \eq{eq-def-pi}.
  The functions $\pi(m_{2},h)$ and $P(m_{1},h)$ are
  given by \eq{eq-for-pi-1RSB} and  \eq{eq-for-P-1RSB} respectively.

  The vanishing on $\lambda_{\rm R}$ signals the Gardner's transition \cite{Ga85}: instability to further breaking of the replica symmetry.

  \subsubsection{$k=\infty$ case: continuous RSB}
  \label{subsubsec-replicon-continuousRSB}

From \eq{eq-replicon-kRSB} we find for $k=\infty$, by which $m_{k} \to 1$,
\beq
\lambda_{\rm R}=
\frac{2}{G(1)^{2}}
-2\frac{\alpha}{p}
\int dh P(1,h)
  \left[
           p(p-1)q^{p-2}
          (\pi(1,h))^{2}
           +(pq^{p-1})^{2}
           (\pi'(1,h))^{2}  \right]
  \eeq
where we used $\pi(x,h)\equiv -f'(x,h)$ defined in \eq{eq-def-pi}
and $G(1)=1-q(1)$ which follows from \eq{eq-def-g-x}.
Now using the exact identity given by \eq{eq-exact-identity-fullRSB-1} which holds for the continuous RSB system, we find it vanishes exactly: $\lambda_{\rm R}=0$. Thus the continuous RSB solution is marginally stable.

\section{Quadratic potential}
\label{sec-quadratic-potential}

Now we are ready to study specific problems with non-linear potentials $V(x)$
and continuous spins.
First let us briefly examine the simplest one,
\begin{equation}
V(x)=\frac{\epsilon}{2} x^{2} \qquad \epsilon>0.
\label{eq-quadratic-potential}
\end{equation}
As we show below it is already a non-trivial problem.
To see this it is useful to expand the interaction part of the
free-energy given by \eq{eq-s-glass-ref} in power series
of the order parameter,
\beqn
&& \left. -\partial_{n} {\cal F}_{\rm int}  \right |_{n=0} \nonumber \\
&& =\partial_{n}  \left[
  n\frac{(\beta \epsilon)^{2}}{2} (1+\delta^{2})
  +\delta^{2} \frac{(\beta \epsilon)^{2}}{2} \sum_{ab} Q_{ab}^{p}
  + \frac{(\beta \epsilon)^{2}}{2} \sum_{ab} Q_{ab}^{2p}
  \right.
  \nonumber \\
  && \left.\left.
  \hspace*{5cm}
   +\frac{2}{3}(\beta \epsilon)^{3} \sum_{abc}Q_{ab}^{p}Q_{bc}^{p}Q_{ca}^{p}
  + \ldots
  \right] \right |_{n=0}
\eeqn
In contrast to the case of the linear potential discussed in
sec \ref{sec-linear}, higher order terms of $Q_{ab}^{p}$ appears.
Thus even with $p=2$, for which system remains RS for the
linear potential case (spherical SK model \cite{kosterlitz1976spherical}),
one can expect that the quadratic potential allows RSB
since the above expression is somewhat
similar to the interaction part of the free-energy of the $2+p$ spherical model
\cite{crisanti2004spherical}, which exhibits various types of RSB.

\begin{figure}[h]
  \includegraphics[width=\textwidth]{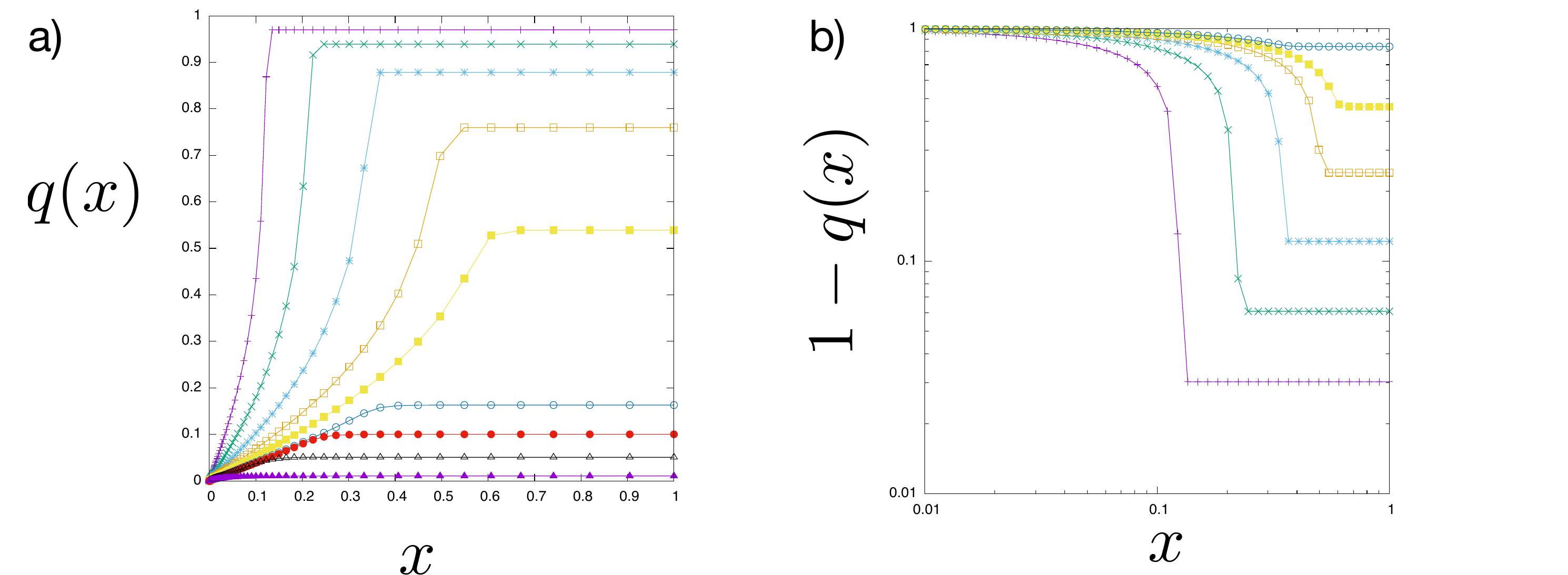}
  \caption{The $q(x)$ function of the quadratic potential
    with $p=2$, $\delta=1.0$, $\alpha=2$ for which
    $T_{\rm c}/\epsilon=\sqrt{2}-1=0.414..$.
    a) linear plot : 
        $T/\epsilon=0.00625,0.0125,0.025,0.05,0.25,0.3,0.35,0.4$
    from top to bottom. The points represent the solution
    of the continuous RSB equation
    (obtained by solving the recursion formulas with $k=200$ steps).
    b) double logarithmic of the function $1-q(x)$ vs $x$.
}
  \label{fig-qx-quadratic}
\end{figure}

In the present paper we do not explore the whole phase diagram
but let us show the existence of continuous RSB for the $p=2$ case.
From \eq{eq-critical-point-rs} we find the critical point,
\beq
\alpha_{\rm c}(\delta)=\left ( 1+ \frac{1}{\beta \epsilon} \right)^{2}|\delta|^{-2}
\eeq
which implies the glass transition temperature,
\beq
T_{\rm c}(\alpha,\delta)=\sqrt{\alpha}|\delta|-1.
\eeq
Above the critical temperature, i.~e. $T> T_{\rm c}$ the
$q=0$ RS solution (liquid phase)
is valid (sec. \ref{section-The-liquid phase-:-q=0-solution}).
One naturally expects continuous glass transition takes place at $T_{\rm c}$.

We solved numerically the continuous RSB equation approximated by $k$-step RSB
(With $k=200$) recursion formulas with appropriate boundary
conditions as explained in sec. \ref{sec-numerical-procedure-kRSB}.
In Fig.~\ref{fig-qx-quadratic} we show the continuous RSB solution
obtained numerically for the case of $p=2$, $\alpha=2$
and $\delta=1.0$ at various temperatures below $T_{\rm c}/\epsilon=\sqrt{2}-1=0.414..$. As expected the glass transition takes place continuously.
Moreover it accompanies continuous RSB as evident in the figure.
The $q(x)$ function has a plateau $q(x)=q_{1}$ for some range $x_{1}< x< 1$.
The plateau height $q_{1}$ is interpreted as the self-overlap of the glassy states or the Edwards-Anderson order parameter $q_{\rm EA}$ \cite{edwards1975theory} while the continuous part at $x < x_{1}$ describes the hierarchical organization of the glassy states
\cite{MPV87}.

Finally let us consider stability of the glass phase  against crystallization.
From the analysis in sec. \ref{sec-stability-against-crystalization} we know
that for the case $p=2$ the flatness of the potential $V'(\delta)=0$
is needed to ensure the locally stability against crystallization
in the disorder-free model.
In the case of the quadratic potential \eq{eq-quadratic-potential}
the condition is met only for $\delta=0$.
However the above results imply $\alpha_{\rm c}(\delta=0) = \infty$.
Thus for the $p=2$ case with the quadratic potential,
we cannot avoid using the disordered model given by \eq{eq-X-disordered} in order to allow the desired glass transition.
The free-energy functional of the disordered model
is given by \eq{eq-free-entropy-Q-and-m} with \eq{eq-F-int-glass-with-m}.
The solution we obtained just above amount to assume $m=0$.
Such a solution is certainly locally stable
for the fully disordered case $\lambda/\sqrt{M}=0$ 
and presumably also for small enough $\lambda$ (see \eq{eq-Hab}). 

\section{Hardcore potential}
\label{sec-hardcore}

We will now focus on the continuous spin system subjected to a more strongly non-linear potential, namely soft/hardcore potential,
\begin{equation}
V(x)= \epsilon x^{2}\theta(-x) \qquad \epsilon>0,
\label{eq-hardcore-potential}
\end{equation}
which becomes a hardcore potential in the limit $\epsilon \to \infty$. Much as in hardspheres\cite{parisi2010mean,charbonneau2014exact} we can expect jamming $q_{\rm EA} \to 1$, i.~e. vanishing of the thermal fluctuation due to tightening of the constraints.
This would happen in two ways: by decreasing $\delta$ or increasing $\alpha$ (connectivity).
In the context of the coloring problems decreasing $\delta$ is analogous to decreasing the number of colors
allowed to use (see Fig.~\ref{fig_rotational_glass} a)).

Note that in the special case $p=1$ and with fully disordered choice $\lambda/\sqrt{M}=0$
in \eq{eq-X-disordered}, the model becomes identical to the perceptron problem \cite{gardner1988space}.
Recent works \cite{franz2016simplest,franz2017universality} have established that the universality class of the jamming in the perceptron model with $\delta < 0$ is the same as that of hardspheres \cite{charbonneau2014exact}. We wish to clarify if the same universality holds for $p \geq 2$ or not.

The soft/hardcore potential defined above is flat such that $V'(x)=0$ for $x>0$.Thus for $\delta > 0$, the supercooled paramagnetic phase and glassy phases are locally stable against crystallization even in the  $p=2$ system (See \eq{eq-stability-para-ising}, \eq{eq-stability-para-continuous} and \eq{eq-Hab}). 
In contrast, if $\delta < 0$ and $p=2$, the quenched disorder is needed to allow the glassy phases.
For $p> 2$, the the supercooled paramagnetic states and glassy states are always locally stable against crystallization.
In the following we study both $\delta >0$ and $\delta < 0$ assuming $m=0$ but we should keep these points in our mind.

Below we closely follow the analysis done on hardspheres in $d \to \infty$ limit
\cite{charbonneau2014exact} and find indeed that many aspects
are quite similar to those found there, especially at jamming.

\subsection{Replica symmetric solution}

Let us first study the RS solution discussed in sec. \ref{sec-RS}
in the case of the hardcore model.

       \subsubsection{Free-energy}
       
       For the
hardcore potential we find the RS free-energy given by \eq{eq-free-energy-RS} as,
\beq
-\beta f_{\rm RS}(q)=\frac{1}{2}\left( \frac{q}{1-q}+\ln (1-q)\right)
+\frac{\alpha}{p}
\int \cD z_{0}
\ln \Theta \left(\frac{\delta -\sqrt{q^{p}}z}{\sqrt{2(1-q^{p})}}\right)
\eeq
where we introduced a function $\Theta(x)$ 
\beq
\Theta(x) \equiv \int_{-\infty}^{x}\frac{dz}{\sqrt{\pi}}e^{-z^{2}}
=\gamma_{1/2} \otimes \theta(x)=\frac{1}{2} (1+{\rm erf}(x)),
\label{eq-def-Theta}
\eeq
with ${\rm erf}(x)$ being the error function,
\beq
{\rm erf}(x)=\frac{2}{\sqrt{\pi}}\int_{0}^{x} dy e^{-y^{2}}=-{\rm erf}(-x),
\eeq
which behaves for $x \to \infty$ as,
\beq
    {\rm erf}(x) = 1-\frac{1}{\sqrt{\pi}}\frac{e^{-x^{2}}}{x} \left (
    1 - \frac{1}{2x^{2}}+\frac{3}{(2x^{2})^{2}}+ \ldots \right).
    \label{eq-erf-asymptotic}
    \eeq
    This implies,
    \beq
    \Theta(x) \simeq
     \left \{
    \begin{array}{cc}
 \frac{1}{2}\frac{e^{-x^{2}}}{(-x)\sqrt{\pi}}\left[1-\frac{1}{2x^{2}}+\frac{3}{(2x^{2})^{2}}+\ldots\right] & x \to -\infty \\
1 & x \to \infty
    \end{array}
    \right.
    \label{eq-theta-asymptotic}
    \eeq

The function ${\cal G}(q)$ defined in \eq{eq-def-function-Gq} becomes,
\beq
    {\cal G}(q)=1-\alpha (1-q)^{2}q^{p-2} 
    \frac{1}{2(1-q^{p})}
    \int \cD z_{0}
    \left. 
    r^{2}(x)
    \right |_{x=\frac{\delta-\sqrt{q^{p}}z_{0}}{\sqrt{2(1-q^{p})}}}
    \label{eq-function-Gq-hardcore}
    \eeq
where we introduced,
\beq
r(x) \equiv \frac{\Theta'(x)}{\Theta(x)}=
\frac{e^{-x^{2}}}{\sqrt{\pi}}/ \Theta(x)
\label{eq-def-r}
\eeq
which behaves asymptotically as,
   \beq
r(x)      
\simeq \left \{
\begin{array}{cc}
-2x \left(1-\frac{1}{2x^{2}}+\frac{3}{(2x^{2})^{2}}+ \ldots \right)^{-1} 
&  x \to -\infty \\
0 & x \to \infty
\end{array}
\right.
       \label{eq-def-r-x}
       \eeq

       \subsubsection{$q=0$ RS solution and its stability}

                  \begin{figure}[t]
       \includegraphics[width=\textwidth]{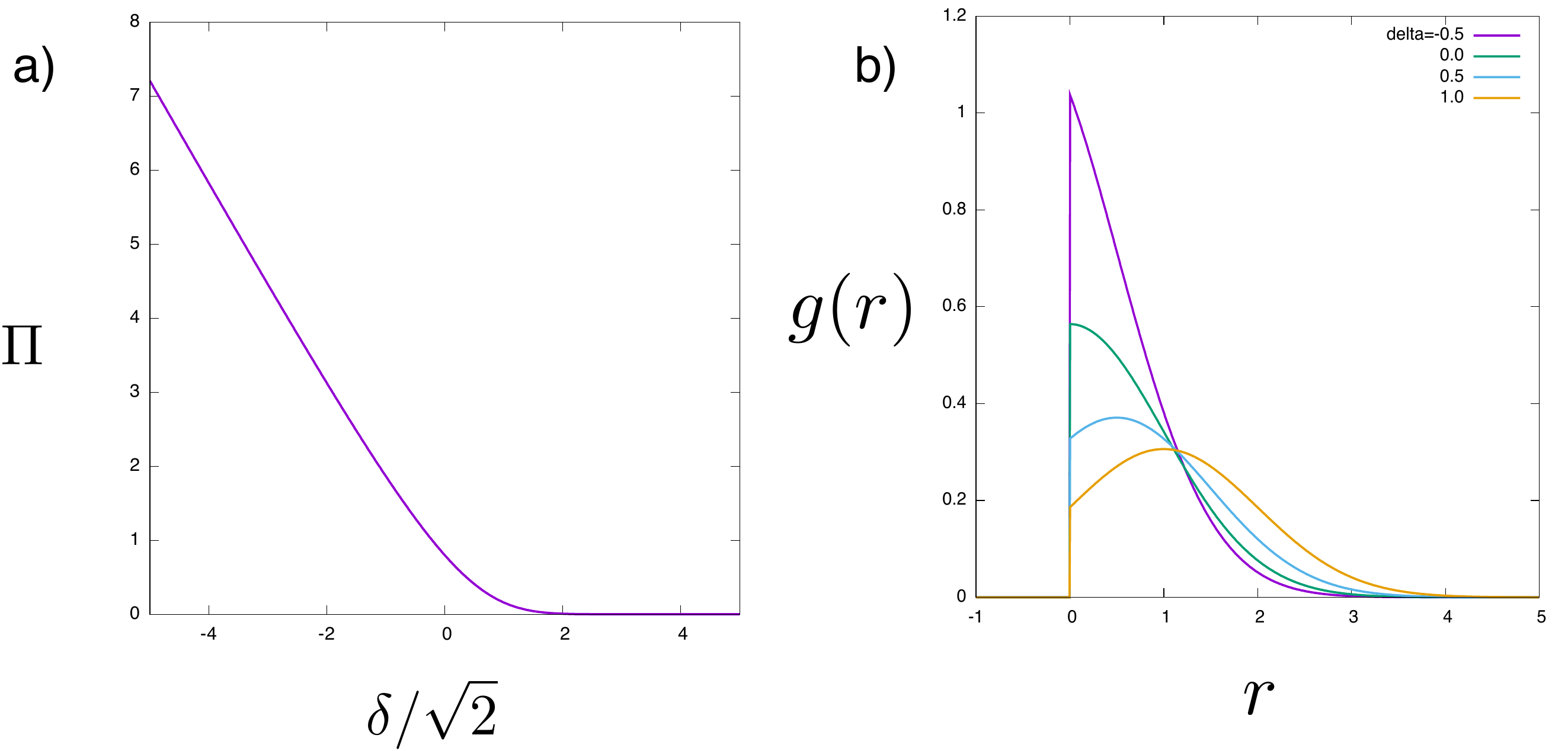}
  \caption{Liquid phase (RS solution with $q=0$) of the hardcore model: a) behavior of the pressure $\Pi$
  and b) the distribution of the gap $g(r)$. Note that these are independent of $p$.}
\label{fig_pressure_g_of_r_rs}
    \end{figure}

       Within the liquid state $q=0$, the $p$-dependence disappears.
       The pressure given by \eq{eq-pressure-RS} is obtained as,
       \beq
       \Pi=\frac{1}{\sqrt{2}}r(\delta/\sqrt{2})=
       \frac{1}{\sqrt{2}}
       \frac{\Theta'(\delta/\sqrt{2})}{\Theta(\delta/\sqrt{2})}
        \label{eq-pressure-RS-hardcore}
    \eeq
    As shown in the left panel of
    Fig. \ref{fig_pressure_g_of_r_rs}, the pressure monotonically increases
    by decreasing $\delta$ as expected.
    We display in the right panel of Fig. \ref{fig_pressure_g_of_r_rs} b) the
    behavior of the distribution of gap  given by \eq{eq-g-of-r-RS} which becomes,
    \beq
    g(r)=\frac{\theta(r)}{\Theta (\delta/\sqrt{2})} \frac{e^{-\frac{(\delta-r)^{2}}{2}}}{\sqrt{2\pi }}
    \eeq
    We see that the peak around $r=0$ develops by decreasing $\delta$ as expected.

    As we found in sec.      \ref{section-The-liquid phase-:-q=0-solution}
    $q=0$ solution is always stable for $p>2$ body interactions. For the $p=2$ case, we find the $q=0$ solution becomes unstable for $\alpha > \alpha_{\rm c}(\delta)$  with       
\beq
\alpha_{\rm c}(\delta)=2r^{-2}(\delta/\sqrt{2})=
2\left(\frac{
  \Theta'(\delta/\sqrt{2})}{\Theta(\delta/\sqrt{2})}\right)^{-2}
\label{eq-critical-point-rs-hardcore}
\eeq
which is obtained from \eq{eq-critical-point-rs}.
The critical line $\alpha_{\rm c}(\delta)$ is displayed in Fig.~\ref{fig-AT-line}.

\subsubsection{Jamming within the RS ansatz}

It is possible to look for glass transition within the RS ansatz
by looking for $q \neq 0$ solution of the RS saddle point equation given by \eq{eq-RS-saddlepoint}, which must solve ${\cal G}(q)=0$.
In sec \ref{sec-RS-p=2-hardcore-glass} we will examine the phase diagram within the
RS ansatz for $p=2$ case. Here we instead focus on the
jamming limit where the EA order parameter saturates $q_{\rm EA}=q \to 1$ signaling vanishing of the thermal fluctuations.

The location of the jamming point can be analyzed as follows.
We find ${\cal G}(q)$ given in \eq{eq-function-Gq-hardcore} becomes in the limit
$q \to 1$,
\beqn
\lim_{q\to 1}{\cal G}(q) &= &1-\frac{\alpha}{2p} \lim_{q \to 1} (1-q)
    \int \cD z_{0}
    \left. 
    r^{2}(x)
    \right |_{x=\frac{\delta-z_{0}}{\sqrt{2(1-q^{p})}}}  \nonumber \\
&=& 1-\frac{\alpha}{p^{2}}\int_{0}^{\infty}\frac{dy}{\sqrt{2\pi}}e^{-(\delta+y)^{2}/2}y^{2} 
    \eeqn
In the last equation we used the asymptotic behavior of the function $r(x)$ given in
\eq{eq-def-r-x}.
Thus we find the jamming line $\alpha=\alpha_{\rm j}(\delta)$,
\beq
\alpha_{\rm j}(\delta)=\frac{p^{2}}{\int_{0}^{\infty}\frac{dy}{\sqrt{2\pi}}e^{-(\delta+y)^{2}/2}y^{2}}
\label{eq-jamming-point-rs-hardcore}
\eeq
which is also displayed in Fig.~\ref{fig-AT-line}.

The pressure given by \eq{eq-pressure-RS} becomes for the hardcore model,
\beq
\Pi=\frac{1}{\sqrt{2}}\int \cD z_{0} \left. r(x)  \right |_{x=\frac{\delta-\sqrt{q^{p}}z_{0}}{\sqrt{2(1-q^{p})}}} \xrightarrow{q \to 1}
\int_{\delta}^{\infty} \cD z_{0} \frac{(z_{0}-\delta)}{\sqrt{1-q^{p}}}
\propto \frac{1}{\sqrt{1-q}}
\label{eq-RS-hardcore-pressure}
\eeq
where we used the asymptotic expansion given by \eq{eq-def-r-x}.
Thus as expected the pressure diverges by jamming (see Fig.~\ref{fig-q-pressure-RS} b)).

Next let us examine the distribution of the gap  given by \eq{eq-g-of-r-RS}
which becomes  for the hardcore model,
\beq
g(r)=\theta(r) \int \cD z_{0}\frac{\gamma_{1-q^{p}}(\delta-\sqrt{q^{p}}z_{0}-r)}{
  \Theta\left(\frac{\delta-\sqrt{q^{p}}z_{0}}{\sqrt{2(1-q^{p})}}\right)}.
\label{eq-RS-hardcore-g-of-r}
\eeq
The behavior in the jamming limit $q \to 1$ can be viewed in the following
two ways  (see Fig.~\ref{fig-q-pressure-RS} c))  much as in the case of hardspheres \cite{charbonneau2014exact},
\begin{enumerate}
\item For {\it fixed} finite $r$, sending $q \to 1$, we find
  \begin{equation}
    \lim_{q \to 1} g(r)=\theta(r)\frac{e^{-\frac{(\delta-r)^{2}}{2}}}{\sqrt{2\pi}}
   \label{eq-gr-scaling-RS-1}
    \end{equation}
  This is because $\gamma_{1-q^{p}}(\delta-\sqrt{q^{p}}z_{0}-r)$
  becomes a delta function in the $q \to 1$ limit and $\lim_{X \to \infty} \Theta(X)=1$.
\item In the vanishing region around $r=0$ parametrized
  as $r=(1-q^{p})\lambda$ we find a different behavior as follows.
  Assuming $q \sim 1$ we find for $r>0$,
  \beqn
  g(r) &\sim &
  \int_{\delta}^{\infty}\frac{dz_{0}}{\sqrt{2\pi}}
  \frac{
        e^{-\frac{(\delta-z_{0}-r)^{2}}{2(1-q^{p})}}
  }{
      2\pi(\sqrt{1-q^{p}})}
  2\left. \sqrt{\pi}|X|e^{X^{2}} \right |_{X=\frac{\delta-\sqrt{q^{p}}}{2\sqrt{1-q^{p}}}}
\nonumber \\
&  \xrightarrow{q \to 1, {\rm fixed} \lambda } &\frac{1}{1-q^{p}}\int_{0}^{\infty} \frac{dy}{\sqrt{2\pi}}
e^{-\frac{(\delta+y)^{2}}{2}} y e^{-\lambda y} \qquad \lambda=\frac{r}{1-q^{p}}
   \label{eq-gr-scaling-RS-2}
  \eeqn
  In the 1st equation we dropped contribution from $\int_{-\infty}^{\delta} dz_{0}..$ which can be neglected compared with the contribution from
  $\int_{\delta}^{\infty} dz_{0}..$
and used the asymptotic behavior of the error function
given ny \eq{eq-erf-asymptotic} which implies $\Theta(-X) \sim \frac{e^{-X^{2}}}{2\sqrt{\pi} X}$ for $X \ggg 1$.
Thus in the jamming limit $q \to 1$
we find diverging peak in the ``contact region'' around $r=0$
whose height diverging as $1/(1-q)$ and the width vanishing as
$1-q$.

\end{enumerate}

\subsection{$k$-step RSB solution}

We study now the $k$-RSB solution discussed in sec. \ref{sec-RSB}
in the case of the hardcore model.

\subsubsection{Inputs}
\label{sec-inputs-hardcore}

Here we present some necessary inputs to study the glass phase and jamming
of the hardcore model within generic $k$-RSB ansatz.
Within the $k$-RSB ansatz, jamming means $q_{\rm EA}=q_{k} \to 1$  (see sec \ref{subsec-Parisi-ansatz}).
With the following inputs, 
1RSB solution can be obtained following sec. \ref{sec-1RSB-equations}
and generic $k$-RSB solution can be obtained following sec. \ref{sec-numerical-procedure-kRSB}.

For the hardcore potential given by \eq{eq-hardcore-potential}
we find
\beqn
f(m_{k+1},h)= - \ln \Theta \left(
\frac{h}{\sqrt{2\Lambda_{k+1}}}
\right)
\label{eq-boundary-f-hardcore}
\eeqn
where $\Theta(x)$ is defined in \eq{eq-def-Theta}.
Then
\beqn
\pi(m_{k+1},h)=
\frac{1}{\sqrt{2\Lambda_{k+1}}}
\frac{\Theta'\left(
  \frac{h}{\sqrt{2\Lambda_{k+1}}}
  \right)}{\Theta\left(
  \frac{h}{\sqrt{2\Lambda_{k+1}}}
  \right)}
\label{eq-boundary-pi-hardcore}
\eeqn
The functions $f(m_{i},h)$ and $\pi(m_{i},h)$ are determined
via recursion formulas given by \eq{eq-recursion-f} and \eq{eq-recursion-pi}
using the boundary values obtained above.

It is useful to study the asymptotic behavior of the functions
$f(m_{i},h)$ and $\pi(m_{i},h)$ in the  limit $h \to -\infty$
both for numerical and analytical purposes.
Using \eq{eq-def-Theta} and 
\eq{eq-erf-asymptotic}
and the recursion formula given by \eq{eq-recursion-f} one finds
for $i=1,2,\ldots,k+1$,
\beq
f(m_{i},h)= 
\left \{
\begin{array}{cc}
  0 & h \to \infty \nonumber \\
  \frac{h^{2}}{2\tilde\Lambda_{i}} & h \to -\infty
\end{array}
\right .
\qquad 
\pi(m_{i},h)= 
\left \{
\begin{array}{cc}
  0 & h \to \infty \nonumber \\
  -\frac{h}{\tilde\Lambda_{i}} & h \to -\infty
\end{array}
\right .
\eeq
where we introduced,
\beq
\tilde\Lambda_{i}\equiv \sum_{j=i}^{k+1}m_{j}\Lambda_{j}.
\label{eq-def-tilde-Lambda}
\eeq
Note that $\tilde\Lambda_{k+1}=m_{k+1}\Lambda_{k+1}=\Lambda_{k+1}=1-q_{k}^{p}$.
In the continuous limit $k \to \infty$ this implies,
\beq
f(x,h)= 
\left \{
\begin{array}{cc}
  0 & h \to \infty \\
  \frac{h^{2}}{2\tilde\Lambda(x)} & h \to -\infty
\end{array}
\right .
\qquad 
\pi(x,h)= 
\left \{
\begin{array}{cc}
  0 & h \to \infty  \\
  -\frac{h}{\tilde\Lambda(x)} & h \to -\infty
\end{array}
\right.
\label{eq-f-pi-large-lambda}
\eeq
with
        \beq
        \tilde{\Lambda}(x)=1-\int_{x}^{1}dy \lambda(y)-x\lambda(x).
        \label{eq-def-tilde-lambda-x}
        \eeq
        with $\lambda(x)=q^{p}(x)$ defined in \eq{eq-def-lambda-x}.

        The above observation suggests us to introduce a function $j(m_{i},h)$
        defined as,
        \beq
        -f(m_{i},h) \equiv
        - \frac{h^{2}}{2\tilde\Lambda_{i}}\theta(-h)+j(m_{i},h).
        \label{eq-def-j}
       \eeq
       From \eq{eq-recursion-f} we find that 
      the function $j(m_{i},h)$ follows a recursion relation,
       \beq
       j(m_{i},h)=\frac{1}{m_{i}}
       \ln \int dy K_{i,i+1}(h,y)e^{m_{i}j(m_{i+1},y)}
       \label{eq-recursion-j}
       \eeq
       with
       \beq
       K_{i,i+1}(y,h)\equiv \frac{1}{\sqrt{2\pi \Lambda_{i}}}
       \exp \left [ -\frac{(h-y)^{2}}{2\Lambda_{i}}
       -\frac{m_{i}}{2}\frac{y^{2}}{\tilde\Lambda_{i+1}}\theta(-y)
       +\frac{m_{i}}{2}\frac{h^{2}}{\tilde\Lambda_{i}}\theta(-h)
       \right]
       \label{eq-def-Kij}
       \eeq
       and the boundary condition,
       \beq
       j(m_{k+1},h)=\ln \Theta\left (\frac{h}{\sqrt{2\Lambda_{k+1}}} \right)
       +\frac{h^{2}}{2\tilde{\Lambda}_{k+1}}\theta(-h)
       \label{eq-boundary-j}
       \eeq
       Correspondingly one finds that \eq{eq-recursion-pi} becomes,
       \beq
       \pi(m_{i},h)=
       \int dy \pi(m_{i+1},y)K_{i,i+1}(h,y)
       e^{m_{i}(j(m_{i+1},y)-j(m_{i},y))}
              \label{eq-recursion-pi-via-j}
       \eeq

       \subsubsection{Rescaled quantities useful close to jamming}

       Let us show below how to modify the numerical algorithm in sec \ref{sec-numerical-procedure-kRSB} to solve the continuous RSB equations close to jamming, where $q_{\rm EA}=q_{k} \to 1$.
To this end let us first introduce several rescaled quantities.

       As mentioned above jamming within the $k$-RSB ansatz means
       \beq
       \Delta \equiv 1-q_{k} \to 0.
       \label{eq-jamming-limit}
       \eeq
       Then it is convenient to
       introduce the following rescaled quantities,
       \beq
       \gamma_{i}=\frac{G_{i}}{\Delta}
       \qquad (i=0,1,2,\ldots,k)
       \eeq
       with $G_{i}$ being defined in \eq{eq-G-q}.
       Note that
       \beq
       \gamma_{k}=1
       \eeq
       since $G_{k}=1-q_{k}=\Delta$. 
       We replace $0=m_{0} < m_{1} < m_{2} < \ldots < m_{k} < m_{k+1}=1$ by,
       \beq
              y_{i}=\frac{m_{i}}{\Delta}
       \qquad (i=0,1,2,\ldots,k,k+1).
       \eeq
       In terms of these we can write (\eq{eq-q-G}),
       \beq
       q_{i}=1-\Delta+\sum_{j=i+1}^{k}\frac{1}{y_{j}}(\gamma_{j}-\gamma_{j-1}) \qquad (i=0,1,2,\ldots,k)
       \label{eq-q-gamma}
       \eeq
       which in turn implies 
       \beq
       \gamma_{i}=\gamma_{i+1}+y_{i+1}(q_{i+1}-q_{i})  \qquad (i=0,1,2,\ldots,k-1)
       \label{eq-gamma-q}
       \eeq

       Let us also introduce,
       \beq
       \hat{f}(y_{i},h)=\Delta f(m_{i},h)   \qquad    \hat{j}(y_{i},h)=\Delta j(m_{i},h)   \qquad     \hat{\pi}(y_{i},h)=\Delta \pi(m_{i},h)     \qquad (i=1,2,\ldots,k+1)
       \label{eq-def-f-pi-hat}
       \eeq
       Then \eq{eq-def-j} becomes
       \beq
       -\hat{f}(y_{i},h)=-\frac{h^{2}}{2}\frac{1}{\hat{\Lambda}_{i}}\theta(-h)+\hat{j}(y_{i},h)
       \label{eq-hat-f-hat-j}
       \eeq
       with
       \beq
       \hat{\Lambda}_{i}\equiv \frac{\tilde{\Lambda_{i}}}{\Delta}=\sum_{j=i}^{k+1}y_{j}\Lambda_{j}
       \qquad \Lambda_{j}=q_{j}^{p}-q_{j-1}^{p}
       \label{eq-lambda-and-tilde-lambda}
       \eeq
       where we used  \eq{eq-def-tilde-Lambda}, \eq{eq-def-Lambda} and \eq{eq-def-lambda}.
       The  recursion given by \eq{eq-recursion-j} and \eq{eq-recursion-pi-via-j} become,
       \beqn
&&               \hat{j}(m_{i},h)=\frac{1}{y_{i}}
       \ln \int dh' K_{i,i+1}(h,h')e^{y_{i}\hat{j}(y_{i+1},h')} \nonumber \\
&&              \hat{\pi}(y_{i},h)=
       \int dh' \hat{\pi}(y_{i+1},h')K_{i,i+1}(h,h')
       e^{y_{i}(\hat{j}(y_{i+1},h')-\hat{j}(y_{i},h'))}
       \qquad 
       \label{eq-recursion-hat-j-and-hat-pi}
       \eeqn
       for $i=0,1,\ldots,k$ while \eq{eq-def-Kij} becomes
       \beq
              K_{i,i+1}(h',h)\equiv \frac{1}{\sqrt{2\pi \Lambda_{i}}}
       \exp \left [ -\frac{(h-h')^{2}}{2\Lambda_{i}}
       -\frac{y_{i}}{2}\frac{(h')^{2}}{\hat\Lambda_{i+1}}\theta(-h')
       +\frac{y_{i}}{2}\frac{h^{2}}{\hat\Lambda_{i}}\theta(-h)
       \right]
       \eeq
       
       The boundary condition given by \eq{eq-boundary-j}
and \eq{eq-boundary-pi-hardcore} become,
        \beqn
&&         \hat{j}(y_{k+1}=1/\Delta,h)=\Delta
\ln \Theta \left (\frac{h}{\sqrt{2\Delta\hat{\Lambda}_{k+1}}} \right)
       +\frac{h^{2}}{2\hat{\Lambda}_{k+1}}\theta(-h) 
        \xrightarrow[]{\Delta \to 0} 0 \nonumber \\
&&    \hat{\pi}(y_{k+1}=1/\Delta,h)
        =\frac{\Delta}{\sqrt{2\Delta\hat{\Lambda}_{k+1}}}
 \left.        \frac{\Theta'(x)}{\Theta(x)} \right |_{x=\frac{h}{\sqrt{2\Delta\hat{\Lambda}_{k+1}}}}
        \xrightarrow[]{\Delta \to 0} 
        -\frac{h}{p}\theta(-h)
       \label{eq-boundary-hat-j-hat-pi-hardcore}
       \eeqn
Here we used \eq{eq-def-f-pi-hat}
and the asymptotic expansions
        \eq{eq-theta-asymptotic} and        \eq{eq-def-r-x}.
We also used       
\eq{eq-lambda-and-tilde-lambda} and \eq{eq-def-Lambda} which imply $\hat{\Lambda}_{k+1}=\tilde{\Lambda}_{k+1}/\Delta=\Lambda_{k+1}/\Delta=(1-q_{k}^{p})/\Delta \xrightarrow[]{\Delta \to 0} p$.

On the other hand the recursion formula given by \eq{eq-recursion-p} becomes
\beq
P(y_{j},h)=e^{-y_{j}\hat{f}(y_{j+1},h)} \gamma_{\Lambda_{j}}\otimes_{h}
\frac{P(y_{j-1},h)}{e^{-y_{j}f(y_{j},h)}} \qquad j=1,2,\ldots,k+1
\label{eq-recursion-p-jamming}
\eeq
and the boundary condition given by \eq{eq-boundary-P} becomes,
\beq
P(y_{0},h)=\frac{1}{\sqrt{2\pi \Lambda_{0}}}e^{-\frac{(\delta-h)^{2}}{2\Lambda_{0}}}
\qquad \Lambda_{0}=q(0)^{p}
\label{eq-boundary-P-jamming}
\eeq

Finally \eq{eq-saddle-point-k-RSB} and \eq{eq-def-kappa} become,
\beqn
\frac{q_{0}}{\gamma_{0}^{2}} &=&\alpha \hat{\kappa}^{0}_{0}   \nonumber \\
\frac{1}{\gamma_{i}}-\frac{1}{\gamma_{0}}&=&
\alpha\left( \sum_{j=0}^{i-1}(y_{j}-y_{j+1})\hat{\kappa}^{0}_{j}+y_{i}\hat{\kappa}^{0}_{i}
\right)
\qquad i=1,2,\ldots,k
\label{eq-saddle-point-k-RSB-jamming} \\
&& \hat{\kappa}^{0}_{i} \equiv q_{i}^{p-1} \int dh P(y_{i},h) \hat{\pi}^{2}(y_{i+1},h)
\qquad i=0,1,2,\ldots,k
\label{eq-def-kappa-jamming}
\eeqn

       \subsubsection{Algorithm to solve the continuous RSB equations close to jamming}
       
       The saddle point equations for a generic finite $k$-RSB ansatz
with some fixed values of $0 < m_{1} < m_{2} < \ldots < m_{k} <1$
can be solved numerically as explained  in sec. \ref{sec-numerical-procedure-kRSB}. We can modify it using the rescaled quantities.

\begin{itemize}
\item[0.] Make some guess for the initial values of $q_{i}$ ($i=0,1,2,\ldots,k$).
  \item[1.] Compute $\Delta$ as $\Delta=1-q_{k}$.
  \item[2.]Given $\Delta$ we have $y_{i}=m_{i}/\Delta$ ($i=1,2,\ldots,k$). Note that $y_{0}=0$ and $y_{k+1}=1/\Delta$.
  Then compute $\gamma_{i}$ for $i=0,1,\ldots,k-1$ using \eq{eq-gamma-q}.
  Note that $\gamma_{k}=1$.
  \item[3.] Compute $\hat{j}(y_{i},h)$ and $\hat{\pi}(y_{i},h)$ recursively
    for $i=k,\ldots,1$ using \eq{eq-recursion-hat-j-and-hat-pi} with the boundary condition given by \eq{eq-boundary-hat-j-hat-pi-hardcore}.
\item[4.] Compute functions $P(m_{i},h)$ recursively for $i=1,\ldots,k+1$
  using  \eq{eq-recursion-p-jamming}
  with the boundary condition given by \eq{eq-boundary-P-jamming}.
\item[5.] Compute $\hat{\kappa}^{0}_{i}$  for $i=0,1,\ldots,k$ using \eq{eq-def-kappa-jamming},  $\gamma_{i}$ for $i=0,1,2,\ldots,k$ using using \eq{eq-saddle-point-k-RSB-jamming}
  and finally $q_{i}$ for $i=0,1,2,\ldots,k$ using \eq{eq-q-gamma}.
  Finally compute $\Lambda_{i}$ and
  $\hat{\Lambda}_{i}$ for $i=1,2,\ldots,k+1$ using \eq{eq-lambda-and-tilde-lambda}.
\item[6.] Return to 1.
\end{itemize}
The above steps 1-6 must be repeated until the solution converges.

\subsubsection{Algorithm to look for the jamming point}
\label{sec-Algorithm-to-look-for-the-jamming point}

We can also look for the $k$-RSB solution for a given, fixed $\Delta$.
This can be seen as the following. In the step 5 of the procedure explained
above we obtain $\gamma_{k}$
using \eq{eq-saddle-point-k-RSB-jamming} but
$\gamma_{k}=1$. Thus
\eq{eq-saddle-point-k-RSB-jamming} for $i=k$ can be considered
as an equation to determine $\alpha=\alpha(\Delta)$.
In particular, the jamming point $\alpha_{\rm j}(\delta)$ via $k$-RSB ansatz can be determined by choosing $\Delta=0$.

       \subsection{Jamming criticality}

       Let us discuss properties of the system approaching the jamming in the case of continuous RSB, i.~.e. $k \to \infty$.
       As mentioned in  sec. \ref{subsec-Parisi-ansatz},
       we expect the $q(x)$ function of the continuous RSB solution has a continuous
              part for some range $x< x_{1}$ and a plateau $q(x)=q(x_{1})=q_{\rm EA}$ for
       $x_{1} < x < 1$. Jamming means $q_{\rm EA}=q(x_{1}) \to 0$ in the continuous RSB.
              For a convenience we define,
       \beq
       \Delta(x)\equiv 1-q(x).
       \label{eq-def-Delta-x}
       \eeq
       Then jamming implies $\Delta_{1}=1-q(x_{1}) \to 0$.
       We discuss below properties of
       the system encoded in the continuous RSB solution in the vicinity  of the core $x \to x_{1}$
       which encodes physical properties of the system in the deepest part of the energy landscape.

In the following we will find results very similar to those found in
the hardspheres in $d \to \infty$ \cite{charbonneau2014exact} where it was shown that continuous RSB solution gives a qualitatively
different result from finite $k$-RSB ansatz
concerning the scaling behavior approaching jamming.

       \subsubsection{Scaling ansatz at the core $x \to x_{1}$
         in the jamming limit $\Delta_{1} \to 1$.}
       \label{sec-scaling-argument}

       Following \cite{charbonneau2014exact} and \cite{franz2016simplest}
       we consider the following scaling ansatz
       at the core $x \to x_{1}$ in the jamming limit $\Delta_{1} \to 0$,
       \beq
       \Delta(x)/\Delta_{1} \simeq (x/x_{1})^{-\kappa}
       \label{eq-scaling-Delta-x}
       \eeq
       with an exponent $\kappa >0$. 

       From \eq{eq-pdeq-P} and \eq{eq-pdeq-pi} we have,
\beqn
\dot{P}(x,h)&=&\frac{\dot{\lambda}(x)}{2} \left[
  P''(x,h)-2x (P(x,h)\pi(x,h))'
  \right] \label{eq-pdeq-P-v2}\\
\dot{\pi}(x,h)&=&-\frac{\dot{\lambda}(x)}{2}
\left[ \pi''(x,h)+2x \pi(x,h)\pi'(x,h)\right]
\label{eq-pdeq-pi-v2}
\eeqn
Based on the asymptotic behavior of the function
$\pi(x,h)$ given in \eq{eq-f-pi-large-lambda} we expect,
\beq
P(x,h) \simeq \frac{1}{\sqrt{2\pi \lambda(x)}}e^{-(\delta -h)^{2}/2\lambda(x)}
\qquad h \to +\infty
\label{eq-scaling-P-largely-positive}
\eeq
and
\beq
\dot{P}(x,h)=\frac{\dot{\lambda}(x)}{2}
    \left[
  P''(x,h)+2\frac{x}{\tilde{\Lambda}(x)} (P'(x,h)h+P(x,h))
  \right] \qquad h \to -\infty
    \eeq
For  $x \to x_{1}$ and $\Delta_{1} \to 0$ we can assume,
\beq
\dot{\lambda}(x)\simeq -p\dot{\Delta}(x) \qquad
\tilde{\Lambda}(x)\simeq p G(x)
\qquad
G(x) \simeq \frac{\kappa}{\kappa-1}x\Delta (x)
       \label{eq-some-relations-lambda-and-Lambda-and-G}
       \eeq
       which follow from \eq{eq-def-lambda-x},
       \eq{eq-def-tilde-lambda-x},
       \eq{eq-def-Delta-x} and
       \eq{eq-scaling-Delta-x}.
       Then assuming $P(x,h)\simeq A(x) e^{B(x)h-C(x)h^{2}/2}$
       for $x \to x_{1}$ one finds,
       $A\propto \Delta^{-(1-1/\kappa)}$, $B\propto \Delta^{-(1-1/\kappa)}$ and $C \propto \Delta^{-2(1-1/\kappa)}$. This implies the following
       scaling form for $x \to x_{1}$,
       \beq
       P(x,h) \sim  \Delta^{-\frac{\kappa-1}{\kappa}}P_{0}(h\Delta^{-\frac{\kappa-1}{\kappa}})
       \qquad h \to -\infty
       \label{eq-scaling-P-largely-negative}
       \eeq
       with some scaling function $P_{0}(x)$.

       To sum up we can expect the following three regimes
       \cite{franz2016simplest} \cite{charbonneau2014exact}
       for $x \to x_{1}$:
       \begin{enumerate}
       \item[(0)] $h \to -\infty$:
       \eq{eq-scaling-P-largely-negative}
and \eq{eq-f-pi-large-lambda} read,
         \beqn
         P(x,h) \sim  \Delta^{-c/\kappa}P_{0}(h\Delta^{-c/\kappa}) \qquad
         \pi(x,h)&\sim  -\frac{h}{\tilde{\Lambda}(x)}
         \label{eq-scaling-P-pi-negative}
         \eeqn
         with
         \beq
         c=\kappa-1
         \label{eq-c-kappa}
         \eeq
       \item[(1)] $h \sim 0$ (intermediate regime)
           \beqn
         P(x,h) \sim \Delta^{-a/\kappa}P_{1}(h\Delta^{-b/\kappa}) \qquad
         \pi(x,h) \sim  \frac{\Delta^{b/\kappa}}{\tilde{\Lambda}(x)}\pi_{1}(h\Delta^{-b/\kappa})
         \label{eq-scaling-intermediate}
         \eeqn
       \item[(2)] $h \to  \infty$:
         \eq{eq-scaling-P-largely-positive}
         ($\lambda(x) \to 1$ for $x \to x_{1}$ and $q(x_{1}) \to 1$)
and \eq{eq-f-pi-large-lambda} implies
                    \beqn
         P(x,h) \sim  P_{2}(h)  \qquad
         \pi(x,h) \sim  0
                  \label{eq-scaling-P-pi-positive}
         \eeqn
         \end{enumerate}
       In the above equations $P_{0}(x)$,$P_{1}(x)$,$\pi_{1}(x)$ and $P_{2}(x)$ are
       some smooth functions and $a$, $b$, $c$, $\kappa$ are some exponents.
       In the following we assume that these exponents are all positive.

Now we can make the following observations:
       \begin{enumerate}
       \item{Matching between (0) and (1)}:
         assuming
         \beqn
         P_{0}(u) &\propto& u^{\theta} \qquad   u \to 0 \\
         P_{1}(u) &\propto& (-u)^{\theta} \qquad u \to -\infty
         \eeqn
         the following relation is needed,
         \beq
         \Delta^{-c/\kappa}(h\Delta^{-c/\kappa})^{\theta} \sim \Delta^{-a/\kappa}(h\Delta^{-b/\kappa})^{\theta}
         \eeq
         which implies
         \beq
         \theta=\frac{c-a}{b-c}.
         \label{eq-scaling-relation-theta}
         \eeq
         We also find
         \beq
         \pi_{1}(u) \sim -u \qquad u \to -\infty
         \eeq
         must hold.
       \item{Matching between (1) and (2)}:
         assuming
\beqn
P_{1}(z) &\propto & z^{-\alpha} \qquad z \to \infty \\
P_{2}(z) &\propto & z^{-\alpha} \qquad  z \to 0
\label{eq-scaling-P-matching}
\eeqn
we find the following relation is needed to eliminate the dependence on $\Delta$,
         \beq
         \alpha=\frac{a}{b}
                  \label{eq-scaling-relation-alpha}
         \eeq
       \item{Analysis on the intermediate regime $h \sim 0$}:
         Plugging \eq{eq-scaling-intermediate} in \eq{eq-pdeq-P-v2}
         and using \eq{eq-some-relations-lambda-and-Lambda-and-G} we find,
         the contribution from the 1st term on the r.h.s. scales are
         $(\Delta^{-b/\kappa})^{2}$ while those from the 2nd term on the r.h.s and
         the term on the l.h.s scales like $\Delta^{-1}$. Thus in order to have a non-trivial
         solution we need,
         \beq
         \frac{b}{\kappa}=\frac{1}{2}.
         \label{eq-b-kappa}
         \eeq
         by which we can eliminate $b$. Now we are left with two exponents
         $a$ and $c=\kappa-1$.
         Furthermore 
         plugging \eq{eq-scaling-intermediate} in \eq{eq-pdeq-P-v2} and
         \eq{eq-pdeq-pi-v2} we find the following two ordinary differential equations,
         \beqn
         \frac{a}{\kappa}P_{1}(z)+\frac{z}{2}P'_{1}(z)
         &=&\frac{p}{2}
P_{1}''(z)-\frac{c}{\kappa}(P_{1}(z)\pi_{1}(z))'
\\
\left(\frac{1}{2}-\frac{c}{\kappa} \right) \pi_{1}(z)-\frac{1}{2}z\pi'_{1}(z)
&=&\frac{p}{2}
\pi_{1}''(z)+\frac{c}{\kappa}\pi_{1}(z)\pi'_{1}(z)
\label{eq-OD-intermediate-regime}
\eeqn
which are subjected to the boundary condition
        \beq
                        P_{1}(z)= 
        \left \{
        \begin{array}{cc}
          (-z)^{\theta} &  z \to -\infty \\
          z^{-\alpha} & z \to \infty
        \end{array}
        \right .
        \qquad 
                \pi_{1}(z)= 
        \left \{
        \begin{array}{cc}
          -z & z \to -\infty  \\
          0 & z \to \infty
        \end{array}
        \right .
        \label{eq-boundary-OD-intermediate-regime}
        \eeq
        One can check that the differential equations
        given by \eq{eq-OD-intermediate-regime}
        with the boundary condition
        given by \eq{eq-boundary-OD-intermediate-regime}
        is consistent with the scaling relations
        for $\theta$ and $\alpha$ 
        given by \eq{eq-scaling-relation-theta} and         \eq{eq-scaling-relation-alpha}.

        Here we notice that the apparent dependence on $p$
in \eq{eq-OD-intermediate-regime} can be formally eliminated by
        the following replacement,
        \beqn
        \frac{z}{\sqrt{p}} \to z \qquad 
        \frac{P_{1}(z)}{\sqrt{p}} \to P_{1}(z)  \qquad
        \frac{\pi_{1}(z)}{\sqrt{p}} \to \pi_{1}(z)
        \label{eq-eliminate-p}
        \eeqn
        This means that if we find a solution for the $p=1$ case,
        the solutions for other values of $p$ can be obtained as well
        using  \eq{eq-eliminate-p} in the reversed manner. 
        Importantly such a solution satisfies the same desired
        asymptotic behaviors
        given by \eq{eq-boundary-OD-intermediate-regime}.
        This implies the universality does not change with $p$.

        However as pointed out in \cite{charbonneau2014exact} the above
        equations do not completely solve the problem.
        We are left with the exponent $a$ undetermined
        while other quantities 
        $P_{1}(z)$, $\pi_{1}(z)$ and the exponent $c$
        can be obtained in a form parametrized by $a$.
        (All other exponents are fixed given $a$ and $c$.)
        Then the final task to fix the value of the exponent $a$
        which can be done using the exact identity given by \eq{eq-exact-identity-fullRSB-2}.
        The latter reads in the limit $x \to x_{1}$ and $q(x_{1})=q(x_{1}) \to 1$
        \beq
        0=(p-1) \int dh T_{1}(h) + \int dh T_{2}(h)
        \label{eq-identity-T1-T2}
        \eeq
        where we defined,
        \beqn
T_{1}(h)& \equiv &[(p-2)G^{2}(x_{1})-2x_{1}G(x_{1})]P(x_{1},h)\pi^{2}(x_{1},h) \nonumber \\
&+&3pG^{2}(x_{1})P(x_{1},h)(\pi'(x_{1},h))^{2} \\
T_{2}(h)& \equiv& p(-2x_{1}G(x_{1}))P(x_{1},h)(\pi'(x_{1},h))^{2} \nonumber \\
&+& p^{2}G^{2}(x_{1})P(x_{1},h)\left [ (\pi''(x_{1},h))^{2}-2x_{1} (\pi'(x_{1},h))^{3}\right].
\eeqn
We notice that the contribution of $(p-1)\int dh T_{1}(h)$
 into   \eq{eq-identity-T1-T2}
 vanishes for the $p=1$ case accidentally but not for $p>1$.
Thus we must carefully examine whether
$\int dh T_{1}(h)$ remain relevant in the jamming limit $\Delta_{1}=\Delta(x_{1}) \to 0$ or not.

We examine contributions of the
integrals $\int dh T_{1}(h)$  and $\int dh T_{2}(h)$ 
from the regimes (1) $h \to -\infty$ and (2) $h \sim 0$. 
In the regime (3) $h \to \infty$
$\pi(x,h) \sim 0$ as \eq{eq-scaling-P-pi-positive}
so we do not need to consider the regime (3).
Using  \eq{eq-scaling-P-pi-negative}, \eq{eq-c-kappa},
and   \eq{eq-some-relations-lambda-and-Lambda-and-G} we find for the regime (0)
$h \to -\infty$,
\beqn
\int_{\rm regime (0)} dh T_{1}(h) & \sim &
-\frac{2}{p^{2}}x_{1}\frac{c}{\kappa}\Delta_{1}^{-(1-c)/\kappa}
  \int_{0}^{\infty}dt P_{0}(-t)t^{2}
\nonumber \\
\int_{\rm regime (0)} dh T_{2}(h) & \sim &  0
\eeqn
where we took leading terms for the jamming limit $\Delta_{1} \to 0$.
Similarly
using  \eq{eq-scaling-intermediate},\eq{eq-b-kappa}
and   \eq{eq-some-relations-lambda-and-Lambda-and-G} we find for the regime
(1) $h \sim 0$,
\beqn
&&  \int_{\rm regime (1)} dh T_{1}(h)  \sim  \Delta^{1/2-a/\kappa}
\int_{-\infty}^{\infty} dz
P_{1}(z) \left[\frac{3}{p}(\pi'_{1}(z))^{2}-\frac{2}{p^{2}}x_{1}\frac{c}{\kappa}\pi_{1}^{2}(z)\right]
\nonumber \\
&& \int_{\rm regime (1)} dh T_{2}(h)  \sim   \nonumber \\
&& \Delta^{-(1+a)/\kappa}
\int_{-\infty}^{\infty} dz P_{1}(z)
\left[ (\pi^{''}(z))^{2}-2\frac{c}{\kappa}\frac{1}{p}
  \left \{
  (\pi'_{1}(z))^{3}+(\pi'(z))^{2}
  \right \}
  \right] 
\eeqn

Collecting the above results we find the most relevant contribution
in the jamming limit $\Delta_{1} \to 0$ is given by
$\int_{\rm regime (1)} dh T_{2}(h)$ as long as the exponents $a$, $c$ are positive.
It means that we must satisfy,
\beq
\int_{-\infty}^{\infty} dz P_{1}(z)
\left[ (\pi^{''}(z))^{2}-2\frac{c}{\kappa}\frac{1}{p}
  \left \{
  (\pi'_{1}(z))^{3}+(\pi'(z))^{2}
  \right \}
  \right] =0
\label{eq-exact-identity-fullRSB-2-jamming-limit}
\eeq
Again we find the apparent $p$ dependence can be formally eliminated by
the replacement given by  \eq{eq-eliminate-p}. 

Based on the above analysis we can conclude that the
critical exponents  and the scaling functions $P_{1}(z)$, and $\pi_{1}(z)$
does not dependent on $p$, i.~.e. super-universal.
The exponents are $a=0.29213..$, $b=0.70787...$,$c = 0.41574...$,
$\alpha= 0.41269..$,$\theta= 0.42311...$ and $\kappa=1.41574...$ \cite{charbonneau2014exact}.

       \end{enumerate}

       \subsubsection{Divergence of the pressure}

       The pressure can be expressed as \eq{eq-pi-krsb} which reads,
\beq
\Pi=\int dh P(x,h)\pi(x,h) 
\eeq
where $x$ can be chosen arbitrary.
Using the scaling ansatz given by \eq{eq-scaling-P-pi-negative} 
and \eq{eq-some-relations-lambda-and-Lambda-and-G} at the core $x \to x_{1}$
and jamming $\Delta_{1} \to 0$
we find contribution from largely negative region of $h$ becomes
\beq
\int_{-\infty}^{0} dh \left ( -\frac{1}{p} \right)\frac{\kappa-1}{\kappa}
\frac{h}{\Delta_{1}}\Delta^{-c/\kappa}P_{0}(h\Delta_{1}^{-c/\kappa})
\sim c_{\rm nt} \Delta^{-1/\kappa}
\qquad c_{\rm nt}=\frac{1}{p}\frac{\kappa-1}{\kappa}\int_{0}^{\infty} dt P_{0}(-t)t.
\eeq
Similarly we can analyze contribution from the region $h \sim 0$ using
\eq{eq-scaling-intermediate}, 
and \eq{eq-some-relations-lambda-and-Lambda-and-G}
\beq
\int dh \frac{1}{p}\frac{h}{\Delta_{1}}\Delta_{1}^{-a/\kappa}P_{1}(h\Delta^{-b/\kappa})
\propto \Delta^{-a/\kappa}.
\eeq
If $a< 1$, which is the case ($a=0.29213...$), the latter gives a only sub-dominant contribution.
To sum up we find, the 'cage size' $\Delta_{1}$ vanishing in the jamming limit $\Pi \to \infty$
as,
\beq
\Delta_{1} \propto \Pi^{-\kappa}
\eeq

\subsubsection{Distribution of gap}
\label{sec-gr-scaling-fullRSB}

       For the hardcore model the distribution of the gap $g(r)$ within
       the $k$-RSB ansatz given by \eq{eq-g-of-r-kRSB} reads,
       \beq
       g(r)=\theta(r)\int \cD z
       \frac{P(m_{k},r-\sqrt{1-q_{k}^{p}}z)}{
         \Theta\left(\frac{r-\sqrt{1-q_{k}^{p}}z}{\sqrt{2(1-q_{k}^{p})}} \right)}
       \eeq

       \begin{enumerate}
       \item For {\it fixed} finite $r$, sending $q_{k}\to 1$, we find,
         \beq
         g(r)=\theta(r)P(m_{k},r)
         \eeq
       where we used $\lim_{X \to \infty}\Theta(X)=1$.
       This is a generalization of the RS ($k=0$) result given by \eq{eq-gr-scaling-RS-1}.

       In the $k \to \infty$ limit, the scaling behavior of $P(x,h)$ close to the core $x \to x_{1}$  as described by \eq{eq-scaling-P-pi-negative} and \eq{eq-scaling-intermediate}
       in the region vanishing in the jamming limit  $\Delta_{1} \to 0$
       implies development of a delta peak $\delta(r)$.
       On the other hand, we have the scaling behavior $P(x,h) \sim h^{-\alpha}$
       for fixed $h \sim  0^{+}$ as given by
       \eq{eq-scaling-P-matching} with $\alpha=a/b$ given by \eq{eq-scaling-relation-alpha}.
       These observations implies,
       \beq
       g(r) \sim \delta(r)+c_{\rm nt}\theta(r)r^{-\alpha},
       \eeq
       where $c_{\rm nt}$ is some numerical factor.

     \item In the vanishing region around $r=0$ parametrized as $r=(1-q_{k}^{p})\lambda$
       we find,
         Assuming $q_{k} \sim 1$ we find for $r>0$,
  \beqn
  g(r)
&  \sim &\frac{1}{1-q_{k}^{p}}\int_{0}^{\infty} dy
  P(m_{k},-y)ye^{-\lambda y} \qquad \lambda=\frac{r}{1-q_{k}^{p}}
  \eeqn
  This is a generalization of the RS ($k=0$) result given by \eq{eq-gr-scaling-RS-1}.

  Now in the $k \to \infty$ limit we have the scaling behavior
  $P(x,h) \sim \Delta^{-c/\kappa}P_{0}(h\Delta^{-c/\kappa})$
  for $h < 0$ in \eq{eq-scaling-P-pi-negative}. Using this for $x \to x_{1}$ we find,
  \beq
  g(r)\sim \frac{1}{p}\frac{1}{\Delta_{1}^{1/\kappa}}\int_{0}^{\infty} dt P_{0}(-t)te^{-\frac{t}{p}\frac{r}{\Delta_{1}^{1/\kappa}}}
  \eeq
  where we used $c=\kappa-1$ and $1-q_{k}^{p} \simeq p\Delta_{1}$ for $\Delta_{1} \to 0$.

Using the above result we can evaluate the fraction of interactions or contacts which is closed. For any small but finite $\epsilon$ we have,
  \beq
  \int_{0}^{\epsilon} dr g(r)=\int_{0}^{\infty} dtP_{0}(-t)
  \int_{0}^{\epsilon t/ (p \Delta_{1}^{1/\kappa})}ds e^{-s}
  \underset{\Delta_{1} \to 0}{\rightarrow} \int_{0}^{\infty} dtP_{0}(-t)
  \eeq
  Thus in the jamming limit, the fraction of closed contact given by \eq{eq-f-closed} can
  be expressed as,
  \beq
  f_{\rm closed}=
\lim_{\epsilon \to 0}  \int_{0}^{\epsilon} dr g(r)
=  \int_{0}^{\infty} dtP_{0}(-t)
\label{eq-fclosed-jamming}
\eeq
Note that here the lower limit of the integration is set to $0$ because of the hardcore constraint.

              \end{enumerate}
       \subsubsection{Isostaticity}

       Let us consider whether isostaticity discussed in sec.
       \ref{sec-pressure-gr-isostaticity}     holds in the jamming limit
       in the present model.
       The condition of isostaticity given by \eq{eq-isostaticity}
       becomes in the $M \to \infty$ limit
         with $\alpha=c/M$ fixed at jamming $\Delta_{1} \to 0$ becomes, 
         \beq
         1=\frac{\alpha}{p}\int_{0}^{\infty} dt P_{0}(-t)
         \label{eq-isostaticity-large-M}
         \eeq
         where we used \eq{eq-fclosed-jamming}.

         Actually using the exact identity
         given by \eq{eq-exact-identity-fullRSB-1} which holds for the continuous RSB
         together with the scaling behavior given by \eq{eq-scaling-P-pi-negative}
         in the $h<0$ region and the relation
         $\tilde{\Lambda}(x)\simeq pG(x)$ given by the 2nd equation of
         \eq{eq-some-relations-lambda-and-Lambda-and-G} which hold close to the core $x \to x_{1}$ at jamming $\Delta_{1} \to 0$ we find,
         \beq
         1=\frac{\alpha(p-1)}{p^{2}}\int_{-\infty}^{0}dh \Delta_{1}^{-c/\kappa}
         P_{0}(h\Delta_{1}^{-c/\kappa})h^{2}
         +\frac{\alpha}{p}\int_{-\infty}^{0}dh \Delta_{1}^{-c/\kappa}
         P_{0}(h\Delta_{1}^{-c/\kappa})
         \underset{\Delta_{1}\to 0}{\rightarrow}
         \frac{\alpha}{p}\int_{0}^{\infty}dt P_{0}(-t)
         \eeq
         Thus we see that the isostaticity holds at jamming.
         Note that the term which is proportional to $p-1$ apparently
         violates the isostaticity but it
         scales as  $\Delta_{1}^{2c/\kappa}$
         and becomes irrelevant in the jamming limit $\Delta_{1} \to 0$
         as long as $c/\kappa > 0$.

         \subsection{Detailed analysis on the $p=2$ case}

         Let us take here the $p=2$ case and study the model more in detail
         to work out the phase diagram and behavior of physical quantities.
         The following  analysis is valid for the disorder-free model in the range $\delta > 0$
         because of the flatness of the potential  as we noted at the beginning of sec \ref{sec-hardcore}.
         We have found that the system exhibit continuous transition to anti-ferromagnetic phase
         for $\delta < 0$ at $T_{\rm c}$  given by \eq{eq-antiferro-softcore}. 
         We have to suppress the anti-ferromagnetic phase using
         the disordered model in order to realize the glassy phases for $\delta < 0$.
         
         \subsubsection{RS solution}
         \label{sec-RS-p=2-hardcore-glass}

         For the $p=2$ case, we find from  \eq{eq-critical-point-rs-hardcore}
         that the paramagnetic solution  $q=0$ becomes unstable at
         the critical point $\alpha_{\rm c}(\delta)$.
         Then we are naturally led to examine the possibility of the $q \neq 0$ solution.
        Within the RS ansatz,  it must solve
       ${\cal G}(q)=0$ (see \eq{eq-RS-saddlepoint}),
       where the function ${\cal G}(q)$ for the hardcore model is given by
           \eq{eq-function-Gq-hardcore}.
Expanding ${\cal G}(q)$ up to order $O(q^{2})$ we find,
\beq
0={\cal G}(q)=1-\frac{\alpha}{\alpha_{\rm c}(\delta)}\left[1-2q+(2-2x_{0}-r_{0})q^{2}\right]
+ O(q^{4})
\label{eq-RS-saddlepoint-nonzero-q-p=2-hardcore}
\eeq
where
\beq
x_{0} \equiv \frac{\delta}{\sqrt{2}} \qquad r_{0} \equiv r(x_{0})
=\frac{\Theta'(x_{0})}{\Theta(x_{0})}
\eeq
The above equation can be solved for $q$ to find,
\beq
q=\frac{1}{2}\epsilon
-\frac{1}{4}\left(1+x_{0}+\frac{r_{0}}{2}\right)\epsilon^{2}+O(\epsilon^{3})
\label{eq-rs-solution-nonzero-q}
\qquad
\epsilon \equiv  \frac{\alpha}{\alpha_{\rm c}(\delta)}-1
\eeq
Thus we find that $q\neq 0$ solution emerges at the critical point
$\alpha=\alpha_{\rm c}(\delta)$, where the $q=0$ solution becomes unstable, and
the EA order parameter $q$ grows
continuously increasing $\alpha$. 

\begin{figure}[h]
  \bc
  \includegraphics[width=0.5\textwidth]{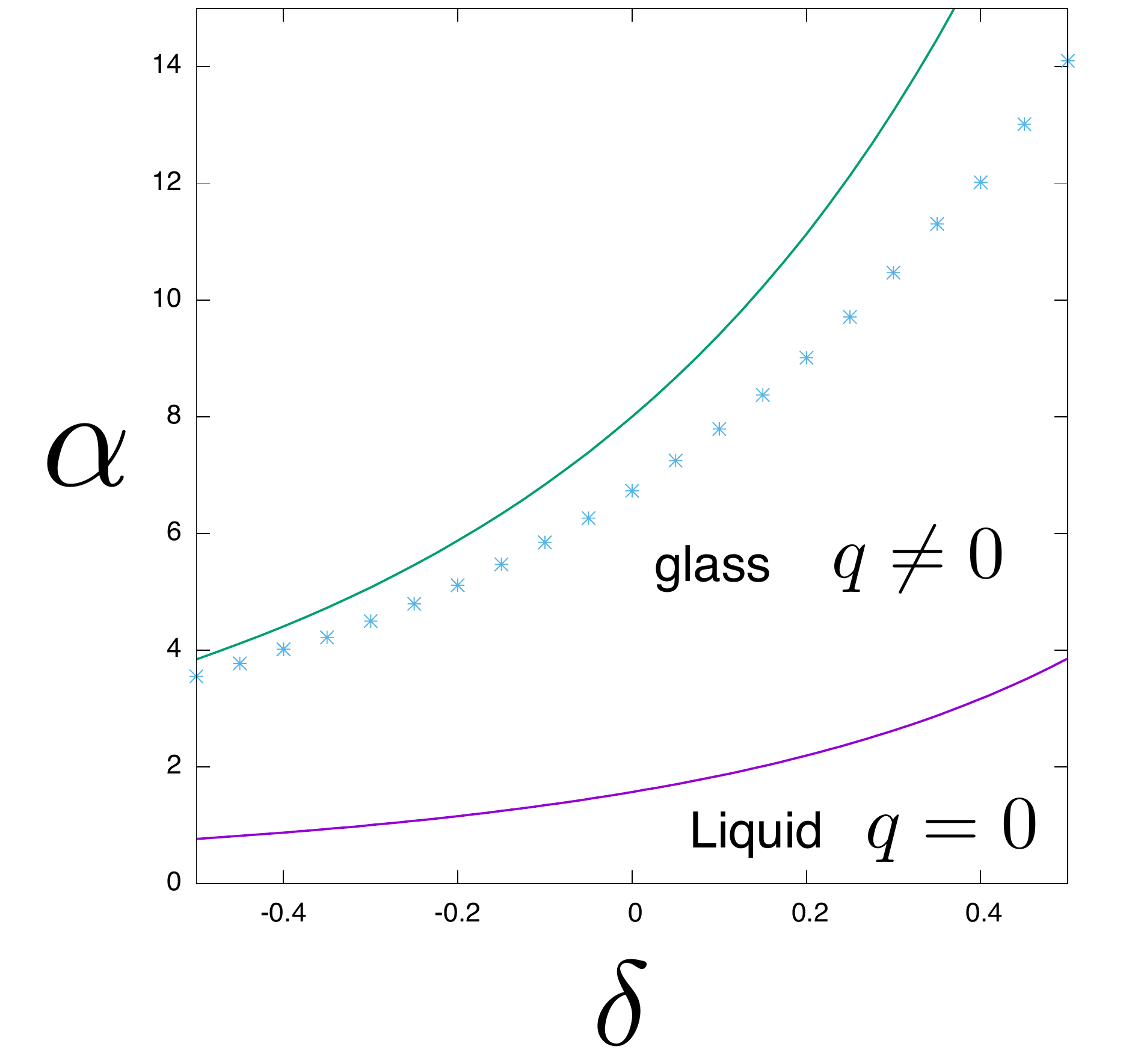}
  \ec
  \caption{The phase diagram  of the $p=2$ body hardcore model within the replica symmetric (RS) ansatz:  $q > 0$ RS solution emerges continuously at the lower curve which represents
    $\alpha=\alpha_{\rm c}(\delta)$ given by \eq{eq-critical-point-rs-hardcore}.
    The value of the order parameter saturates $q \to 1$ approaching the upper solid line
    $\alpha=\alpha_{\rm j}(\delta)$ given by \eq{eq-jamming-point-rs-hardcore}, which is the jamming line within the RS ansatz.
    The lower curve coincides with the AT line above which the RS solution becomes unstable.
    The dotted line is the jamming line obtained by the continuous RSB solution which is discussed later.
  }
\label{fig-AT-line}
\end{figure}

In Fig.~\ref{fig-AT-line} we show the phase diagram for the $p=2$ hardcore model
within the RS ansatz. The glass transition line $\alpha=\alpha_{\rm c}(\delta)$ is given by
\eq{eq-critical-point-rs-hardcore}. The jamming line $\alpha=\alpha_{\rm j}(\delta)$
is given by \eq{eq-jamming-point-rs-hardcore}.

    \begin{figure}[h]
        \includegraphics[width=\textwidth]{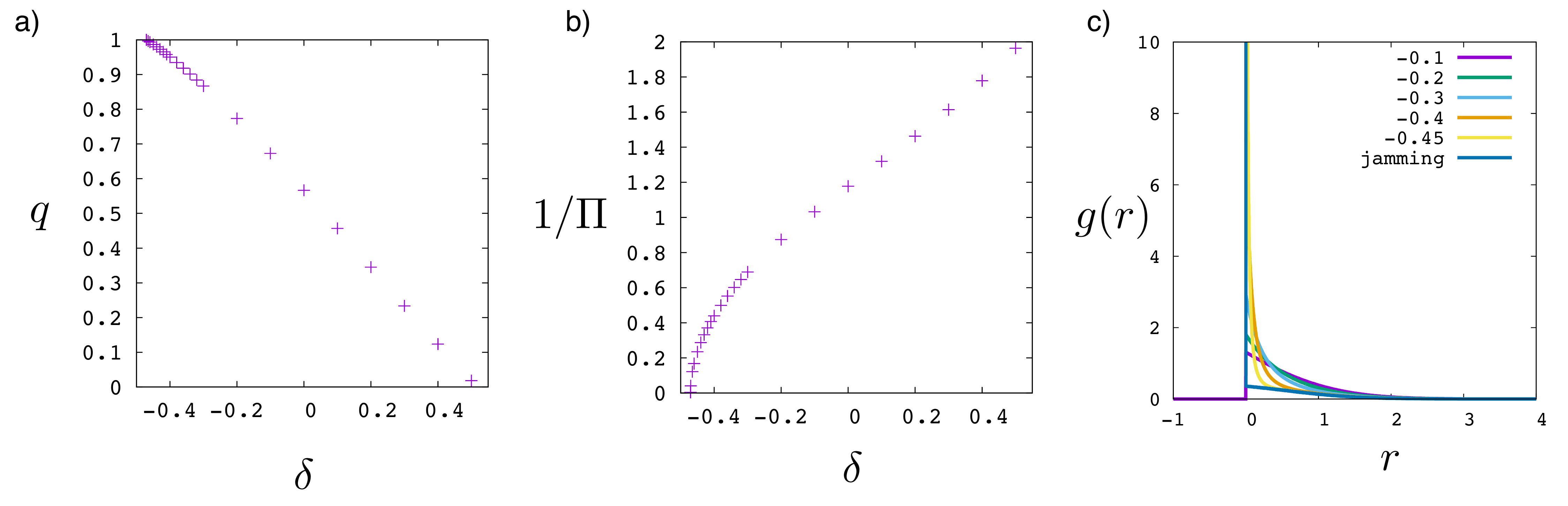}
        \caption{Glass phase of the hardcore model within the RS ansatz (here we choose $\alpha=4$): a) behavior of the order parameter $q$, b) inverse of the pressure $\Pi$, c) the distribution of the gap $g(r)$.
          The line labeled as {\it jamming} is for
          $\delta_{\rm j} \sim -0.47065$.}
\label{fig-q-pressure-RS}
    \end{figure}

    In Fig.~\ref{fig-q-pressure-RS}, we display
an example of a set of solutions of the RS saddle point equation given by \eq{eq-RS-saddlepoint}
with \eq{eq-function-Gq-hardcore} for a $\alpha=4$ with varying $\delta$.
As shown in the panel  a), the glass order parameter $q$ emerges continuously
at the critical point $\delta_{\rm c} \sim 0.51$ (determined by $\alpha_{\rm c}(\delta_{c})=4$,
see Fig.~\ref{fig-AT-line}) and increases by decreasing $\delta$ and saturates to $q=1$
approaching the jamming point $\delta_{\rm j} \sim -0.47065$
(determined by $\alpha_{\rm j}(\delta)=4$,
see Fig.~\ref{fig-AT-line}) There we also show the behavior of the
pressure given by \eq{eq-RS-hardcore-pressure} which diverges approaching the jamming
and evolution of $g(r)$ given by \eq{eq-RS-hardcore-g-of-r}
which develops a diverging contact peak at $r=0$ approaching
the jamming.

Finally let us examine the stability of this solution. From the result
reported in appendix  \ref{subsec-replicon-RS} we find
the replicon eigenvalue $\lambda_{R}$ of the RS solution as,
\beqn
\lambda_{\rm R}&=& \frac{2}{(1-q)^{2}} {\cal R}(q)\\
       {\cal R}(q)&=&1-\frac{\alpha}{2(1+q)^{2}}
       \int \cD z_{0}
         \left (
         r^{2}(x)(1-q^{2})
       +
       4q^{2}
         x^{2}r^{2}(x)+xr^{3}(x)+\frac{r^{4}(x)}{4}
         \right)_{x=\frac{\delta-\sqrt{q^{p}}z_{0}}{\sqrt{2(1-q^{p})}}}
                \qquad 
         \label{eq-replicon-rs-nonzero-q}
       \\
       &=& 1-\frac{\alpha}{\alpha_{\rm c}(\delta)}\left [1-2q
         +\frac{1}{2} (5r^{2}_{0}+16r_{0}x_{0}+12x^{2}_{0}+2)q^{2}+O(q^{3})\right]
       \label{eq-replicon-rs-nonzero-q-expansion}
       \eeqn
       In the last equation we made an expansion in series of $q$ which can be obtained by using $r'(x)=-2xr(x)-r^{2}(x)$ which follows from \eq{eq-def-r}.
       Comparing  the function ${\cal G}(q)$
       in \eq{eq-RS-saddlepoint-nonzero-q-p=2-hardcore}
       and ${\cal R}(q)$
       in        \eq{eq-replicon-rs-nonzero-q-expansion}
we notice that they are identical up to $O(q)$ but different 
in the $O(q^{2})$ terms. Using \eq{eq-rs-solution-nonzero-q} in \eq{eq-replicon-rs-nonzero-q-expansion}
we find up to $O(\epsilon^{2})$,
\beq
\lambda_{\rm R}= \frac{2}{(1-q)^{2}}A(x_{0})\epsilon^{2}\qquad
A(x)=\frac{1}{4}-\frac{x}{2}-\frac{r(x)}{4}-\frac{5}{8}r^{2}(x)-2xr(x)-\frac{3}{2}x^{2}
\eeq
It turns out that $A(x)$ is a monotonically decreasing function of $x$.
Using the asymptotic behavior of the function $r(x)$ given in \eq{eq-def-r-x}
one can find $\lim_{x\to -\infty}A(x)=-1/4$. Thus we find that the replicon eigenvalue is definitely negative meaning that the RS solution is unstable for $\alpha > \alpha_{\rm c}(\delta)$.
We also checked numerically, solving ${\cal G}(q)=0$  for $q$ and evaluating
$\lambda_{\rm R}$ (\eq{eq-replicon-rs-nonzero-q}) that this is indeed the case in the whole regime of $\alpha > \alpha_{\rm c}(\delta)$.
Thus the replica symmetry must be broken for $\alpha > \alpha_{\rm c}(\delta)$.

Remarkably the situation is very similar to the Sherrington-Kirkpatrick (SK) model for the spin glasses \cite{kirkpatrick1978infinite,de1978stability,MPV87}.  To summarize we find 
the liquid solution described by the $q=0$ RS solution which 
becomes unstable approaching the critical point
$\alpha_{\rm c}(\delta)$ where all eigenvalues of the Hessian matrix vanish.
It immediately means divergence of the so called
spin-glass susceptibility and 
negative divergence of non-linear compressibility
$d^{2}p/d\delta^{2}$
much as the spinglass transition of the SK model \cite{binder1986spin,MPV87}.
The line $\alpha=\alpha_{\rm c}(\delta)$ is the equivalent of the
d'Almeida-Thouless (AT) line \cite{de1978stability}.
Beyond the transition point, going into the glass phase,  we have to consider breaking of the replica symmetry.\cite{parisi1980sequence,parisi1979infinite,parisi1980order}.

    \subsubsection{RSB solution}

\begin{figure}[h]
  \includegraphics[width=\textwidth]{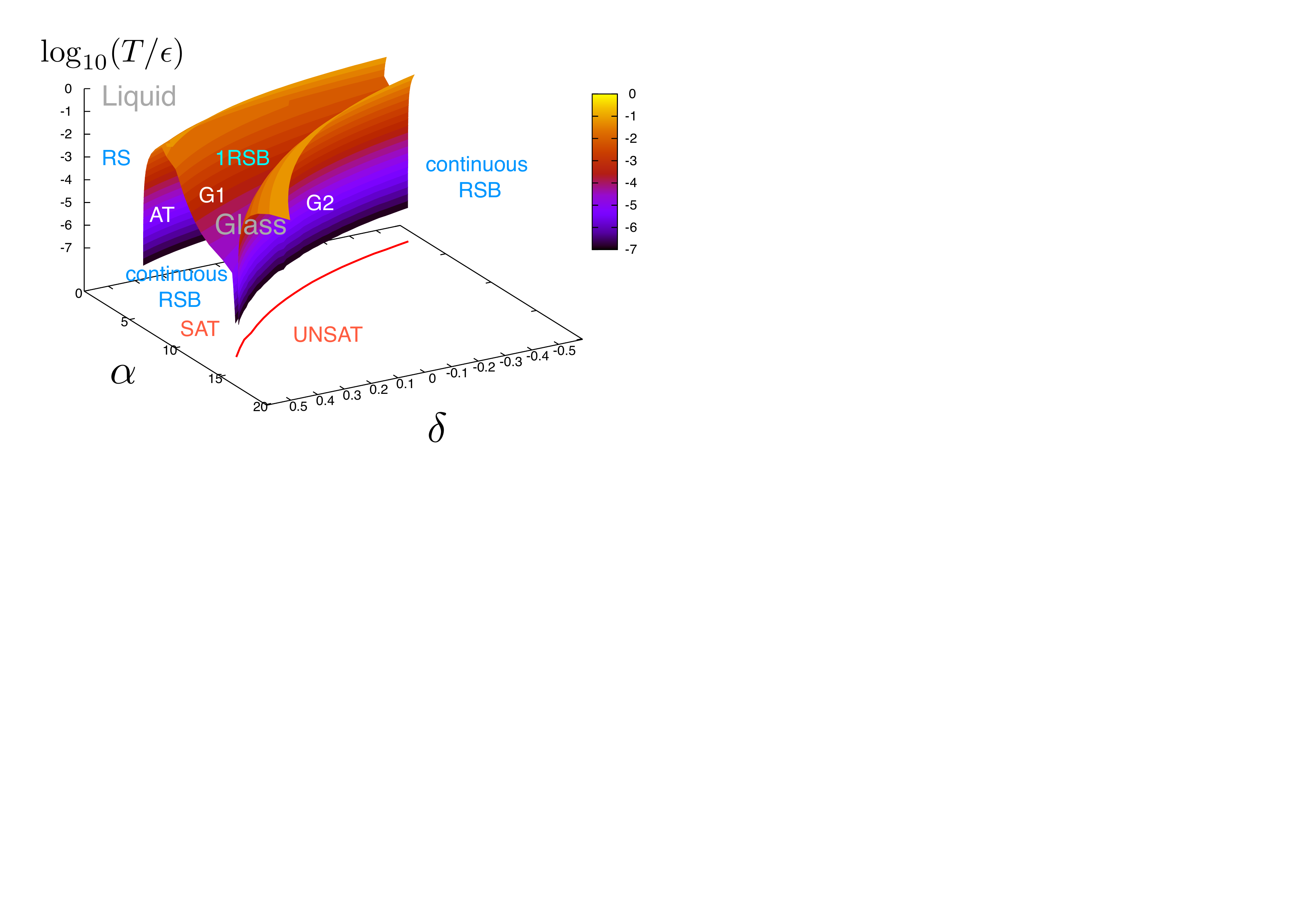}
  \caption{Phase diagram of the soft/hardcore model ($p=2$).
    On the plane AT separating the liquid (RS) and glass (RSB)
    the d'Almeida-Thouless (AT) instability occurs. The 1RSB solution becomes unstable below the two planes G1 and G2 on which the Gardner transition occurs.
    The G1 plane separates from the AT plane at finite temperatures.
    The red line on the bottom 
    represents the jamming line $\alpha=\alpha_{\rm j}(\delta)$ at $T=0$.
    }
  \label{fig-ATplane-Gplane-Jline}
\end{figure}

    Finally let us study the glass phase of the $p=2$ hardcore model
    using the RSB ansatz. Here we use the softcore potential
    given by \eq{eq-soft-hardcore-potential} and extend the analysis to finite temperatures.
    Using \eq{eq-critical-point-rs} and evaluating the integral numerically
    using the softcore potential we can easily find the plane
    $\alpha=\alpha_{\rm c}(\delta, T)$ where the AT instability $\lambda_{R}=0$ occurs.
    The result is shown in Fig.~\ref{fig-ATplane-Gplane-Jline}
    where the AT plane is indicated as 'AT'.
    The zero temperature limit of it agrees with the AT line of the hardcore model shown in Fig.~\ref{fig-AT-line}
    as it should be. The AT plane separates the liquid phase (paramagnet) with $q=0$ on its left hand side and glass
    the glass phase on the right hand side.

    Next let us examined the 1RSB ansatz on the glass side of the AT plane.
    We solved the 1RSB equations numerically following the scheme explained
    in sec. \ref{sec-1RSB-equations} to obtain $q(x_{1})$ assuming $q_{0}=0$.
    Note that $q_{0}=0$ always solves the saddle point equation
    for $p > 1$ as we mentioned in sec. \ref{subsec:variational-equations}.
    We found $q(x_{1})$ emerges continuously starting from the AT plane, as expected.
    Then we evaluated the complexity $\Sigma^{*}(m_{1})$ numerically (see sec. \ref{subsubsec-1RSB})
    and determined $m_{1}$ where the complexity vanishes. We examined the stability of the 1RSB solution
    by evaluating the replicon eigenvalue $\lambda_{R}$ of the
    1RSB solution given by \eq{eq-replicon-1RSB}.
    As shown in  Fig.~\ref{fig-ATplane-Gplane-Jline}.
    we have two planes indicated as 'G1' and 'G2' where the replicon eigenvalue vanishes suggesting
    the Gardner's transition \cite{Ga85}. We found 'G1' plane merges with the AT plane at higher temperatures
    as can be seen in the figure.
    The 1RSB solution is stable above these Gardner planes but become unstable below them.

\begin{figure}[h]
    \includegraphics[width=\textwidth]{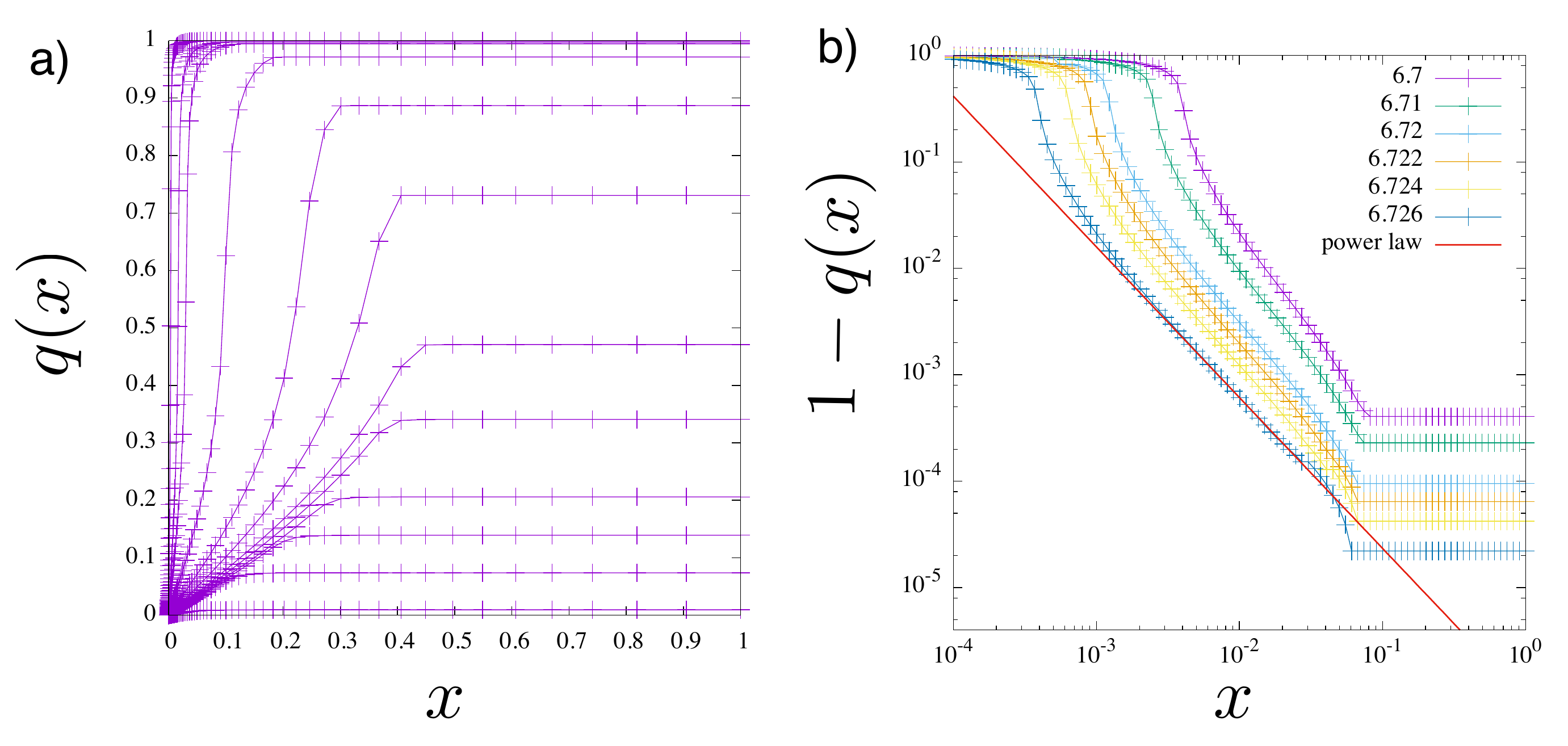}
  \caption{The $q(x)$ function of the hardcore model
    with $p=2$, $\delta=0$ for which $\alpha_{\rm c}=1.5708..$ and $\alpha_{\rm j}=6.732..$.
    a) $\alpha=1.6,1.8,2.0,2.2,2.4,2.6,2.8,3.0,4.0,5.0,6.0,6.5,6.6,6.7$ from the bottom to the top.
    b) The straight line represents
    the power law fit $a x^{-\kappa}$ with $\kappa=1.4157$,
    the same exponent as that for the hardspheres\cite{charbonneau2014exact}.
    }
  \label{fig-qx-hardcore}
\end{figure}

    Below the Gardner planes and on the right on the glass side of the AT plane we naturally expect continuous RSB. 
    Indeed we obtain the continuous RSB solution (approximated by $k=200$ RSB) as shown in Fig.~\ref{fig-qx-hardcore}
    where we show some examples of the $q(x)$ functions obtained at $T=0$ (hardcore limit).
    We obtained the result using the scheme explained in sec. \ref{sec-numerical-procedure-kRSB} together with the
    inputs for the hardcore case shown in sec. \ref{sec-inputs-hardcore}.
    As can be seen in Fig.~\ref{fig-qx-hardcore} a), we found the continuous RSB solution with nonzero $q(x)$
    function emerges continuously starting from the AT line $\alpha=\alpha_{\rm c}(\delta)$
    as expected.
    
    Using the scheme explained in sec.~\ref{sec-Algorithm-to-look-for-the-jamming point} we obtained the jamming line
    $\alpha=\alpha_{\rm j}(\delta)$ of the hardcore model and the result is displayed in Fig.~\ref{fig-ATplane-Gplane-Jline}
    at the bottom.
    It is also shown in Fig.~\ref{fig-AT-line} where we can see that the RS ansatz overestimates the jamming line.
    Quite interestingly we find in Fig.~\ref{fig-ATplane-Gplane-Jline} that the two Gardner planes 'G1' and 'G2'
    merges onto the the jamming line in the zero temperature limit. The geometry of the phase diagram is very different from that
    of the hardspheres \cite{biroli2016breakdown} where there is only one Gardner line.
    We analyzed the criticality of jamming $q(x_{1})\to 1$ of the hardcore model ($\epsilon \to \infty$), i.~e.
    $\alpha \to \alpha^{-}_{\rm J}(\delta)$ at $T=0$.
    As shown in Fig.~\ref{fig-qx-hardcore} b), we find power law behavior
    $1-q(x) \propto x^{-\kappa}$ with the expected exponent $\kappa=1.4157..$.
  This confirms the scaling argument presented in
    sec. \ref{sec-scaling-argument} establishing that the jamming of the present model belong to the same universality
    class as that of the hardspheres \cite{charbonneau2014exact}.

The liquid phase $q=0$ at $T=0$ can be regarded as an easy SAT region where the space of the solutions to satisfy the hard constraints ($\epsilon \to \infty$), i. e. the manifold of the ground states are continuously connected.
The glass phase at $T=0$ with $0 < q(x_{1}) < 1$  can be regarded as hard SAT phase where the manifold of the ground states splits
into clusters. The major difference with respect to the case of usual discrete coloring  \cite{zdeborova2007phase}
is that the transition is continuous and the clustering is hierarchical reflecting the continuous RSB.
The region $\alpha > \alpha_{\rm j}(\delta)$ is the UNSAT region where the hard constraint cannot be satisfied.

\section{Conclusions}
\label{sec-Conclusions}

In the present paper we developed a family of exactly solvable large $M$-component vectorial Ising/continuous spin systems with $p$-body interactions which exhibit glass transitions by the self-generated randomness. We also established a connection between the disorder-free model and a completely disordered spin glasses model by constructing a model which interpolates the two limits. We showed that the supercooled paramagnetic states and glassy states are locally stable against crystallization under certain conditions, namely either 1) $p>2$ or 2) the interaction potential $V(x)$ has a flatness. In those cases the quenched disorder is unnecessary to enable the glassy phases. Otherwise the quenched disorder is needed to enable glass transitions suppressing the crystalline phases. We developed a replica formalism to solve the problems exactly in the $M \to \infty$ limit.

We applied the scheme to explicitly analyze the continuous spin models with the two types of non-linear potentials: the quadratic and soft/hardcore potential. In both cases we found continuous RSB so that the free-energy landscape is marginal as indicated by vanishing of the replicon eigenvalue $\lambda_{\rm R}=0$. Interestingly this happens even with the $p=2$ continuous spin model in  contrast to the case of the linear potential. However there is an important difference between the two models that the criticality approaching jamming exists in the hardcore model but not in the quadratic model. This is evident in Fig.~\ref{fig-qx-quadratic} where one can see that the $q(x)$ function of the quadratic model does not develop any power law behaviors approaching jammming $q(x_{1}) \to 1$ ($T \to 0$) in sharp contrast to that of the hardcore model shown in Fig.~\ref{fig-qx-hardcore}. Critical jamming implies {\it mechanical} marginality which is reflected, for instance, as avalanche like responses to perturbations \cite{wyart2005rigidity,wyart2012marginal,le2010avalanches,muller2015marginal,franz2017mean}. Possible relation between the landscape marginality and mechanical marginality is an interesting open question \cite{wyart2005rigidity,wyart2012marginal,muller2015marginal,franz2015universal}. In the hardcore model, using the continuous RSB solution we found that the isostaticity holds at jamming and the universality of it turned out to be the same as hardspheres for all values of $p$ establishing the superniversality. This observation extend the result on the perceptron \cite{franz2016simplest,franz2017universality} which corresponds to $p=1$.

Although we limited ourselves to the case $M \to \infty$ in the present paper, systematic $1/M$ expansions are possible. Such an approach has been conducted in the case of $p=2$ continuous spin model with the linear potential, where  RSB does not take place \cite{aspelmeier2004generalized,moore20121}. This would be an alternative, analytically tractable approach to analyze systems on tree-like lattices of finite connectivity (Bethe lattice) where mean-field approach should remain valid. So far such systems remained hard to be analyzed by the the replica approach \cite{mottishaw1987stability} so that the cavity approach is usually preferred which however is limited to 1RSB at the moment \cite{mezard2001bethe}. An advantage of the $1/M$ expansion approach is that one can analyze the system almost as easily as the globally coupled systems so that one can construct continuous RSB explicitly as we have done in the $M \to \infty$ limit, which may become necessary deep in the glassy phases \cite{Ga85}. It will be interesting to study, for example, how the nature of jamming change if $M$ becomes finite. To what extent such mean-field results would remain useful for finite dimensional systems where the lattices are no more like trees, remains of course as an outstanding open problem.

We expect our results provide a useful basis to formulate theoretical approaches to study glass transitions of rotational degree of freedoms. For example it is natural to study glass transitions of particulate systems with rotational degrees of freedom such as patchy colloids and ellipsoids (Fig.~\ref{fig_rotational_glass} c)) extending the present work. Another interesting problem is to study the apparently disorder-free spinglass transitions realized in some frustrated magnets \cite{schiffer1995frustration,gingras1997static} (Fig.~\ref{fig_rotational_glass} b)). Continuous constrained satisfaction problems such as the continuous coloring problem (Fig.~\ref{fig_rotational_glass} a)) and related statistical inference problems can also be studied in similar frameworks.

\section*{Acknowledgements}
The author thanks Silvio Franz, Atsushi Ikeda, Florent Krzakala, Yuliang Jin, Kota Mitsumoto, Pierrefrancesco Urbani, Shuta Yokoi and Francesco Zamponi for useful discussions.



\paragraph{Funding information}
This work was supported by KAKENHI (No. 25103005  ``Fluctuation \& Structure'' and No. 50335337)
from MEXT, Japan.

\begin{appendix}

          \section{Density functional approach}
\label{appendix_density_functional}

In this appendix we discuss an alternative derivation of the free-energy functional given by  \eq{eq-F-functional-glass}
using a density functional approach closely following the study on the hardspheres \cite{kurchan2012exact,kurchan2013exact,charbonneau2014exact}.

\subsection{Spin liquid}

Let us introduce 'spin' density defined as,
\beq
N\rho(\vS)=\sum_{i=1}^{N}
\delta(\vS-\vS_{i})
\eeq
where $\delta(\vS)$ is the delta function in the spin-space
which satisfies $\intS d\vS \delta(\vS)=1$.
Let us also introduce the Mayer function,
\beq
f(\vS_{1},\vS_{2},\ldots,\vS_{p})=
e^{-\beta V(\vS_{1},\vS_{2},\ldots,\vS_{p})}-1
=-1+
\exp\left [-\beta V \left(
\delta -\frac{1}{\sqrt{M}}
\sum_{\mu=1}^{M}\prod_{l=1}^{p}(S_{l})^{\mu}
\right)
\right].
\eeq

A convenient strategy is to write the free-energy as,
\beq
e^{-\beta F}=\int {\cal D}[\rho(\Vec{S})]e^{-\beta {\cal F}[\rho(\vS)]}
\eeq
where we introduced a functional  ${\cal F}[\rho]$,
\beq
e^{-\beta {\cal F}[\rho(\vS)]} \equiv 
\intS \prod_{i=1}^{N} d{\bf S}_{i}
\prod_{<i_{1},i_{2},\ldots,i_{p}>}e^{-\beta V(S_{i_{1}},S_{i_{2}},\ldots,S_{i_{p}})}
\delta\left[\rho(\Vec{S})-N^{-1}\sum_{i=1}^{N}\delta(\vS-\vS_{i})\right]
\eeq
with $\int {\cal D}[\rho(\vS)]$ being a functional integration
over $\rho(\vS)>0$ and $\delta [ \ldots]$ is a functional delta function.

To obtain the functional ${\cal F}[\rho]$
one can follow the standard step of the liquid theory
\cite{hansen1990theory}: one defines first a free-energy
$F[\phi(\vS)]$ of the system with modified Hamiltonian
$H=\sum_{<i_{1},i_{2},\ldots,i_{p}>}V(\vS_{i_{1}},\vS_{i_{1}},\ldots,\vS_{i_{p}})+\intS d\vS\rho(\vS)\phi(\vS)$ then perform a Legendre transformation
to obtain a free-energy as functional of the spin density ${\cal F}[\rho(\vS)]=F[\phi(\vS)]-\intS d\vS\rho(\vS)\phi(\vS)$.
As the result one finds,
\beq
-\beta \frac{{\cal F}[\rho(\vS)]}{N}=
-\intS d\vS \rho(\vS)\ln \rho (\vS)
+\frac{c}{p}\intS d\vS_{1}d\vS_{2}\cdots d\vS_{p}
\rho(\vS_{1})\rho(\vS_{2})\cdots \rho(\vS_{p})
f(\vS_{1},\vS_{2},\ldots,\vS_{p}).
\label{eq-free-energy-functional-1replica}
\eeq
The free-energy $F$ is obtained by minimizing
the variational free-energy functional ${\cal F}[\rho(\vS)]$.
The 1st term on the r.h.s of \eq{eq-free-energy-functional-1replica}
represents the entropic (paramagnetic) part of the free-energy.
The 2nd term is the 1st virial correction due to interactions.
The reason for the absence of the higher order terms,
all of which are represented as 1 particle irreducible (1PI) diagrams
such as a triangle, a square, e.t.c. \cite{hansen1990theory},
is the tree-like geometry of the lattices that we consider.

\subsection{Replicated spin liquid}

In principle all stable and metastable states of the system, including
liquid (paramagnetic) state $\rho_{\rm liq}(\Vec{S})$,
crystalline state $\rho_{\rm crystal}(\Vec{S})$
and glassy states $\rho_{\alpha}(\Vec{S})$ ($\alpha=1,2,\ldots,$),
would be found as local minima of the free-energy functional given by \eq{eq-free-energy-functional-1replica}. In the present paper we focus on the properties of glassy states which emerge from supercooled paramagnetic state.

A useful way to analyze the properties of glassy states is the replica approach.
We consider replicated spin liquid of $n$ replicas
labeled as $a=1,2,\ldots,n$ obeying the Hamiltonian,
\beq
H_{n}=\sum_{a=1}^{n}\sum_{<i_{1},i_{2},\ldots,i_{p}>}
V(\vS^{a}_{i_{1}},\vS^{a}_{i_{1}},\ldots,\vS^{a}_{i_{p}})
\eeq
For convenience we introduce a short hand notation
\beq
\ovS=(\vS^{1},\vS^{2},\ldots,\vS^{n})
\label{eq-def-ovS}
\eeq
where $\vS$s themselves are $M$ component spin vectors.
Introducing replicated spin density
\beq
N\rho(\Vec{\overline{S}})=\sum_{i=1}^{N}
\prod_{a=1}^{n}
\delta(\vS^{a}-\vS_{i}^{a})
\eeq
which is normalized such that $\int d\ovS \rho(\Vec{\overline{S}})=1$, we find,
\beq
-\beta F= \left. \partial_{n} Z_{n}\right|_{n=0}
\label{eq-free-energy}
\qquad 
Z_{n}=\int {\cal D}[\rho(\ovS)]
e^{-\beta {\cal F}_{n}[\rho(\ovS)]}
\eeq
with the variational replicated free-energy functional defined as,
\beq
-\beta \frac{{\cal F}_{n}[\rho(\ovS)]}{N}=
-\intS d\ovS \rho(\ovS)\ln \rho (\ovS)
+\frac{c}{p}\intS d\ovS_{1}d\ovS_{2}\cdots d\ovS_{p}
\rho(\ovS_{1})\rho(\ovS_{2})\cdots \rho(\ovS_{p})
f_{n}(\ovS_{1},\ovS_{2},\ldots,\ovS_{p}).
\label{eq-free-energy-functional-nreplica}
\eeq
where $d\ovS=\prod_{a=1}^{n}d\vS^{a}$
and we introduced a replicated Mayer function,
\beq
f_{n}(\ovS_{1},\ovS_{2},\ldots,\ovS_{p})
=-1+e^{-\beta\sum_{a=1}^{n}V(\vS^{a}_{1},\vS^{a}_{2},\ldots,\vS^{a}_{p})}
=-1+
\prod_{a=1}^{n}
\exp \left[ -\beta 
V\left(
\delta -\frac{1}{\sqrt{M}}
\sum_{\mu=1}^{M}\prod_{l=1}^{p}(S^{a})_{l}^{\mu}
\right)
\right].
\label{eq-def-replicated-mayer}
\eeq

\subsection{Glass order parameter functional}

We look for glassy metastable states which keep the statistical rotational invariance of the 
liquid (paramagnetic) state.
To this end it is natural to consider the overlap matrix given by \eq{eq-def-Qab} as the order parameter,
\beq
\hat{Q}_{ab}=Q_{ab} \qquad 
Q_{ab}=\lim_{N \to \infty}\frac{1}{N}\sum_{i=1}^{N}\frac{1}{M}\sum_{\mu=1}^{M}(S^{a})_{i}^{\mu}(S^{b})_{i}^{\mu}
\eeq
for $a,b=1,2,\ldots,n$.
Note  that $Q_{aa}=1$ due to the normalization of the spins $|\vS_{i}^{a}|^{2}=M$. 

\subsubsection{Variational free-energy}
\label{subsubsec-Variational-free-energy}

Based on the above discussion we expect that 
$\rho(\ovS)$ of the the glassy states, which keeps the statistical rotational invariance of the liquid,
is parametrized solely by the overlap matrix $\hat{Q}$,
\beq
\rho(\ovS)=\rho(\hat{Q}).
\eeq
Since the system is regular and every vertex is exactly equivalent to each other
in our system, it is natural to expect the order parameter does not fluctuate in space.

Similarly we anticipate that the replicated Mayer function can be parametrized as,
\beq
f_{n}(\ovS_{1},\ovS_{2},\ldots,\ovS_{p})=
f_{n}(\hat{Q}_{1},\hat{Q}_{2},\ldots,\hat{Q}_{p}).
\label{eq-replicated-Mayer-Q}
\eeq
so that the variational free-energy functional
given by \eq{eq-free-energy-functional-nreplica} as a whole can be cast into the following rotationally invariant form,
\beqn
-\beta \frac{{\cal F}_{n}[\rho(\hat{Q})]}{N}&=&
-\int d \hat{Q} J(\hat{Q}) \rho(\hat{Q})\ln \rho (\hat{Q})
+\frac{c}{p}  \int \prod_{l=1}^{p} \{ d\hat{Q}_{l}J(\hat{Q}_{l})\rho(\hat{Q}_{l})\}
f_{n}(\hat{Q}_{1},\hat{Q}_{2},\ldots,\hat{Q}_{p})\nonumber \\
&+&\lambda \left ( \int d\hat{Q}J(\hat{Q})\rho(\hat{Q})-1\right)
\label{eq-free-ene-q-v0}
\eeqn
where $d\hat{Q}=\prod_{a<b} dQ_{ab}$. Here $J(\hat{Q})$
is the Jacobian (see below)
and the parameter  $\lambda$ in the last term of \eq{eq-free-ene-q-v0} is
a Lagrange multiplier to enforce the normalization of the spin density.
Note that in the 2nd integral on the r.~h.~s. of \eq{eq-free-ene-q-v0}
we assumed a simply factorized Jacobian $\prod_{l=1}^{p}J(\hat{Q})$
disregarding possible cross-correlations of spins at different sites $l=1,2,\ldots,p$. We comment on the validity of this assumption later.

The Jacobian $J$ is defined as
\beq
J(\hat{Q})\equiv \int d\ovS \prod_{a \leq b}
\delta \left(Q_{ab}-\frac{1}{M}\sum_{\mu=1}^{M}(S^{a})^{\mu}(S^{b})^{\mu}\right)
\label{eq-def-jacobian}
\eeq
Here we have replaced the constrained integral $\intS d\ovS$ by an unconstrained integral
$\int d\ovS \equiv \prod_{\mu=1}^{M} \prod_{a=1}^{n} \int_{-\infty}^{\infty}d(S^{a})^{\mu}$.
This is made possible by setting
\beq
Q_{aa}=1
\label{eq-qaa}
\eeq
for all $a$ in \eq{eq-def-jacobian} so that the normalization condition of the spins
$|\vS^{2}|=M$ is enforced.
Then one can evaluate the Jacobian to find (see Eq.(17) and (78) of  \cite{kurchan2012exact}),
\beq
J(\hat{Q}) =C_{n+1,M} e^{\frac{1}{2}(M-(n+1))\ln {\rm det}\hat{Q}}
\label{eq-jacobian}
\eeq
with $C_{n,M}$ being a numerical prefactor which behaves
for $M \gg 1$ as,
\beq
\ln C_{n,M}=\frac{M}{2}(n-1)\ln (2\pi e)-\frac{M}{2}(n-1)\ln M \qquad M \gg 1.
\label{eq-factor-c-large-M}
\eeq

Minimization of the variational free-energy given by \eq{eq-free-ene-q-v0}
with respect to $\rho(\hat{Q})$, 
\beq
0=\frac{\delta}{\delta \rho(\hat{Q})}
\beta \frac{{\cal F}_{n}[\rho(\hat{Q})]}{N}
\eeq
yields,
\beq
\ln \rho(\hat{Q})=\lambda-1+
c
\int \prod_{l=1}^{p-1} \{ d\hat{Q}_{l}J(\hat{Q}_{l})\rho(\hat{Q}_{l})\}
f_{n}(\hat{Q},\hat{Q}_{1},\ldots,\hat{Q}_{l-1}).
\label{eq-saddle-point-rho}
\eeq
In addition, normalization of the spin density implies,
\beq
1 = \int d\hat{Q}J(\hat{Q})\rho(\hat{Q})
=\int d\hat{Q}\exp\left(\ln J(\hat{Q})+\ln \rho(\hat{Q})\right)
\label{eq-integral-normalization}
\eeq
with
\beq
\ln J(\hat{Q})+\ln \rho(\hat{Q})=\ln C_{n,M}+
  \frac{1}{2}(M-n)\ln {\rm det}\hat{Q}
  +\lambda-1+
  c
\int \prod_{l=1}^{p-1} \{ d\hat{Q}_{l}J(\hat{Q}_{l})\rho(\hat{Q}_{l})\}
f_{n}(\hat{Q},\hat{Q}_{1},\ldots,\hat{Q}_{l-1})
 \label{eq-integral-normalization-2}
\eeq
where we used \eq{eq-jacobian} and  \eq{eq-saddle-point-rho}.

\subsubsection{$M \to \infty$ limit}
\label{subsubsec-large-M}

In the present paper we limit ourselves with the $M \to \infty$ limit
which greatly simplifies the analysis. The first advantage of the $M \to \infty$ limit is that the integrals over $\hat{Q}$ can be done by saddle point method.
For example in \eq{eq-integral-normalization} the saddle point value $\hat{Q}^{*}$ is determined by,
\beq
0=\left. \frac{\delta}{\delta \hat{Q}} \left( \ln J(\hat{Q})+\ln \rho(\hat{Q})
\right)
\right |_{\hat{Q}=\hat{Q}^{*}}
\label{eq-saddle-point-q}
\eeq
Importantly the integrals over $\hat{Q}$ in the
variational free-energy functional
given by \eq{eq-free-ene-q-v0} can also be evaluated by the saddle point method
in the $M \to \infty$ limit and the saddle point should be exactly the same as the one 
given by \eq{eq-saddle-point-q}. This is because in the free-energy
functional given by \eq{eq-free-ene-q-v0}, only the factor $J(\hat{Q})\rho(\hat{Q})$ is
exponentially large in $M$.\cite{kurchan2012exact}

Here let us comment on the validity of our assumption
used in \eq{eq-free-ene-q-v0} that fluctuations of
spins at different sites $l=1,2,\ldots,p$ are uncorrelated which allowed
us to assume a simply factorized Jacobian $\prod_{l=1}^{p}J(\hat{Q})$
in the 2nd integral of \eq{eq-free-ene-q-v0}.
Actually more generally we should write the Jacobian as,
\beqn
K(\{\hat{Q}_{l}\},\{\hat{P}_{ll'}\})&\equiv &\int \prod_{l=1}^{p} d\ovS_{l} \prod_{a \leq b}\prod_{l}
\delta \left((Q_{l})_{ab}-\frac{1}{M}\sum_{\mu=1}^{M}((S_{l})^{a})^{\mu}((S_{l})^{b})^{\mu}\right) \nonumber \\
&&\times \prod_{l<l'}\delta \left((P_{ll'})_{ab}-\frac{1}{M}\sum_{\mu=1}^{M}((S_{l})^{a})^{\mu}((S_{l'})^{b})^{\mu}\right)
\label{eq-def-jacobian2}
\eeqn
where $P_{l,l'}$ represents cross-correlation of the fluctuation of spins
are different sites $l$ and $l'$.
Then similarly to \eq{eq-integral-normalization} we may consider
the normalization of the spin density,
\beq
1=\int \prod_{l}d\hat{Q}_{l}\prod_{l<l'}d\hat{P}_{ll'}
\prod_{l=1}^{p}\rho(\{\hat{Q}_{l}\})K(\{\hat{Q}_{l}\},\{\hat{P}_{ll'}\}).
\eeq
which implies, similarly to \eq{eq-saddle-point-q},
\beqn
0&=&\left. \frac{\delta}{\delta \hat{Q}_{l}} \left( \ln K(\{\hat{Q}_{l}\},\{\hat{P}_{ll'}\})+\sum_{l=1}^{p}\ln \rho(\hat{Q}_{l})
\right)
\right |_{\hat{Q}=\hat{Q}^{*}} \qquad l=1,2,\ldots,p \nonumber \\
0&=&\left. \frac{\delta}{\delta \hat{P}_{l,l'}} \left( \ln K(\{\hat{Q}_{l}\},\{\hat{P}_{ll'}\})+\sum_{l=1}^{p}\ln \rho(\hat{Q}_{l})
\right)
\right |_{\hat{Q}=\hat{Q}^{*}} \qquad l< l'
\label{eq-saddle-point-Q-and-P}
\eeqn
The explicit form of $K(\{\hat{Q}_{l}\},\{\hat{P}_{ll'}\})$ can be worked
out similarly to \eq{eq-jacobian} (see Eq.(40) and (78) of  \cite{kurchan2012exact}) and one finds the 2nd equation of \eq{eq-saddle-point-Q-and-P} yields $\hat{P}_{l,l'}=0$ meaning that the cross-correlation between spin fluctuations
at different sites vanish (see Eq.(62) and (63) of  \cite{kurchan2012exact}).
Thus we can use the factorized form for the Jacobian.

Now inspecting \eq{eq-integral-normalization-2}, 
it is evident that a sensible choice for the scaling
of the connectivity to obtain nontrivial result in the $M \to \infty$ limit is
\beq
c=\alpha M
\label{eq-c-scaling}
\eeq
parametrized by $\alpha>0$. Then using 
\eq{eq-integral-normalization-2}
and \eq{eq-c-scaling}
the saddle point equation given by \eq{eq-saddle-point-q} becomes
\beq
0=\left. \frac{\delta}{\delta \hat{Q}}
\left( \frac{1}{2} \ln {\rm det} \hat{Q}
+\alpha f_{n}(\hat{Q},\hat{Q^{*}},\ldots,\hat{Q^{*}})
\right)
\right |_{\hat{Q}=\hat{Q}^{*}}
=\left. \frac{\delta}{\delta \hat{Q}}
\left( \frac{1}{2} \ln {\rm det} \hat{Q}+\frac{\alpha}{p}
 f_{n}(\hat{Q},\hat{Q},\ldots,\hat{Q})
\right)
\right |_{\hat{Q}=\hat{Q}^{*}}
\label{eq-saddle-point-q-large-M}
\eeq
The value of the Lagrange multiplier $\lambda$ is fixed by  \eq{eq-integral-normalization},
which requires vanishing of $O(M)$ terms in $\ln \rho(\hat{Q^{*}})+\ln J(\hat{Q^{*}})$,
\beq
0=\lambda-1+M\left[\frac{1}{2}n\ln (2\pi e)-\frac{1}{2}n\ln M
  +\frac{1}{2} \ln {\rm det}\hat{Q^{*}}+
  \alpha f_{n}(\hat{Q^{*}},\hat{Q^{*}},\ldots,\hat{Q^{*}})\right]
\eeq
where we used \eq{eq-factor-c-large-M}, \eq{eq-integral-normalization-2}
and dropped sub-leading terms.
Using this result together with \eq{eq-saddle-point-rho},
we find the saddle point value of the variational free-energy given by \eq{eq-free-ene-q-v0} in the $M \to \infty$ limit as,
\begin{eqnarray}
-\beta \frac{{\cal F}_{n}[\rho(\hat{Q^{*}})]}{N}&=&
-\ln \rho(\hat{Q^{*}})+M\frac{\alpha}{p}f_{n}(\hat{Q^{*}},\hat{Q^{*}},\ldots,\hat{Q^{*}})
=1-\lambda-M\frac{\alpha}{p}
(p-1) f_{n}(\hat{Q^{*}},\hat{Q^{*}},\ldots,\hat{Q^{*}}) \nonumber \\
& = & M \left[
  \frac{1}{2}n\ln \left(\frac{2\pi e}{M}\right)+
  \frac{1}{2}\ln {\rm det} \hat{Q^{*}}
  +\frac{\alpha}{p} f_{n}(\hat{Q^{*}},\hat{Q^{*}},\ldots,\hat{Q^{*}})
  \right]
\label{eq-free-energy-n-replicas}
\end{eqnarray}
Note that if we regard $\hat{Q^{*}}$ in the above expression
as a variational parameter, we
find a variational equation which is exactly the same as
\eq{eq-saddle-point-q-large-M}.

Next let us examine the interaction part
of the free-energy to extract the explicit form of the replicated Mayer function.
The interaction part of the free-energy of the replicated system 
reads as (see \eq{eq-free-energy-functional-nreplica}, \eq{eq-def-jacobian}),
\begin{eqnarray}
&&\frac{c}{p}\intS d\ovS_{1}d\ovS_{2}\cdots d\ovS_{p}
\rho(\ovS_{1})\rho(\ovS_{2})\cdots \rho(\ovS_{p})
f_{n}(\ovS_{1},\ovS_{2},\ldots,\ovS_{p}) \nonumber \\
&=& \frac{c}{p} \left [
-1+  \int \prod_{l=1}^{p} \{ d\hat{Q}_{l} J(\hat{Q}_{l})\rho(\hat{Q}_{l})\}
\left \langle
\prod_{a=1}^{n}
\exp \left( -\beta V
\left(
\delta -\frac{1}{\sqrt{M}}
\sum_{\mu=1}^{M}\prod_{l=1}^{p}(S^{a})_{l}^{\mu}
\right)\right)
\right \rangle_{\hat{Q}} \right] \nonumber \\
&=& \frac{c}{p} \left [
  -1
 + \int \prod_{l=1}^{p} \{ d\hat{Q}_{l}J(\hat{Q_{l}})\rho(\hat{Q}_{l})\}
  \int \prod_{a=1}^{n}
  \left\{ \frac{d\kappa_{a}}{2\pi}e^{i\kappa_{a}\delta}
{\cal Z}_{\kappa_{a}}
\right\} \left \langle
\exp\left(\sum_{a=1}^{n}\frac{-i\kappa_{a}}{\sqrt{M}}
\sum_{\mu=1}^{M}
      \prod_{l=1}^{p}
      (S^{a})_{l}^{\mu}\right)
      \right\rangle_{\hat{Q}}
            \right]
  \nonumber \\
  &&
\label{eq-interaction-free-ene}
\end{eqnarray}
where we introduced a Fourier transform,
\beq
Z_{\kappa} \equiv \int dx e^{-i\kappa x} e^{-\beta V(x)}
\eeq
and a short hand notation,
\begin{eqnarray}
  \langle \cdots \rangle_{\hat{Q}}
    \equiv
  \int \prod_{l=1}^{p} \left \{ d\ovS_{l} \frac{1}{J(\hat{Q_{l}})} \prod_{a \leq b}
\delta
  \left((Q_{l})_{ab}-\frac{1}{M}\sum_{\mu=1}^{M}(S_{l}^{a})^{\mu}(S_{l}^{b})^{\mu}\right)  
  \right \} \cdots
\label{eq-av-with-Q}
\end{eqnarray}
In the last equation $Q_{aa}=1$ (\eq{eq-qaa}) to enforce the normalization of the spins $|(\vS^{a})^{2}|=M$ and $\int d\ovS$ is an unconstrained integral. 

For $M \gg 1$ we can evaluate the last factor in
\eq{eq-interaction-free-ene} by performing $1/\sqrt{M}$ expansion,
\begin{eqnarray}
  &&
\ln \left  \langle
       \exp\left(\sum_{a=1}^{n}\frac{-i\kappa_{a}}{\sqrt{M}}
  \sum_{\mu=1}^{M}\prod_{l=1}^{p}(S^{a})_{l}^{\mu}\right)
\right  \rangle_{\hat{Q}} 
     \nonumber \\
&=& \ln \left [
1+
     \sum_{a=1}^{n}\frac{-i\kappa_{a}}{\sqrt{M}}\sum_{\mu=1}^{M}
\prod_{l=1}^{p} \left \langle (S_{l}^{a})^{\mu} \right \rangle_{\hat{Q}}
+\frac{1}{2}
\sum_{a=1}^{n}\sum_{b=1}^{n}(-i\kappa_{a})(-i\kappa_{b})\frac{1}{M}
\prod_{l=1}^{p}
\sum_{\mu,\nu=1}^{M}
\left \langle
(S_{l}^{a})^{\mu}(S_{l}^{b})^{\nu}
 \right \rangle_{\hat{Q}}
+ \ldots \right ]
\nonumber \\
&&  \xrightarrow[M \to \infty]{}
\frac{1}{2} \sum_{a,b=1}^{n}(-i\kappa_{a})(-i\kappa_{b})
\prod_{l=1}^{p}(Q_{l})_{ab}
=     \frac{1}{2}
\sum_{a,b=1}^{n}(-i\kappa_{a})(-i\kappa_{b})
\prod_{l=1}^{p}(Q_{l})_{ab}        \qquad
\end{eqnarray}
Here we used the fact that in the $M \to \infty$ limit,
different components of the spins $S^{\mu}$ become independent from each other.
This can be checked by introducing integral representation of
the $\delta$ function in \eq{eq-av-with-Q} which can be evaluated by the saddle point method in the $M \to \infty$ limit.

To sum up we find the replicated Mayer function in the $M \to \infty$ limit as,
\begin{eqnarray}
&&  f_{n}(\hat{Q}_{1},\hat{Q}_{2},\ldots,\hat{Q}_{p})
     \nonumber \\
&& = -1+\int \prod_{a=1}^{n} \left \{\frac{d\kappa_{a}}{2\pi}e^{i\kappa_{a} \delta}
  \right \}
    \exp \left(
  \frac{1}{2}\sum_{a,b=1}^{n}\prod_{l=1}^{p}(Q_{l})_{ab} 
(-i\kappa_{a})(-i\kappa_{b})
   \right) \prod_{a=1}^{n}
 \int dh_{a} e^{-i\kappa_{a} h_{a}} e^{-\beta V(h_{a})}
  \nonumber \\
    &=&  -1+
  \int \prod_{a=1}^{n} dh_{a}
  \left \{
 \exp \left(
  \frac{1}{2}\sum_{a,b=1}^{n}\prod_{l=1}^{p}(Q_{l})_{ab} 
\frac{\partial^{2}}{\partial h_{a}\partial h_{b}}
\right)
\prod_{a=1}^{n}\delta(\delta-h_{a})
\right \}
\prod_{a} e^{-\beta V(h_{a})}
  \nonumber \\
  &=&  -1+\left. 
    \exp \left(
  \frac{1}{2}\sum_{a,b=1}^{n}
  \prod_{l=1}^{p} (Q_{l})_{ab}
  \frac{\partial^{2}}{\partial h_{a}\partial h_{b}} \right)
  \prod_{a=1}^{n}  e^{-\beta V(\delta+h_{a})}  \right |_{\{h_{a}=0\}}
  \label{eq-replicated-Mayer-result}
\end{eqnarray}
The last equation is obtained by repeating integrations by parts.

Collecting the above results we find the thermodynamic free-energy given by \eq{eq-free-energy}as,
\beq
-\beta \frac{F}{NM}= -\beta f[\hat{Q}^{*}]
  \label{eq-free-energy-Q}
\eeq
with the variational free-energy (more precisely free-entropy),
\begin{eqnarray}
  -\beta f[\hat{Q}]&=&\left. \partial_{n} s_{n}[\hat{Q}]\right |_{n=0} \nonumber \\
    s_{n}[\hat{Q}]& \equiv&   \frac{1}{2}\ln {\rm det} \hat{Q}
  - \frac{\alpha}{p} {\cal F}_{\rm int}[\hat{Q}] \nonumber \\
-{\cal F}_{\rm int}[\hat{Q}] &\equiv & 
  \left.    \exp \left(
    \frac{1}{2}\sum_{a,b=1}^{n}  (Q_{ab})^{p}
  \frac{\partial^{2}}{\partial h_{a}\partial h_{b}} \right)
 \prod_{a=1}^{n}  e^{-\beta V(\delta+h_{a})}  \right |_{\{h_{a}=0\}}    
 \label{eq-free-entropy-Q}
\end{eqnarray}
where we dropped off irrelevant constants.
The saddle point $\hat{Q}^{*}$ is determined by,
\beq
\left. \frac{\delta s [\hat{Q}]}{\delta Q_{ab}} \right |_{\hat{Q}=\hat{Q^{*}}}=0
\eeq
for all $a \neq b$.

\subsubsection{Gaussian ansatz}

Finally let us note that one can check that the above result
can be reproduced by assuming an Gaussian ansatz for the replicated spin density,
\beq
\rho_{\rm Gaussian}(\Vec{S})=\frac{e^{-\frac{1}{2}\sum_{a,b=1}^{n}(Q^{-1})_{ab}\sum_{\mu=1}^{M}(S^{a})^{\mu}(S^{b})^{\mu}}}{\sqrt{2\pi ({\rm det} \hat{Q})^{M}}}
\eeq

The situation is essentially the same as that of 
hardspheres in the large dimensional limit \cite{kurchan2012exact,kurchan2013exact,charbonneau2014exact}.

\section{Eigenvalues of the stability matrix}
\label{sec-Hessian}

Here we analyze the Hessian matrix $M_{a \neq b, c \neq d}$ of the free-energy around the saddle points.
It is a matrix of size $n(n-1) \times n(n-1)$ defined as,
\beq
M_{a \neq b, c \neq d} \equiv -\frac{\partial^{2} s[\hat{Q}]}{\partial Q_{a < b}\partial Q_{c < d}}
\label{eq-Hessian}
\eeq
where $s_{n}[\hat{Q}]$ is the free-entropy defined in  \eq{eq-free-entropy-Q}
which reads,
\begin{eqnarray}
  s_{n}[\hat{Q}]& \equiv&   \frac{1}{2}\ln {\rm det} \hat{Q}
  - \frac{\alpha}{p} {\cal F}_{\rm int}[\hat{Q}] \\
-{\cal F}_{\rm int}[\hat{Q}] &\equiv & 
  \left.    \exp \left(
    \frac{1}{2}\sum_{a,b=1}^{n}  (Q_{ab})^{p}
  \frac{\partial^{2}}{\partial h_{a}\partial h_{b}} \right)
 \prod_{a=1}^{n}   e^{-\beta V(\delta+h_{a})}  \right |_{\{h_{a}=0\}} .
\end{eqnarray}

The Hessian matrix can be naturally written as
sum of the contribution from the entropic part
and interaction part of the free-energy,
\begin{eqnarray}
  M_{a \neq b, c \neq d} &=&M^{\rm ent.}_{a \neq b, c \neq d} +M^{\rm int.}_{a \neq b, c \neq d}
  \label{eq-m-ent-int}
\end{eqnarray}
with
\beqn
M^{\rm ent.}_{a \neq b, c \neq d}  &=&
  -\frac{\partial^{2} }{\partial Q_{a < b}\partial Q_{c< d}}  \frac{1}{2}\ln {\rm det} \hat{Q}=Q^{-1}_{ac}Q^{-1}_{bd}+Q^{-1}_{ad}Q^{-1}_{bc} \label{eq-m-ent} \\
M^{\rm int.}_{a \neq b, c \neq d} &=&\frac{\alpha}{p}
    \frac{\partial^{2} }{\partial Q_{a< b}\partial Q_{c< d}}
         {\cal F}_{\rm int}[\hat{Q}] \nonumber  \\
         & =& \frac{\alpha}{p} \left[
         p(p-1)Q_{a<b}^{p-2}(\delta_{ac}\delta_{bd}+\delta_{ad}\delta_{bc})
         \frac{\partial^{2}}{\partial h_{a}\partial h_{b}}
         \right. \nonumber \\
  &&  \hspace*{2cm} \left.    +p^{2}Q_{a<b}^{p-1}Q_{c<d}^{p-1}
         \frac{\partial^{4}}{\partial h_{a}\partial h_{b}\partial h_{c}\partial h_{d}}
         \right ] \left.
         {\cal F}_{\rm int}[\hat{Q},\{h_{a}\}]
                  \right |_{\{h_{a}=0\}}
         \label{eq-m-int}
         \eeqn
         with
         \beq
        - {\cal F}_{\rm int}[\hat{Q},\{h_{a}\}]
         \equiv
         \exp \left( \frac{1}{2}\sum_{e,f=1}^{n} Q_{ef}^{p}\frac{\partial ^{2}}{\partial h_{e}\partial h_{f}}\right)
         \prod_{a=1}^{n}   e^{-\beta V(\delta + h_{a})}
         \label{eq-cal-F-h}
       \eeq

       \subsection{RS ansatz}
       \label{sec-Hessian-RS}

         Here we analyze the eigenvalues of the Hessian matrix for the
         case of the replica symmetric (RS) solution
characterized by the order parameter matrix of the form given by \eq{eq-RS-ansatz},
which reads as,
\beq
\hat{Q}^{\rm RS}=(1-q)\delta_{ab}+q
\label{eq-RS-ansatz2}
\eeq
The replica symmetry implies the following matrix structure,
\beq
M_{a \neq b, c \neq d}=M_{1}\frac{\delta_{ac}\delta_{bd}+\delta_{ad}\delta_{bc}}{2}
+M_{2}\frac{\delta_{ac}+\delta_{ad}+\delta_{bc}+\delta_{bd}}{4}
+M_{3}
\eeq
from which the eigenvalues of the  Hessian matrix are obtained as \cite{de1978stability,kurchan2013exact},
\begin{eqnarray}
\lambda_{R}&=&M_{1} \\
\lambda_{L}&=&n(n-1)M_{3}+(n-1)M_{2}+M_{1} \xrightarrow[n \to 0]{}M_{1}-M_{2} \\
\lambda_{A}&=&\frac{1}{2}(n-2)M_{2}+M_{1}  \xrightarrow[n \to 0]{}M_{1}-M_{2} 
\end{eqnarray}
         The factors $M_{i}$'s can be decomposed into
         the entropic and interaction parts, $M_{i}=M^{\rm ent}_{i}+M^{\rm int}_{i}$
         like  \eq{eq-m-ent-int}.

\subsubsection{Contribution form the entropic part}

First let us examine the entropic part.
The replica symmetric matrix given by \eq{eq-RS-ansatz2} can be easily inverted to find,
  \beq
(\hat{Q}^{\rm RS})^{-1}_{ab}=\hat{q}\delta_{ab}+\tilde{q}
  \eeq
  with
  \beqn
  \hat{q}&=&\frac{1}{1-q} \\
  \tilde{q}&=&-\frac{q}{1+(n-2)q-(n-1)q^{2}}\xrightarrow[]{n\to 0}-\frac{q}{(1-q)^{2}}
  \eeqn
Using this in \eq{eq-m-ent}, we obtain the entropic contributions as,
  \begin{eqnarray}
   M^{\rm ent.}_{1} &=&\lim_{n \to 0}2(\hat{q})^{2}=\frac{2}{(1-q)^{2}} \\
   M^{\rm ent.}_{2} &=&\lim_{n \to 0}4\hat{q}\tilde{q}=-4\frac{q}{(1-q)^{3}} \\
   M^{\rm ent.}_{3} &=&\lim_{n \to 0}2(\tilde{q})^{2}=2\frac{q^{2}}{(1-q)^{4}}
    \end{eqnarray}

\subsubsection{Contribution form the interaction part}

  Next let us examine the interaction  part \eq{eq-m-int}.
  Within the RS ansatz we find,
  \beqn
\lim_{n \to 0}  M^{\rm int.}_{a \neq b, c \neq d} 
  & =& \lim_{n \to 0}
\exp \left( \frac{q^{p}}{2}\sum_{e,f=1}^{n}
\frac{\partial ^{2}}{\partial h_{e}\partial h_{f}}\right)
  \frac{\alpha}{p} \left[
         p(p-1)q^{p-2}(\delta_{ac}\delta_{bd}+\delta_{ad}\delta_{bc})
         \frac{\partial^{2}}{\partial h_{a}\partial h_{b}} \right. \nonumber \\
&&     \hspace*{3cm} \left.   +  p^{2}q^{2(p-1)}
         \frac{\partial^{4}}{\partial h_{a}\partial h_{b}\partial h_{c}\partial h_{d}}
         \right ] \left.
\prod_{a=1}^{n} g(\delta+h_{a})
\right |_{\{h_{a}=0\}}
\label{eq-hessian-int-RS}
                  \eeqn
where  we used
\beqn
-{\cal F}_{\rm int}[\hat{Q}^{\rm RS},\{h_{a}\}]&=&\exp \left( \frac{1}{2}\sum_{e,f=1}^{n}
((1-q^{p})\delta_{ab}+q^{p})\frac{\partial ^{2}}{\partial h_{e}\partial h_{f}}\right)
\prod_{a=1}^{n}   e^{-\beta V(\delta + h_{a})} \nonumber \\
&=&
\exp \left( \frac{q^{p}}{2}\sum_{e,f=1}^{n}
\frac{\partial ^{2}}{\partial h_{e}\partial h_{f}}\right)
\prod_{a=1}^{n} g(\delta+h_{a})
\eeqn
In the last equation we introduced a shorthanded notation
of the quantity defined in \eq{eq-initial-condition-recursion-g},
\beq
g(h) \equiv g(m_{k+1},h)= \gamma_{1-q_{k}^{p}}\otimes  e^{-\beta V(h)} 
\label{eq-def-g-short}
\eeq
For a convenience let us also introduce a related shorthanded notation
(See \eq{eq-def-f-g})
\beq
f(h)\equiv f(m_{k+1},h)=-\frac{1}{m_{k+1}}\log g(m_{k+1},h)
=-\log g(h) 
\eeq
where $m_{k+1}=1$. Note that $k=0$ for the RS case.

By taking derivatives we find,
\beqn
&& \lim_{n \to 0} \left. 
       \exp \left( \frac{q^{p}}{2}\sum_{e,f=1}^{n}
       \frac{\partial ^{2}}{\partial h_{e}\partial h_{f}}\right)
       \frac{\partial^{2}}{\partial h_{a}\partial h_{b}}
       \prod_{a=1}^{n} g(\delta + h_{a}) \right |_{h_{a}=0}
       \nonumber \\
&&    \hspace*{3cm}   = \left.
       \exp \left( \frac{q^{p}}{2}
       \frac{\partial ^{2}}{\partial h^{2}} \right)
       \left (
  \frac{g'(h)}
       {g(h)}\right)^{2}\right |_{h=\delta}
       = \gamma_{q^{p}}\otimes\left (
  \frac{g'(\delta)}
       {g(\delta)}\right)^{2} 
       \eeqn
       in the last equation we used \eq{eq-formula1}.
Similarly we obtain,
\beqn
&& \lim_{n \to 0} \exp \left( \frac{q^{p}}{2}\sum_{e,f=1}^{n}
       \frac{\partial ^{2}}{\partial h_{e}\partial h_{f}}\right)
\left.  \frac{\partial^{4}}{\partial h_{a}\partial h_{b}\partial h_{c}\partial h_{d}}
       \prod_{a=1}^{n} g(\delta + h_{a})
\right |_{\{h_{a}=0\}} \nonumber \\
&=& \gamma_{q^{p}}\otimes \left \{
(\delta_{ac}\delta_{bd}+\delta_{ad}\delta_{bc})
 \left (
  \frac{g^{''}(\delta)}
       {g(\delta)}\right)^{2}
       \nonumber \right. \\
&&   \left.  +[\delta_{ac}+\delta_{bc}+\delta_{ad}+\delta_{bd}
       -2(\delta_{ad}\delta_{bc}+\delta_{ac}\delta_{bd})]
 \left[
  \frac{g^{''}(\delta)}
       {g(\delta)}
      \left(   \frac{g'(\delta)}
           {g(\delta)}\right)^{2}\right] \right. \nonumber \\
&& \left.  +[1-(\delta_{ac}+\delta_{bc}+\delta_{ad}+\delta_{bd})+(\delta_{ad}\delta_{bc}+\delta_{ac}\delta_{bd})] \left (
  \frac{g'(\delta)}
       {g(\delta)}\right)^{4} \right \}
       \eeqn
       From the above result we find the contributions by the interaction part as,
         \begin{eqnarray}
           -M^{\rm int.}_{1} &=& \frac{2\alpha}{p}  \left[
           p(p-1)q^{p-2}
             \gamma_{q^{p}}\otimes \left (\frac{g'(\delta)}{g(\delta)}\right)^{2}
             \right. \nonumber \\
&&             +(pq^{p-1})^{2} \gamma_{q^{p}}  \otimes\left\{\left (\frac{g''(\delta)}{g(\delta)} \right)^{2}
                          \right.
             \left. \left. 
             -2     \frac{g^{''}(\delta)}
       {g(\delta)}
      \left(   \frac{g'(\delta)}
           {g(\delta)}\right)^{2}
             +             \left (
                \frac{g'(\delta)}{g(\delta)}\right)^{4}
                \right\}             \right] \nonumber \\
           & =& \frac{2\alpha}{p}  \left[
             p(p-1)q^{p-2}
             \gamma_{q^{p}}\otimes (f'(\delta))^{2}
             +(pq^{p-1})^{2} \gamma_{q^{p}}  \otimes (f''(\delta))^{2}
             \right] \nonumber \\
           -M^{\rm int.}_{2} &=& \frac{4\alpha}{p}  
             (pq^{p-1})^{2} \gamma_{q^{p}}  \otimes \left\{
               \frac{g''(\delta)}
       {g(\delta)}
      \left(   \frac{g'(\delta)}
           {g(\delta)}\right)^{2}
             - \left (\frac{g'(\delta)}{g(\delta)}\right)^{4}
             \right\}
                          =
             \frac{4\alpha}{p} (pq^{p-1})^{2} \gamma_{q^{p}}  \otimes
             (-f''(\delta)(f'(\delta))^{2}) \nonumber \\
            - M^{\rm int.}_{3} &=&\frac{\alpha}{p} (pq^{p-1}))^{2}
            \gamma_{q^{p}}\otimes \left (\frac{g'(\delta)}{g(\delta)}\right)^{4}
            = \frac{\alpha}{p} (pq^{p-1}))^{2}
            \gamma_{q^{p}}\otimes (f'(\delta))^{4}
            \label{eq-M1-M2-M3-int-RS}
         \end{eqnarray}

         \subsubsection{Replicon eigenvalue}
         \label{subsec-replicon-RS}

Summing up the above results we find the replicon eigenvalue which is responsible for the
RSB instability of the RS ansatz as,
\beq
\lambda_{\rm R}=
\frac{2}{(1-q_{0})^{2}}
-2\frac{\alpha}{p}
\int dh \frac{e^{-\frac{h^{2}}{2q^{p}}}}{\sqrt{2\pi q^{p}}}
  \left[
           p(p-1)q^{p-2}
          (f'(\delta-h))^{2}
           +(pq^{p-1})^{2}
           (f''(\delta-h))^{2}  \right]
  \label{eq-replicon-RS}
\eeq

\subsection{$k$-RSB ansatz}
\label{subsec-Hessian-kRSB}

         Next let us analyze the case of $k$-step replica symmetry breaking  solution with the ansatz given by \eq{eq:parisi-matrix}.
         Within the $k$-RSB ansatz, $n$ replicas are divided into
         $n/m_{1}$ groups of size $m_{1}$ and each of the latter is divided into $m_{1}/m_{2}$ groups of size $m_{2}$, and so on. Finally we find $n/m_{k}$ groups of size $m_{k}$.
         Within each of the groups of size $m_{k}$, the replica symmetry remains.
As we did in the 1-RSB case, here
we only analyze stability of the replica symmetry within such a
most inner-core group.
Thus we just consider the Hessian matrix 
$M_{a \neq b, c \neq d}$ given by \eq{eq-Hessian} assuming that all indexes
$a$,$b$,$c$,$d$ are in the same most-inner core replica group of size $m_{k}$,
which we denote as ${\cal C}$ in the following.

\subsubsection{Contributions from the interaction term}

Let us first examine the contributions from the interaction term.
Within the $k$-RSB ansatz, the interaction part of
the Hessian matrix $M^{\rm int}_{a \neq b, c \neq d}$ given by \eq{eq-m-ent-int}
for $a$,$b$,$c$,$d$ in the same most-inner core replica group ${\cal C}$
becomes, using \eq{eq-Fint-k-RSB} and \eq{eq-Fint-recursive-construction},
  \beqn
&-&\lim_{n \to 0}  M^{\rm int., {\cal C}}_{a \neq b, c \neq d } 
   = \lim_{n \to 0}\prod_{l=0}^{k}
\exp \left( \frac{\Lambda_{l}}{2}\sum_{e,f=1}^{n}I_{ef}^{m_{l}}
\frac{\partial ^{2}}{\partial h_{e}\partial h_{f}}\right)
\left (\prod_{a \notin {\cal C}} g(\delta+h_{a}) \right) \nonumber \\
&&   \frac{\alpha}{p} \left[
         p(p-1)q_{k}^{p-2}(\delta_{ac}\delta_{bd}+\delta_{ad}\delta_{bc})
         \frac{\partial^{2}}{\partial h_{a}\partial h_{b}} \right.
         \left.
         +  p^{2}q_{k}^{2(p-1)}
         \frac{\partial^{4}}{\partial h_{a}\partial h_{b}\partial h_{c}\partial h_{d}}
         \right ] \left.
\prod_{a \in {\cal C}} g(\delta+h_{a})
\right |_{\{h_{a}=0\}} \nonumber \\
  & =& \lim_{m_{0} \to 0} \gamma_{\Lambda_{0}}\otimes \left \{
g^{m_{0}/m_{1}-1}(m_{1},\delta)\gamma_{\Lambda_{1}} \otimes \left \{
g^{m_{1}/m_{2}-1}(m_{2},\delta)\gamma_{\Lambda_{2}} \otimes \left \{
\cdots
\right. \right. \right. \nonumber \\
&&  \left. 
\cdots g^{m_{k-1}/m_{k}-1}(m_{k},h) \gamma_{\Lambda_{k}} \otimes \left \{
g^{m_{k}}(m_{k+1},h)
\left[
  S_{1}(h)\frac{\delta_{ac}\delta_{bd}+\delta_{ad}\delta_{bc}}{2}
  \right.  \right.\right.
  \nonumber \\
  &&
  \left.  \left.\left.
\hspace*{7cm}+S_{2}(h)\frac{\delta_{ac}+\delta_{ad}+\delta_{bc}+\delta_{bd}}{4}
+S_{3}(h)\right] \right \} 
\right |_{h=\delta} \nonumber \\
&=&\int dh P(m_{k},h)
\left[
  S_{1}(h)\frac{\delta_{ac}\delta_{bd}+\delta_{ad}\delta_{bc}}{2}
+S_{2}(h)\frac{\delta_{ac}+\delta_{ad}+\delta_{bc}+\delta_{bd}}{4}
+S_{3}(h)\right] 
\eeqn
In the last equation we used \eq{eq-p0j-expansion} derived
in appendix \ref{sec-appendix-Poj} and \eq{eq-def-P}.
In the last equation we introduced,
            \begin{eqnarray}
           S_{1}(h) &=& \frac{2\alpha}{p}  \left[
           p(p-1)q^{p-2}
          (f'(h))^{2}
           +(pq^{p-1})^{2}
           (f''(h))^{2}  \right]  \label {eq-S1} \\
           S_{2}(h) &=& \frac{4\alpha}{p}  
           (pq^{p-1})^{2}
           (-f''(h)(f'(h))^{2})
             \\
            S_{3}(h) &=&\frac{\alpha}{p} (pq^{p-1})^{2}
            (f'(h))^{4}
            \end{eqnarray}
            Thus we find the contributions from the interaction term as,
            \begin{eqnarray}
           -M^{\rm int.}_{1} &=& \int dh P(m_{k},h)S_{1}(h) \label{eq-M1-S1}\\
           -M^{\rm int.}_{2} &=& \int dh P(m_{k},h)S_{2}(h)\\
                      -M^{\rm int.}_{3} &=& \int dh P(m_{k},h)S_{3}(h)
            \end{eqnarray}
                  The above formula
                  reduces to the RS one given by \eq{eq-hessian-int-RS}
                  for $k=0$ case as it should.

                   \subsubsection{Contributions from the entropic term}

                  Next let us examine the entropic contribution.
                  To this end it is useful to note first that
the entropic contribution to the $k$-RSB free-energy
can also be expressed in a recursive manner exploiting the hierarchical structure of the order parameter,
                  \beqn
&&                  \frac{1}{2}\ln {\rm det}\hat{Q}=-\ln I(\hat{Q}) \\
&&                  I(\hat{Q})\equiv\left.
                  \int \prod_{a=1}^{n} d\phi_{a}
                  e^{-\frac{1}{2}\sum_{a,b=1}^{n}\phi_{a}Q_{ab}\phi_{b}+\sum_{a}h_{a}\phi_{a}}
                  \right |_{h_{a}=0}  \nonumber \\
&&                  =
\prod_{l=0}^{k}e^{-\frac{\Lambda_{l}}{2}\sum_{ab}I^{m_{l}}_{ab}\frac{\partial^{2}}{\partial h_{a}\partial h_{b}}}
\prod_{a=1}^{n}g_{\rm e}(m_{k+1},h_{a}) \qquad 
\label{eq-entropic-hierarchical}
                  \eeqn
                  where  (see \eq{eq:parisi-matrix})
                  \beq
                  \Lambda_{0}=q_{0} \qquad
                  \Lambda_{i}=q_{i}-q_{i-1}
                  \eeq
                  and 
                  \beq
                  g_{\rm e}(m_{k+1},h)\equiv \int \frac{d\phi}{\sqrt{2\pi}}
                  e^{-\frac{1}{2}(1-q_{k})\phi^{2}+h\phi}
                  =\frac{e^{\frac{h^{2}}{2(1-q_{k})}}}{\sqrt{1-q_{k}}}
                  \label{eq-boundary-g-e}
                  \eeq
                  Comparing the above expressions with \eq{eq-Fint-k-RSB} we find the entropic term is expressed very similarly as the interaction term.
                  We just need to put $p=1$
                  in \eq{eq-def-lambda} (see also  \eq{eq-def-Lambda})
                  and replace $g(m_{k+1},h)$ by $g_{\rm e}(m_{k+1},h)$ defined above. Again we can define a family of functions
                  $g_{\rm e}(m_{l},h)$ for $l=0,1,\ldots,k$ through
                  \eq{eq-recursion-g-0}
                  with the boundary condition 
                  given by \eq{eq-boundary-g-e}. Then we can write 
                  $\ln I(\hat{Q})=g_{\rm e}(m_{0},0)$.
                  In addition we can introduce  $f_{\rm e}(m_{i},h)\equiv -(1/m_{i})\ln g_{\rm e}(m_{i},h)$ (\eq{eq-def-f-g}) and
                  $P^{\rm e}_{i,j}(y,h) \equiv \delta f_{\rm e}(m_{i},y)/\delta f_{\rm e}(m_{j},h)$ (\eq{eq-def-pij})
                  and $P^{\rm e}(m_{j},h)\equiv P^{\rm e}_{0,j+1}(0,h)$ (\eq{eq-def-P}).

                  We have to note however that sign in front of
                  $\Lambda_{l}$ in \eq{eq-entropic-hierarchical} is {\it negative}.
                  Thus we have to understand the operator $\otimes$
                  which appears in equations like
\eq{eq-recursion-g-0}
                  not in the Gaussian convolution form
                  given by \eq{eq-formula2} but 
                  in the original differential
                  form  given by \eq{eq-formula1}. 
                  
                  Using the above results we can write the entropic part of the sub-matrix of the Hessian matrix associated with a most inner core group ${\cal C}$ as,
                    \beqn
                   \lim_{n \to 0}  M^{\rm ent., {\cal C}}_{a \neq b, c \neq d }&=&
                    -\lim_{n \to 0}\frac{\partial^{2}}{\partial Q_{a\neq b}\partial Q_{c\neq d}}\frac{1}{2}\ln {\rm det}\hat{Q}\nonumber \\
                    &=&\lim_{n \to 0}Q^{-1}_{a\neq b}Q^{-1}_{c\neq d}
                    +\lim_{n \to 0}\frac{\partial^{2}}{\partial Q_{a\neq b}\partial Q_{c\neq d}} I(\hat{Q})
                    \eeqn

                                                    Note that the 1st term on the r.h.s of the last equation contributes only to $M_{3}^{\rm ent.}$. 
For the replicon mode we need $M^{\rm ent.}_{1}$ which is obtained as,
\beq
M^{\rm ent.}_{1}=\frac{2}{(1-q_{k})^{2}}.
\eeq
This can be obtained using \eq{eq-M1-S1} and \eq{eq-S1}
with the following modifications:  $p \to 1$,
$f''(h)  \to f''_{\rm e}(m_{k+1},h)=-1/(1-q_{k})$ which can be
obtained from  \eq{eq-boundary-g-e}, and $-\alpha/p \to 1$.

\subsubsection{Replicon eigenvalue}

Summing up the above results we find the replicon eigenvalue
which is responsible for the RSB instability of
a most-inner core replica group in the $k$-RSB ansatz as,
\beq
\lambda_{\rm R}=
\frac{2}{(1-q_{k})^{2}}
-2\frac{\alpha}{p}
\int dh P(m_{k},h)
  \left[
           p(p-1)q^{p-2}
          (f'(m_{k+1},h))^{2}
           +(pq^{p-1})^{2}
           (f''(m_{k+1},h))^{2}  \right]
  \label{eq-replicon-kRSB}
\eeq

            \section{Derivation of \eq{eq-dev-f-lambda} }
            \label{sec-dev-f-lambda}

            Here we show the derivation of \eq{eq-dev-f-lambda}.
            Let us begin with the case $1 \le i=j \le k$.
            Using the recursion formula given by \eq{eq-recursion-f} we find,
\begin{eqnarray}
  \partial_{\lambda_{i}} f(m_{i},y)=
  e^{m_{i}f(m_{i},y)}\int {\cal D}z_{i}
  e^{-m_{i}f(m_{i+1},\Xi_{i})}  \partial_{\lambda_{i}} f(m_{i+1},\Xi_{i})
\end{eqnarray}
where $\Xi_{i}=y-\sqrt{\lambda_{i}-\lambda_{i-1}}z_{i}$ and with
$\Xi_{i+1}=\Xi_{i}-\sqrt{\lambda_{i+1}-\lambda_{i}}z_{i+1}$ we find,
\begin{eqnarray}
  \partial_{\lambda_{i}} f(m_{i+1},y)=
  e^{m_{i+1}f(m_{i+1},y)}\int {\cal D}z_{i+1}
  e^{-m_{i+1}f(m_{i+2},\Xi_{i+1})}
   \partial_{\lambda_{i}} f(m_{i+2},\Xi_{i+1})
\end{eqnarray}
Then by noting that $\Xi_{i+1}=y-\sqrt{\lambda_{i}-\lambda_{i-1}}z_{i}-\sqrt{\lambda_{i+1}-\lambda_{i}}z_{i+1}$ we find,
\beq
\partial_{\lambda_{i}}f(m_{i+2},\Xi_{i+1})
=f'(m_{i+2},\Xi_{i+1})
\left(\frac{1}{2} \frac{z_{i+1}}{\sqrt{\lambda_{i+1}-\lambda_{i}}}
-\frac{1}{2} \frac{z_{i}}{\sqrt{\lambda_{i}-\lambda_{i-1}}}
\right)
\eeq
where the dash represents the partial derivative with respect to the 2nd argument $\partial_{h}f(x,h)=f'(x,h)$.

Collecting the above results,
we find for $i=0,2,\ldots,k$,
\begin{eqnarray}
\partial_{\lambda_{i}}f(m_{i},y) &=&\frac{1}{2}
(m_{i+1}-m_{i})e^{m_{i}f(m_{i},y)}
\int {\cal D} z_{i} e^{-m_{i}f(m_{i+1},\Xi_{i})}
\left(  f'(m_{i+1},\Xi_{i})\right)^{2}\nonumber \\
&=& \frac{1}{2} (m_{i+1}-m_{i})
\int dh P_{i,i+1}(y,h) \left(f'(m_{i+1},h)\right)^{2}
\label{eq-dev-f-lambda-one-step}
\end{eqnarray}
To derive the 1st equation 
we performed integrations by parts.
In 2nd equation we used the identity given by \eq{eq-pij-one-step}.
One can naturally
generalize the above analysis and find for $1 \le i \le  j \le k$,
\beq
\partial_{\lambda_{j}}f(m_{i},y)
=\int dh \frac{\delta f(m_{i},y)}{\delta f(m_{j},h)}
\partial_{\lambda_{j}}f(m_{j},h)
=\frac{1}{2} (m_{j}-m_{j+1})
\int dh
P_{ij+1}(y,h)
\left(f'(m_{j+1},h)\right)^{2}.
\eeq
which is the desired result given by \eq{eq-dev-f-lambda}.
In the last equation we used \eq{eq-def-pij},
\eq{eq-dev-f-lambda-one-step} and the identity given by \eq{eq-pij-chain-rule}.

\section{Expansion of $P_{o,j}(h,y)$}
\label{sec-appendix-Poj}

Using the recursion formula given by \eq{eq-recursion-g}
(see also the expansion displayed in \eq{eq-Fint-recursive-construction})
we find,
  \beqn
 \frac{\delta g(m_{0},h)}{\delta g(m_{j},y)}
  &=& \left. \frac{m_{0}}{m_{1}}
  \gamma_{\Lambda_{0}}\otimes\left \{ g^{m_{0}/m_{1}-1}(m_{1},h) \frac{\delta g(m_{1},h)}{\delta g(m_{j},y)} \right \} \right |_{h=\delta}\nonumber \\
  &=& \left. \frac{m_{0}}{m_{1}}  \frac{m_{1}}{m_{2}} 
  \gamma_{\Lambda_{0}}\otimes\left \{ g^{m_{0}/m_{1}-1}(m_{1},h)
  \gamma_{\Lambda_{1}} \otimes \left \{
g^{m_{1}/m_{2}-1}(m_{2},h)
\frac{\delta g(m_{2},h)}{\delta g(m_{j},y)} \right \}\right \} \right |_{h=\delta}\nonumber  \\
&=& \ldots \nonumber \\
&=& \frac{m_{0}}{m_{1}}  \frac{m_{1}}{m_{2}} \cdots  \frac{m_{j-1}}{m_{j}} 
  \gamma_{\Lambda_{0}}\otimes\left \{ g^{m_{0}/m_{1}-1}(m_{1},h)
  \gamma_{\Lambda_{1}} \otimes \left \{
g^{m_{1}/m_{2}-1}(m_{2},h)\gamma_{\Lambda_{2}} \otimes \cdots  \right. \right. \nonumber \\
&&  \left. \left. \cdots \gamma_{\Lambda_{j-1}}\otimes\left \{
g^{m_{j-1}/m_{j}-1}(m_{j},h)
\delta(h-y)
\right \}\right \}\right \}
\label{eq-del-F-del-g1}
\eeqn
On the other hand using \eq{eq-def-pij} and \eq{eq-def-f-g}
we find,
  \beq
  \frac{\delta g(m_{0},h) }{\delta g(m_{j},y)}
  =\frac{m_{0}}{m_{j}}\frac{g(m_{0},\delta)}{g(m_{j},y)} P_{0,j}(\delta,y).
  \label{eq-del-F-del-g2}
  \eeq
Combining the above results we find,
         \beqn
         P_{0,j}(h,y)
         &=& 
  \gamma_{\Lambda_{0}}\otimes\left \{ g^{m_{0}/m_{1}-1}(m_{1},h)
  \gamma_{\Lambda_{1}} \otimes \left \{
g^{m_{1}/m_{2}-1}(m_{2},h)\gamma_{\Lambda_{2}} \otimes \cdots  \right. \right. \nonumber \\
&&  \left. \left. \cdots \gamma_{\Lambda_{j-1}}\otimes\left \{
g^{m_{j-1}/m_{j}}(m_{j},h)
\delta(h-y)
\right \}\right \}\right \}.
\label{eq-p0j-expansion}
\eeqn
In the last equation we have took the limit $m_{0}=n \to 0$ so that
$\lim_{m_{0} \to 0}g(m_{0},h) \to 1$.

\end{appendix}



 \bibliographystyle{SciPost_bibstyle} 
\bibliography{ref_yoshino}

\nolinenumbers

\end{document}